%% file: main.tex
\newif\ifreport
\reporttrue

\documentclass{lmcs}
\pdfoutput=1

\usepackage{lastpage}
\lmcsdoi{21}{2}{20}
\lmcsheading{}{\pageref{LastPage}}{}{}%
{Jan.~25,~2024}{Jun.~05,~2025}{}

\input{macro}

\crefname{exa}{Example}{Examples}

\usetikzlibrary{positioning}

\crefname{defi}{Definition}{Definitions}
\Crefname{defi}{Definition}{Definitions}
\crefname{cor}{Corollary}{Corollaries}
\Crefname{cor}{Corollary}{Corollaries}
\crefname{lem}{Lemma}{Lemma}
\Crefname{lem}{Lemma}{Lemma}
\crefname{prop}{Proposition}{Propositions}
\Crefname{lem}{Proposition}{Propositions}

\begin{document}

\title[Relating Reversible Petri Nets and Reversible Event Structures]
{Relating Reversible Petri Nets and Reversible Event Structures,
  categorically}

   \thanks{This work has been supported by the Italian MUR PRIN
    2022 project \emph{DeKLA}, the INdAM-GNCS project CUP\_E55F22000270001
    \emph{Propriet\`a Qualitative e Quantitative di Sistemi Reversibili}, and
    the European Union - NextGenerationEU program Research and Innovation
    Program PE00000014 \emph{SEcurity and RIghts in the CyberSpace} (SERICS),
    projects \emph{Secure and TRaceable Identities in Distributed
      Environments} (STRIDE) and \emph{Securing softWare frOm first
      PrincipleS} (SWOPS), the EU H2020 RISE programme under the  Marie
  Sk\l{}odowska-Curie grant agreement 778233, UBACyT projects 20020170100544BA
  and 20020170100086BA}

\author[H. Melgratti]{Hern\'an Melgratti\lmcsorcid{0000-0003-0760-0618}}[a]

\author[C. A. Mezzina]{Claudio Antares
  Mezzina\lmcsorcid{0000-0003-1556-2623}}[b]

\author[G. M. Pinna]{G. Michele Pinna\lmcsorcid{0000-0001-8911-1580}}[c]

\address{ICC - Universidad de Buenos Aires - Conicet, Argentina}
\address{Dipartimento di Scienze Pure e Applicate, Universit\`a di Urbino, Italy}
\address{ Universit\`a di Cagliari, Italy}

\begin{abstract}
  Causal nets (CNs) are Petri nets where causal dependencies are modelled via
  inhibitor arcs. They play the role of occurrence nets when representing the
  behaviour of a concurrent and distributed system, even when reversibility is
  considered. In this paper we extend CNs to account also for asymmetric
  conflicts and study (i) how this kind of nets, and their reversible
  versions, can be turned into a category; and (ii) their relation with the
  categories of reversible asymmetric event structures.
\end{abstract}

\maketitle

\input{intro}

\input raes

\input racn

\input coreflection

\input applications
\input conclusions

\subsection*{Acknowledgments} We would like to thanks the reviewers for their useful criticisms that helped us to improve the paper.

\bibliography{biblio}
\bibliographystyle{alphaurl}
\appendix
\section{Appendix}
\input{appendix}

\end{document}

%% file: macro.tex
\usepackage[normalem]{ulem}
\usepackage{eqparbox}
\usepackage{extarrows}
\usepackage{latexsym, stmaryrd, graphicx, color}
\usepackage{tikz}
\usetikzlibrary{arrows,arrows.meta,shapes,snakes,automata,backgrounds,petri,
decorations.pathmorphing,positioning,fit}
\usepackage{proof}
\usepackage{mathtools}
\usepackage{amsfonts,amssymb}
\usepackage{listings}

\usepackage{caption}
\usepackage{subcaption}
\usepackage[hypertexnames=false]{hyperref}
\usepackage{cleveref}
\usepackage{xspace}
\usepackage{thm-restate}
\usepackage{xcolor}
\usepackage{listings}
\usepackage{enumitem}
\usepackage[utf8]{inputenc}

\makeatletter
\@ifpackageloaded{stix}{%
}{%
  \DeclareFontEncoding{LS2}{}{\noaccents@}
  \DeclareFontSubstitution{LS2}{stix}{m}{n}
  \DeclareSymbolFont{stix@largesymbols}{LS2}{stixex}{m}{n}
  \SetSymbolFont{stix@largesymbols}{bold}{LS2}{stixex}{b}{n}
  \DeclareMathDelimiter{\lBrace}{\mathopen} {stix@largesymbols}{"E8}%
                                            {stix@largesymbols}{"0E}
  \DeclareMathDelimiter{\rBrace}{\mathclose}{stix@largesymbols}{"E9}%
                                            {stix@largesymbols}{"0F}
}
\makeatother

\newcommand{\zero}{\ensuremath{\mathsf{0}}}

\newcommand{\trans}[1]{\ensuremath{\,[\/{#1}\/\rangle}\,}
\newcommand{\pre}[1]{\ensuremath{{}^{\bullet}{#1}}}
\newcommand{\post}[1]{\ensuremath{{#1}^{\bullet}}}

\newcommand{\hist}[1]{\ensuremath{\lfloor #1 \rfloor}}
\newcommand{\histtwo}[2]{\ensuremath{\lfloor #1 \rfloor_{#2}}}

\newcommand{\nat}{\ensuremath{\mathbb{N}}}

\newcommand{\flt}[1]{\ensuremath{[\![{#1}]\!]}}
\newcommand{\pes}{\nuevamacro{\textsc{pes}}}
\newcommand{\aes}{\nuevamacro{\textsc{aes}}}
\newcommand{\raes}{\nuevamacro{r\textsc{aes}}}

\newcommand{\multisetenum}[1]{\lBrace{#1\rBrace}}
\newcommand{\multisetcomp}[2]{\lBrace{#1} \mid {#2}\rBrace}
\newcommand{\setenum}[1]{\{#1\}}
\newcommand{\setcomp}[2]{\{{#1} \mid {#2}\}}
\newcommand{\card}[1]{\ensuremath{|#1|}}
\newcommand{\reachMark}[1]{\ensuremath{\mathcal{M}_{#1}}}

\newcommand{\firseq}[2]{\ensuremath{\mathcal{R}^{#1}_{#2}}}
\newcommand{\states}[1]{\ensuremath{\mathsf{St}(#1)}}

\newcommand{\lead}[1]{\ensuremath{\mathit{lead}(#1)}}
\newcommand{\start}[1]{\ensuremath{\mathit{start}(#1)}}
\newcommand{\fs}{\textsf{fs}}

\newcommand{\MC}[1]{\ensuremath{\sim}}

\newcommand{\Conf}[2]{\ensuremath{\mathsf{Conf}_{{#2}}(#1)}}

\newcommand{\inib}[1]{\ensuremath{{}^{\circ}{#1}}}

\newcommand{\len}[1]{\ensuremath{\mathit{len}(#1)}}

\newcommand{\un}[1]{\underline{#1}}

\newcommand{\anR}{U}
\newcommand{\anr}{u}

\newcommand{\rcn}{r\textsc{cn}}

\newcommand{\Inib}[2]{\ensuremath{\!~^{\circ}{#1}_{#2}}}

\newcommand{\cnconf}{{\natural}}

\newcommand{\rpes}{{r\pes}\xspace}

\newcommand{\peses}{{\pes{}s}\xspace}
\newcommand{\inet}{{\sc{ipt}}\xspace}
\newcommand{\rcns}{{\rcn}s\xspace}
\newcommand{\es}{\nuevamacro{\text{\sc es}}}
\newcommand{\eses}{\nuevamacro{\text{\sc es}s}}

\newcommand{\nuevamacro}[1]{{#1}\xspace}

\newcommand{\acn}{\nuevamacro{\text{\sc acn}}}

\newcommand{\rpacn}{\nuevamacro{r\text{\sc acn}}}
\newcommand{\racn}{\nuevamacro{r\text{\sc acn}}}
\newcommand{\racns}{\nuevamacro{r\text{\sc acn}s}}

\newcommand{\aeses}{\nuevamacro{\text{\sc aes}s}}
\newcommand{\raeses}{\nuevamacro{r\textsc{aes}s}}

\newcommand{\acns}{\nuevamacro{\text{\sc acn}s}}
\newcommand{\paca}{\nuevamacro{\text{\sc acn}}}
\newcommand{\pacas}{\nuevamacro{\text{\sc acn}s}}
\newcommand{\aca}{\nuevamacro{\text{\sc acn}}}
\newcommand{\acas}{\acns}

\newcommand{\pn}{\nuevamacro{\text{\sc pn}}}
\newcommand{\pns}{\nuevamacro{\text{\sc pn}s}}

\newcommand{\ca}{\nuevamacro{\text{\sc cn}}}

\newcommand{\parcn}{\nuevamacro{r\text{\sc acn}}}
\newcommand{\parcns}{\nuevamacro{r\text{\sc acn}s}}
\newcommand{\inets}{\nuevamacro{\text{\sc ipt}s}}

\newcommand{\toberemoved}[1]{}

\makeatletter
\DeclareRobustCommand{\cev}[1]{%
  \mathpalette\do@cev{#1}%
}
\newcommand{\do@cev}[2]{%
  \fix@cev{#1}{+}%
  \reflectbox{$\m@th#1\overrightarrow{\reflectbox{$\fix@cev{#1}{-}\m@th#1#2\fix@cev{#1}{+}$}}$}%
  \fix@cev{#1}{-}%
}
\newcommand{\fix@cev}[2]{%
  \ifx#1\displaystyle
    \mkern#23mu
  \else
    \ifx#1\textstyle
      \mkern#23mu
    \else
      \ifx#1\scriptstyle
        \mkern#22mu
      \else
        \mkern#22mu
      \fi
    \fi
  \fi
}

\makeatother

\newcommand{\fwdset}[1]{\overline{{T_{#1}}}}
\newcommand{\bwdset}[1]{\underline{{T_{#1}}}}
\newcommand{\afwd}{{t}}
\newcommand{\abwd}{{\underline t}}

\newcommand{\arcn}[2]{{#1^{#2}}}
\newcommand{\resarcn}[3]{{#1^{#2}_{#3}}}

\newcommand{\composizione}[2]{#1;#2}
\newcommand{\mrflt}[1]{\flt{#1}}

\newcommand{\acntoaes}[1]{\mathcal{E}_{#1}}
\newcommand{\aestoacn}[1]{\mathcal{N}_{#1}}
\newcommand{\pprec}{\ensuremath{\prec\!\!\prec}}

\newcommand{\rcnprevent}{\textcolor{black}{\leadsto}}
\newcommand{\preventedby}{\textcolor{black}{\ \mbox{\reflectbox{$\leadsto$}}\ }}

\lstdefinelanguage{customc}{
  belowcaptionskip=1\baselineskip,
  breaklines=true,
  xleftmargin=\parindent,
  language=C,
  showstringspaces=false,
  basicstyle=\footnotesize\ttfamily,
  keywordstyle=\bfseries\color{green!40!black},
  commentstyle=\itshape\color{purple!40!black},
  identifierstyle=\color{blue},
  stringstyle=\color{orange},
}

\lstdefinestyle{INLINE}{
}

%% file: intro.tex

\section{Introduction}

\begin{figure}[t]
   \begin{subfigure}{.15\textwidth}
      \vspace*{.1cm}
      \scalebox{0.70}{\input{figures/exintro-es.tex}}
    \caption{\pes $\mathsf{P}$}\label{fig:exintro-es}
  \end{subfigure}
  \begin{subfigure}{.35\textwidth}
      \vspace*{.1cm}
      \centerline{\scalebox{0.70}{\input{figures/exintro-conf.tex}}}
  \caption{Transition system of $P$}\label{fig:exintro-conf}
  \end{subfigure}
  \quad
   \begin{subfigure}{.18\textwidth}
      \centerline{\scalebox{0.70}{\input{figures/exintro-aes-es.tex}}}
    \caption{\aes $\mathsf{G}$}\label{fig:exintro-aes-es}
  \end{subfigure}
  \begin{subfigure}{.18\textwidth}
      \centerline{\scalebox{0.70}{\input{figures/exintro-aes.tex}}}
  \caption{\aes $\mathsf{G}'$}\label{fig:exintro-aes}
  \end{subfigure}

  \vspace*{1cm}
     \begin{subfigure}{.35\textwidth}
      \vspace*{.1cm}
      \centerline{\scalebox{0.70}{\input{figures/exintro-conf-aes.tex}}}
     \caption{Transition system of $\mathsf{G}'$}\label{fig:exintro-aesconf}
  \end{subfigure}
  \begin{subfigure}{.20\textwidth}
      \centerline{\scalebox{0.70}{\input{figures/exintro-raes.tex}}}
  \caption{\raes{} $\mathsf{H}$}\label{fig:exintro-raes}
  \end{subfigure}
   \begin{subfigure}{.40\textwidth}
      \centerline{\scalebox{0.70}{\input{figures/exintro-rconf.tex}}}
  \caption{Transition system of ${\mathsf{H}}$}\label{fig:exintro-rconf-raes}
  \end{subfigure}
  \caption{A simple \pes{}, \aes{}, and \raes{} along with their associated
    transition systems.}
\end{figure}
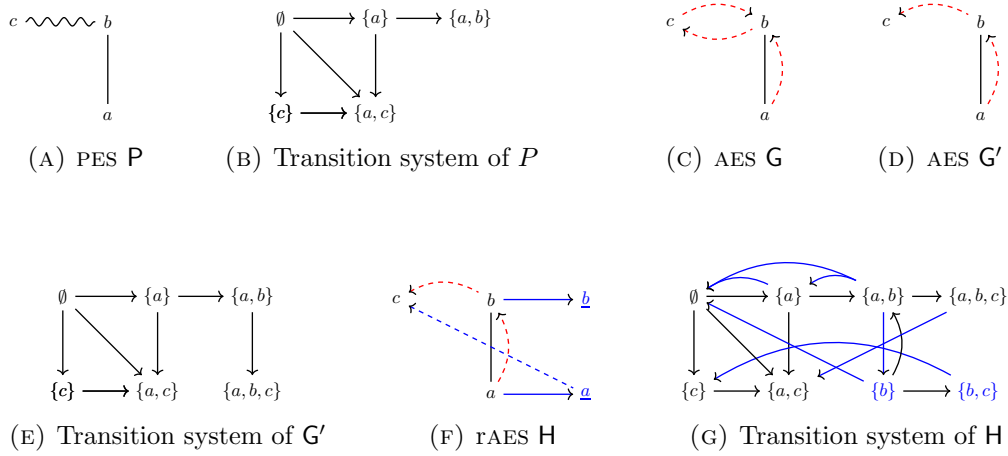

\emph{Event structures} (\eses) \cite{Win:ES} are a denotational formalism to
describe the behaviour of concurrent systems as a set of event occurences and
constraints on such events, regulating how such events can happen. \emph{Prime
  event structures} ($\peses$) are among the simplest form of event structures.
A Prime event structure describes a computational process as a set of events
whose occurrence is constrained by two relations: \emph{causality} and
(symmetric) \emph{conflicts}. A simple \pes is depicted in
Fig.~\ref{fig:exintro-es}, where causality ($<$) is drawn with straight lines
(to be read from bottom to top) and conflicts ($\#$) with curly lines. In this
case, $b$ causally depends on $a$ (i.e., $a < b$) meaning that $b$ cannot
occur if $a$ does not occur first; additionally, $b$ and $c$ are in conflict
(i.e., $b \# c$) meaning that $b$ and $c$ are mutually exclusive and cannot
occur in the same execution of the process.
The behaviour of a \pes can be understood in terms of a transition system
defined over {\em configurations} (i.e., sets of events), as illustrated in
Fig.~\ref{fig:exintro-conf}. For instance, the transition
$\emptyset \rightarrow \{a,c\}$ indicates that the initial state $\emptyset$
(i.e., no event has been executed yet) may evolve to $\{a,c\}$ by concurrently
executing $a$ and $c$.
Neither $\{b\}$ nor $\{a,b,c\}$ are configurations because $b$ cannot occur
without $a$; and $b$ and $c$ cannot happen in the same run.

In order to accommodate asymmetries that may arise, e.g., in shared-memory
concurrency, \emph{Asymmetric Event Structures} (\aeses) relax the notion of
conflicts by considering instead \emph{weak causality}.
Intuitively, an event $e$ weakly causes the event $e'$ (written
$e \nearrow e'$) if $e'$ can happen after $e$ but $e$ cannot happen after
$e'$. This can be considered as an asymmetric conflict because $e'$ forbids
$e$ to take place, but not the other way round.
%
\input{shared_memory}

However, symmetric conflicts can be recovered by making a pair of conflicting
events to weakly cause each other. For instance, the \pes $\mathsf{P}$ in
Fig.~\ref{fig:exintro-es} can be rendered as the \aes $\mathsf{G}$ in
Fig.~\ref{fig:exintro-aes-es}, where weak causality is depicted with red,
dashed arrows. Now the conflict between $b$ and $c$ is represented as
$b \nearrow c$ and $c \nearrow b$ (the additional weak causal
  dependency $a \nearrow b$ is required by consistency with the causality
  relation).
Unsurprisingly, the transition system associated with $\mathsf{G}$ coincides
with that of $\mathsf{P}$ in Fig.~\ref{fig:exintro-conf}.
Differently, the \aes $\mathsf{G}'$ relaxes the conflict between $b$ and $c$
by making it asymmetric: we keep $b \nearrow c$ but drop $c \nearrow b$.
Hence, $c$ can be added to the configuration $\setenum{a, b}$ but $b$ cannot
be added to $\setenum{a, c}$, as rendered in the transition system in
Fig.~\ref{fig:exintro-aesconf}.

Since their introduction~\cite{Win:ES}, \eses have played a central role in
the development of denotational semantics for concurrency models; in
particular, for Petri nets (\pns).
It is well-known that different classes of \eses correspond to different
classes of \pns~\cite{Win:ES,BCM01IC,flowEvent}. Moreover, different relations
in \eses translate into different operational mechanisms of \pns.
Causality and conflicts are typically modelled in nets via places
\emph{shared} among transitions. However, shared places fall short when
translating other kinds of dependencies, such as weak causality, which
requires contextual arcs~\cite{BCM01IC}.

Reversible computing has been studied in the 70's for its promise of achieving
low-energy computation \cite{landaurer}. Lately \cite{revbook,wg2} is gaining
interests for its application in different fields: from modelling bio-chemical
reactions, to improve parallel discrete events simulation, to model fault
tollerant primitives and finally to debugging.
In a model of reversible computation we can distinguish two different flows of
computation, i.e., in addition to the description of the standard {\em
  forward} execution, there is a {\em backward} flow that expresses the way in
which the effects of the forward computation can be undone. If in a sequential
setting it is pretty clear what is the backward flow of a computation (i.e. it
is the exact inverse of the forward one), and the only challenge is on how to
record computational history \cite{Leeman86,YokoyamaG07}. In a concurrent
setting it is more complicated. Different proposal have arisen in the last
years trying to cope with different models: \emph{causal-consistent
  reversibility} \cite{rccs} has been shown to be the right notion to model
primitives for fault-handling systems such as transactions
\cite{DanosK05,LaneseLMSS13} or roll-back recovery schemas
\cite{VassorS18,MezzinaTY25,Vidal23}, or to reverse-debug concurrent systems;
while \emph{out-of-causal} order reversibility \cite{PhillipsUY13} has been
shown as being able to model bio-chemical reactions and compensations.
Different models of concurrent computation~\cite{revbook} are available today:
RCCS~\cite{rccs}, CCSK~\cite{ccsk}, rho$\pi$~\cite{rhopi}, R$\pi$~\cite{rpi},
reversible Petri nets~\cite{revpt}, reversible event structures~\cite{rpes},
to name a few.
As expected, the attention has turned to the question of how these models
relate to each other (see
e.g.~\cite{LaneseMM21,MedicMPY20,lics,MelgrattiMP24,GraversenPY21,tocl}).
This work addresses such goal by revisiting, in the context of reversibility,
the connection between {\em Event Structures} (\eses)~\cite{Win:ES} and {\em
  Petri Nets} (\pns) established by Winskel~\cite{NielsenPW79}.

On the event structure side, we focus on the \emph{reversible Asymmetric Event
  Structures} (\raeses) introduced in~\cite{rpes}, which are a reversible
counterpart of \emph{Asymmetric Event Structures} (\aeses)~\cite{BCM01IC},
which in turn are a generalisation of {\em Prime Event Structures}
(\peses)~\cite{NielsenPW79}.

In \raeses, the backward flow is described in terms of a set of reversing
events, each of them representing the undoing of some event of the forward
flow.
The reversal of an  event $e$ is denoted as $\underline e$.
  This notation is utilised in defining the set of reversible events and in describing the relationships associated with the undoing of events.
For instance, in the \aes $\mathsf{G'}$ in Fig.~\ref{fig:exintro-aes}, with
$\{\underline a, \underline b\}$ we indicate that $a$ and $b$ are reversible,
while $c$ is not.
Two relations, dubbed \emph{reverse causation} ($\prec$) and \emph{prevention}
($\triangleleft$), describe the backward flow and regulate the way in which
reversing events occur:
$\prec$ prescribes the events required for the undoing while $\triangleleft$
stipulates those that preclude it. The \raes $\mathsf{H}$ in
Fig.~\ref{fig:exintro-raes} consists of the forward flow defined by
$\mathsf{G'}$ extended with a backward flow represented by blue arrows: solid
arrows correspond to reverse causation and dashed ones to prevention.
Then, {$a \prec \underline{a}$} says that {$\underline{a}$} can be executed
(meaning $a$ can be undone) only when $a$ has occurred (the pair
{$b \prec \underline{b}$} is analogous).
The prevention arc {$\underline{a}\triangleleft c$} states that {$a$} can be
reversed only if {$c$} has not occurred.
The transition system associated with $\mathsf{H}$ is in
Fig.~\ref{fig:exintro-rconf-raes}: transitions corresponding to the forward
flow (i.e., the ones in black mimicking those in
Fig.~\ref{fig:exintro-aesconf}) add events to configurations; on the contrary,
reversing transitions (in blue) remove events from configurations.
For instance, the transition $\{a,b\} \rightarrow \{a\}$ accounts for the fact
that $b$ is reversible and can always be undone.
Note that $a$ can be reversed both in $\{a\}$ and $\{a,b\}$ but not in
$\{a,c\}$ since the prevention relation (i.e., $\underline{a}\triangleleft c$)
forbids $a$ to be reversed if $c$ has already occurred.
Interestingly, $a$ can be reversed in $\{a,b\}$ leading to the configuration
$\{b\}$, which is not reachable by the forward flow (black arrows). This is
known as out-of-causal order reversibility \cite{PhillipsUY13}.

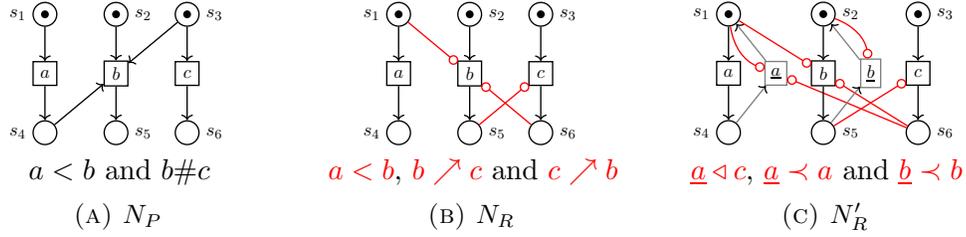
\begin{figure*}[bt]
  \begin{subfigure}{.30\textwidth}
    \centerline{\scalebox{0.70}{\input{figures/exintroabis.tex}}}
    \centerline{\textcolor{black}{$a < b$} and $b \# c$}
    \caption{$N_P$\label{ex:introa}}
  \end{subfigure}
  \begin{subfigure}{.30\textwidth}
    \centerline{\scalebox{0.70}{\input{figures/exintrobbis.tex}}}

    \centerline{\textcolor{red}{$a < b$}, \textcolor{red}{$b\nearrow c$} and
      \textcolor{red}{$c\nearrow b$}}
    \caption{$N_R$\label{ex:introb}}
  \end{subfigure}
  \begin{subfigure}{.30\textwidth}
    \centerline{\scalebox{0.70}{\input{figures/exintroc.tex}}}
    \centerline{\textcolor{red}{$\underline a \triangleleft c$},
      \textcolor{red}{$\un a\prec a$} and \textcolor{red}{$\un b\prec b$}}
    \caption{$N'_R$\label{ex:introc}}
  \end{subfigure}
  \caption{An occurrence net $N_P$, its associated causal net $N_R$, and its
    reversible causal net $N'_R$.}
  \label{fig:exintro}
\end{figure*}

Reversible \eses have introduced further questions about the required features
of their operational counterpart, suggesting that shared places are not always
a suitable choice (\cite{PhilippouP22a,PhilippouP22,lics,tocl}). It has been
recently shown that the operational model behind (reversible) \peses can be
recovered as a subclass of contextual Petri nets, called {\em (reversible)
  Causal Nets} (\rcns), in which causality is modelled via inhibitor
arcs~\cite{lics} rather than relying on a shared place in which one transition
produces a token to be consumed by another one. Inhibitor arcs neither produce
nor consume tokens but check for the \emph{absence} of them.
This idea is rendered by the nets in Fig.~\ref{fig:exintro}.
We recall that a \pn gives an operational description of a computation in
terms of {\em transitions} (depicted as boxes) that consume and produce {\em
  tokens} (bullets) in different {\em places} (circles). According to the
black arrows in Fig.~\ref{ex:introa}, the transition $a$ consumes a token from
$s_1$ producing a token in $s_4$; similarly the transition $b$
consumes from $s_4$, $s_2$ and $s_3$ producing in $s_5$. Note that $b$ and $c$
are in mutual exclusion (i.e., conflict) because they both compete for the
shared resource (i.e., token) in $s_3$. The arc connecting $s_4$ to $b$
indicates that $b$ cannot be fired if $s_4$ does not contains any token;
consequently, $b$ can happen only after $a$ has produced the token in $s_4$.
The causal relation between $a$ and $b$ arises because of $s_4$.
$N_P$ in Fig.~\ref{ex:introa} is the \emph{classical} operational counterpart
of the \es $P$ in Fig.~\ref{fig:exintro-es}.
However, we note that both the causalities and conflicts in $N_P$ can be
alternatively represented by using inhibitor arcs instead of shared places, as
shown in Fig.~\ref{ex:introb}. The inhibitor arc (depicted as a red line
ending with a circle) between $s_1$ and $b$ models $a < b$, whereas the
inhibitor arcs $(s_5,c)$ and $(s_6,b)$ represent the symmetric conflict
between $b$ and $c$.
The net in Fig.~\ref{ex:introb} can be made reversible by adding a reversing
transition for each reversible event, as shown in Fig.~\ref{ex:introc} (in
gray). The added transitions $\underline a$ and $\underline b$ respectively
reverse the effects of $a$ and $b$: each of them consumes (produces) the
tokens produced (resp., consumed) by the associated forward transition.
Inhibitor arcs are also used for modelling reverse causation and prevention
as, e.g., the inhibitor arc connecting $\underline a$ with $s_6$ stands for
$\underline a\triangleleft c$.

In this paper we generalise the ideas presented in~\cite{lics} so to be able
to deal with \raeses (and not just with the subclass of reversible \peses).
This is achieved by using inhibitor arcs to represent, not only causal
dependencies, but also (symmetric) conflicts. In this way, all the
dependencies between transitions are modelled \emph{uniformly} by using
inhibitor arcs.
Concretely, we identify a subclass of \rcn{s}, dubbed {\em reversible
  Asymmetric Causal Nets} (\racns), which are the operational counterpart of
\raeses.
We show that the correspondence is tight by following the long tradition of
comparing concurrency models in categorical terms
\cite{NielsenPW79,Win:ES,Win:PNAMC}.
To do so we first turn \raeses and \racns into categories by providing
suitable notions of morphisms, then we introduce two functors relating these
categories and finally we show that the functor that associates \raeses to
\racns is the left adjoint of the one that that give \racns out of \raeses.
Besides establishing a correspondence between \raeses and \racns, this allows
us to reinterpret the results in~\cite{lics} categorically.

This paper is an extended and enhanced version of \cite{forte}. In this
presentation, we have refined the concept of morphisms for causal nets,
specifically by relaxing the requirements on inhibitor arcs that capture
causality when conflicting transitions are collapsed in the morphism's image.
Additionally, we include technical details omitted in the conference paper,
and we furnish complete proofs of the results. Additional examples and
explanations have been added. Also, \Cref{sec:app} is completely new. Finally,
the whole presentation has been thoroughly refined and improved.

\paragraph*{Structure of the paper}
This paper is structured as follows: in \Cref{sec:raes} we recall Asymmetric
Event Structures and their reversible variants, along with their corresponding
categories. In \Cref{sec:rcn} Asymmetric Causal Nets, reversible Asymmetric
Causal Nets and their categories are presented. The main results about
categorical connection are showns in \Cref{sec:adjunctions}. \Cref{sec:app}
discusses about some applications of our results. Finally, \Cref{sec:conc}
concludes the paper.

%% file: shared_memory.tex
Let us consider the following shared memory scenario:
 \[
\begin{array}{l@{\qquad}||@{\qquad}l@{\qquad\qquad}llr}
\textbf{Thread 1} & \textbf{Thread 2} & \textbf{Possible Values}\\
1:\mathtt{lock(m);} &1:\mathtt{lock(m);} & \mathtt{x=0;\, y =1}\\
 2: \mathtt{if(y == 0)}	& 	2: \mathtt{y++;} & \mathtt{x=1;\, y =1}\\
  3: \;\;\mathtt{then\;x =1;} 	& 1:\mathtt{unlock(m);}\\
  4:\mathtt{unlock(m);}
\end{array}
\]
in which two  variables $x$ and $y$ are shared among two threads.
Let us suppose, for simplicity, that the two variables are initialised to $0$.
Assuming strong consistency, the actions of the two threads can be scheduled into two different ways leading to  two different final configurations:
either $x = 0$ and $y=1$ or both variables are equal to $1$. If we represent
the behaviour of the above snippet as an event structure, we can identify two
relevant events: the assignment of  $1$ to $x$ and the increment of $y$.
Let us call such events $x_1$ and $y_1$. We can observe that if $x_1$ happens then $y_1$ can happen, while if $y_1$ happens, then $x_1$ cannot. And this means that the $x_1$ weakly causes $y_1$,  that is $x_1 \nearrow y_1$.

%% file: figures/exintro-es.tex
\scalebox{0.9}{\begin{tikzpicture}
\usetikzlibrary{decorations.pathmorphing}

\tikzset{snake it/.style={decorate, decoration=snake}}
\tikzstyle{cau}=[-,thick]
\tikzstyle{conf}=[snake it,thick]
\tikzstyle{transition}=[rectangle, draw=none,thick,minimum size=5mm]
\node[transition] (b) at (0,2)  {$b$}
;
\node[transition] (a) at (0,0) {$a$}
edge[cau] (b)
;
\node[transition] (c) at (-2,2)  {$c$}
edge[conf] (b);
\end{tikzpicture}}

%% file: figures/exintro-conf.tex
\scalebox{0.9}{\begin{tikzpicture}
\usetikzlibrary{decorations.pathmorphing}

\tikzset{snake it/.style={decorate, decoration=snake}}
\tikzstyle{tr}=[->,thick]
\tikzstyle{transition}=[rectangle, draw=none,thick,minimum size=5mm]
\node[transition] (ab) at (4,2)  {$\{a,b\}$}
;
\node[transition] (ac) at (2,0)  {$\{a,c\}$}
;
\node[transition] (a) at (2,2){$\{a\}$}
edge[tr] (ac)
edge[tr] (ab)
;
\node[transition] (c) at (0,0) {$\{c\}$}
edge[tr] (ac)
;
\node[transition] (empty) at (0,2)  {$\emptyset$}
edge[tr] (a)
edge[tr] (c)
edge[tr] (ac)
;
\node[transition] (c) at (0,0) {$\{c\}$}
edge[tr] (ac)
;

\end{tikzpicture}}

%% file: figures/exintro-aes-es.tex
\scalebox{0.9}{\begin{tikzpicture}
\usetikzlibrary{decorations.pathmorphing,arrows.meta}

\tikzset{snake it/.style={decorate, decoration=snake}}
\tikzstyle{cau}=[-,thick]
\tikzstyle{conf}=[snake it,thick]
\tikzstyle{wcau}=[dashed,->,thick,draw=red]
\tikzstyle{cauw}=[dashed,<-,thick,draw=red]
\tikzstyle{transition}=[rectangle, draw=none,thick,minimum size=5mm]
\node[transition] (b) at (0,2)  {$b$}
;
\node[transition] (a) at (0,0) {$a$}
edge[cau] (b)
edge[wcau, bend right = 30] (b)
;
\node[transition] (c) at (-2,2)  {$c$}
edge[wcau, bend left = 30] (b)
edge[cauw, bend right = 30] (b);
\end{tikzpicture}}

%% file: figures/exintro-aes.tex
\scalebox{0.9}{\begin{tikzpicture}
\usetikzlibrary{decorations.pathmorphing,arrows.meta}

\tikzset{snake it/.style={decorate, decoration=snake}}
\tikzstyle{cau}=[-,thick]
\tikzstyle{conf}=[snake it,thick]
\tikzstyle{wcau}=[dashed,->,thick,draw=red]
\tikzstyle{cauw}=[dashed,<-,thick,draw=red]
\tikzstyle{transition}=[rectangle, draw=none,thick,minimum size=5mm]
\node[transition] (b) at (0,2)  {$b$}
;
\node[transition] (a) at (0,0) {$a$}
edge[cau] (b)
edge[wcau, bend right = 30] (b)
;
\node[transition] (c) at (-2,2)  {$c$}
edge[cauw, bend left = 30] (b)
;
\end{tikzpicture}}

%% file: figures/exintro-conf-aes.tex
\scalebox{0.9}{\begin{tikzpicture}
\usetikzlibrary{decorations.pathmorphing}

\tikzset{snake it/.style={decorate, decoration=snake}}
\tikzstyle{tr}=[->,thick]
\tikzstyle{rt}=[<-,thick]
\tikzstyle{transition}=[rectangle, draw=none,thick,minimum size=5mm]
\node[transition] (ab) at (4,2)  {$\{a,b\}$}
;
\node[transition] (ac) at (2,0)  {$\{a,c\}$}
;
\node[transition] (a) at (2,2){$\{a\}$}
edge[tr] (ac)
edge[tr] (ab)
;
\node[transition] (c) at (0,0) {$\{c\}$}
edge[tr] (ac)
;
\node[transition] (empty) at (0,2)  {$\emptyset$}
edge[tr] (a)
edge[tr] (c)
edge[tr] (ac)
;
\node[transition] (c) at (0,0) {$\{c\}$}
edge[tr] (ac)
;
\node[transition] (abc) at (4,0) {$\{a,b,c\}$}
edge[rt] (ab)
;
\end{tikzpicture}}

%% file: figures/exintro-raes.tex
\scalebox{0.9}{\begin{tikzpicture}
\usetikzlibrary{decorations.pathmorphing,quotes}

\tikzset{snake it/.style={decorate, decoration=snake}}
\tikzstyle{rprec}=[->,thick,draw=blue]
\tikzstyle{precr}=[<-,thick,draw=blue]
\tikzstyle{prevr}=[dashed,->,thick, draw=blue]
\tikzstyle{rprev}=[dashed,<-,thick, draw=blue]
\tikzstyle{cauw}=[dashed,<-,thick,draw=red]
\tikzstyle{wcau}=[dashed,->,thick,draw=red]

\tikzstyle{transition}=[rectangle, draw=none,thick,minimum size=5mm]
\tikzstyle{rtransition}=[rectangle, draw=none,thick,minimum size=5mm, color = blue]
\tikzstyle{cau}=[-,thick]

\node[transition] (b) at (0,2)  {$b$}
;
\node[transition] (a) at (0,0) {$a$}
edge[cau] (b)
edge[wcau, bend right = 30] (b)
;
\node[rtransition] (rb) at (2,2)  {$\underline{b}$}
edge[precr] (b)
;
\node[rtransition] (ra) at (2,0)  {$\underline{a}$}
edge[precr] (a)
;
\node[transition] (c) at (-2,2)  {$c$}
edge[cauw, bend left = 30] (b)
edge[rprev] (ra)
;
\end{tikzpicture}}

%% file: figures/exintro-rconf.tex
\scalebox{0.9}{\begin{tikzpicture}
\usetikzlibrary{decorations.pathmorphing}

\tikzset{snake it/.style={decorate, decoration=snake}}
\tikzstyle{tr}=[->,thick]
\tikzstyle{rt}=[<-,thick]
\tikzstyle{rtr}=[->,draw=blue,thick]
\tikzstyle{rrt}=[<-,draw=blue,thick]
\tikzstyle{transition}=[rectangle, draw=none,thick,minimum size=5mm]
\tikzstyle{rtransition}=[rectangle, draw=none,thick,minimum size=5mm, color = blue]
\node[transition] (ab) at (4,2)  {$\{a,b\}$}
;
\node[transition] (ac) at (2,0)  {$\{a,c\}$}
;
\node[rtransition] (bc) at (6,0)  {$\{b,c\}$}
;
\node[transition] (a) at (2,2){$\{a\}$}
edge[tr] (ac)
edge[tr] (ab)
;
\node[transition] (c) at (0,0) {$\{c\}$}
edge[tr] (ac)
edge[rrt, bend left] (bc)
;
\node[transition] (empty) at (0,2)  {$\emptyset$}
edge[tr] (a)
edge[tr] (c)
edge[tr] (ac)
;
\node[rtransition] (b) at (4,0)  {$\{b\}$}
edge[tr,bend right] (ab)
edge[rtr] (empty)
edge[tr] (bc)
;
\node[transition] (abc) at (6,2) {$\{a,b,c\}$}
edge[rt] (ab)
edge[rtr] (ac)
;
\draw[rtr] (ab)--(b);
\draw[rtr] (ab)  to[bend right] (empty);
\draw[rtr] (ab)  to[bend right] (a);
\draw[rtr] (a)  to[bend right] (empty);

\end{tikzpicture}}

%% file: figures/exintroabis.tex
\scalebox{0.9}{\begin{tikzpicture}
\tikzstyle{inhibitorred}=[o-, draw=red,thick]
\tikzstyle{inhibitorblu}=[o-, draw=blue,thick]
\tikzstyle{pre}=[<-,thick]
\tikzstyle{post}=[->,thick]
\tikzstyle{readblue}=[-, draw=blue,thick]
\tikzstyle{transition}=[rectangle, draw=black,thick,minimum size=5mm]
\tikzstyle{place}=[circle, draw=black,thick,minimum size=5mm]
\node[place,tokens=1] (p1) at (0,2.5) [label=left:$s_1$] {};
\node[place,tokens=1] (p3) at (1.5,2.5) [label=right:$s_2$] {};
\node[place,tokens=1] (p5) at (3,2.5) [label=right:$s_3$] {};
\node[place] (p2) at (0,0) [label=left:$s_4$] {};
\node[place] (p4) at (1.5,0) [label=right:$s_5$] {};
\node[place] (p6) at (3,0) [label=right:$s_6$] {};
\node[transition] (a) at (0,1.25)  {$a$}
edge[pre] (p1)
edge[post](p2);

\node[transition] (b) at (1.5,1.25) {$b$}
edge[pre] (p3)
edge[pre] (p5)
edge[pre] (p2)
edge[post] (p4)
;
\node[transition] (c) at (3,1.25)  {$c$}
edge[pre] (p5)
edge[post] (p6);
\end{tikzpicture}}

%% file: figures/exintrobbis.tex
\scalebox{0.9}{\begin{tikzpicture}
\tikzstyle{inhibitorred}=[o-, draw=red,thick]
\tikzstyle{inhibitorblu}=[o-, draw=blue,thick]
\tikzstyle{pre}=[<-,thick]
\tikzstyle{post}=[->,thick]
\tikzstyle{readblue}=[-, draw=blue,thick]
\tikzstyle{transition}=[rectangle, draw=black,thick,minimum size=5mm]
\tikzstyle{place}=[circle, draw=black,thick,minimum size=5mm]
\node[place,tokens=1] (p1) at (0,2.5) [label=left:$s_1$] {};
\node[place,tokens=1] (p3) at (1.5,2.5) [label=right:$s_2$] {};
\node[place,tokens=1] (p5) at (3,2.5) [label=right:$s_3$] {};
\node[place] (p2) at (0,0) [label=left:$s_4$] {};
\node[place] (p4) at (1.5,0) [label=right:$s_5$] {};
\node[place] (p6) at (3,0) [label=right:$s_6$] {};
\node[transition] (a) at (0,1.25)  {$a$}
edge[pre] (p1)
edge[post](p2);

\node[transition] (b) at (1.5,1.25) {$b$}
edge[pre] (p3)
edge[post] (p4)
edge[inhibitorred] (p1)
edge[inhibitorred] (p6)
;
\node[transition] (c) at (3,1.25)  {$c$}
edge[pre] (p5)
edge[post] (p6)
edge[inhibitorred] (p4)
;
\end{tikzpicture}}

%% file: figures/exintroc.tex
\scalebox{0.9}{\begin{tikzpicture}
\tikzstyle{inhibitorred}=[o-, draw=red,thick]
\tikzstyle{inhibitorblu}=[o-, draw=blue,thick]
\tikzstyle{prerev}=[<-,thick,draw=gray]
\tikzstyle{postrev}=[->,thick,draw=gray]
\tikzstyle{pre}=[<-,thick]
\tikzstyle{rev}=[-, draw=gray,thick]
\tikzstyle{post}=[->,thick]
\tikzstyle{readblue}=[-, draw=blue,thick]
\tikzstyle{transition}=[rectangle, draw=black,thick,minimum size=5mm]
\tikzstyle{place}=[circle, draw=black,thick,minimum size=5mm]
\node[place,tokens=1] (p1) at (0,2.5) [label=left:$s_1$] {};
\node[place,tokens=1] (p3) at (2,2.5) [label=right:$s_2$] {};
\node[place,tokens=1] (p5) at (4,2.5) [label=right:$s_3$] {};
\node[place] (p2) at (0,0) [label=left:$s_4$] {};
\node[place] (p4) at (2,0) [label=right:$s_5$] {};
\node[place] (p6) at (4,0) [label=right:$s_6$] {};
\node[transition] (a) at (0,1.25)  {$a$}
edge[pre] (p1)
edge[post](p2);

\node[rev] (ra) at (1,1.25) {$\underline{a}$}
edge[prerev] (p2)
edge[inhibitorred] (p6)
edge[postrev] (p1)
edge[inhibitorred, bend left =30] (p1)
;

\node[transition] (b) at (2,1.25) {$b$}
edge[pre] (p3)
edge[post] (p4)
edge[inhibitorred] (p1)
edge[inhibitorred] (p6)

;
\node[rev] (rb) at (3,1.25) {$\underline{b}$}
edge[prerev] (p4)
edge[postrev] (p3)
edge[inhibitorred, bend right =30] (p3)
;

\node[transition] (c) at (4,1.25)  {$c$}
edge[pre] (p5)
edge[post] (p6)
edge[inhibitorred] (p4)
;

\end{tikzpicture}}

%% file: raes.tex

\section{Reversible Asymmetric Event Structures}\label{sec:raes}
In this section we recall the basics of {\em Asymmetric Event Structures}
(\aeses)~\cite{BCM01IC} and their reversible version introduced
in~\cite{rpes,GPY:categories}.

\input{asymmes}
\input{revasymmes}
\input{coprodraes}

%% file: asymmes.tex

\subsection{Asymmetric Event Structures}\label{sec:aes}
An \aes consists of a set of events and two relations: \emph{causality} ($<$)
and \emph{weak causality} or \emph{precedence} ($\nearrow$). If $e$ weakly
causes $e'$, written $e \nearrow e'$, then $e$ cannot occur after $e'$; i.e.,
if both events occur in a computation, then $e$ precedes $e'$. In this case we
say that $e'$ is in an \emph{asymmetric} conflict with $e$. Events $e$ and
$e'$ are in {\em (symmetric) conflict}, written $e \# e'$, iff $e\nearrow e'$
and $e'\nearrow e$; intuitively, they cannot take place in the same
computation.

\begin{defi}\label{de:aes}
  An \emph{Asymmetric Event Structure} (\aes) is a triple
  $\mathsf{G} = (E, <, \nearrow)$ where
  \begin{enumerate}
  \item \label{def:aes-countable} $E$ is a set of \emph{events};

  \item \label{def:aes-finite-causes} $<\ \subseteq E\times E$ is an
    irreflexive partial order, called \emph{causality}, defined such that
    $\forall e\in E$. $\hist{e} = \setcomp{e'\in E}{e' \leq e}$ is finite; and
  \item $\nearrow\ \subseteq E\times E$, called {\em weak causality}, is
    defined such that for all $e, e'\in E$:
    \begin{enumerate}
    \item \label{def:aes-reflect-causality}
      $e < e'\ \Rightarrow\ e \nearrow e'$;
    \item\label{def:aes-acyclic-causality}
      ${\nearrow} \cap {(\hist{e}\times \hist{e})}$ is acyclic; and
    \item \label{def:aes-conflict-inh}
      if $e \# e'$ and $e' < e''$ then $e \# e''$.
    \end{enumerate}
  \end{enumerate}
\end{defi}
\noindent 
Each event can be attributed to a finite set of causes, as specified in Condition~\ref{def:aes-finite-causes}.
Moreover, weak causality is consistent with causality: if $e$ is a cause of
$e'$, then $e$ is also a weak cause of $e'$ (as stated in
Condition~\ref{def:aes-reflect-causality}).
Additionally, circular dependencies among the causes of an event are precluded by
Condition~\ref{def:aes-acyclic-causality}.
Finally, (symmetric) conflicts ($\#$) must be inherited along causality, as
required by Condition~\ref{def:aes-conflict-inh}.

\begin{exa}\label{ex:aes}
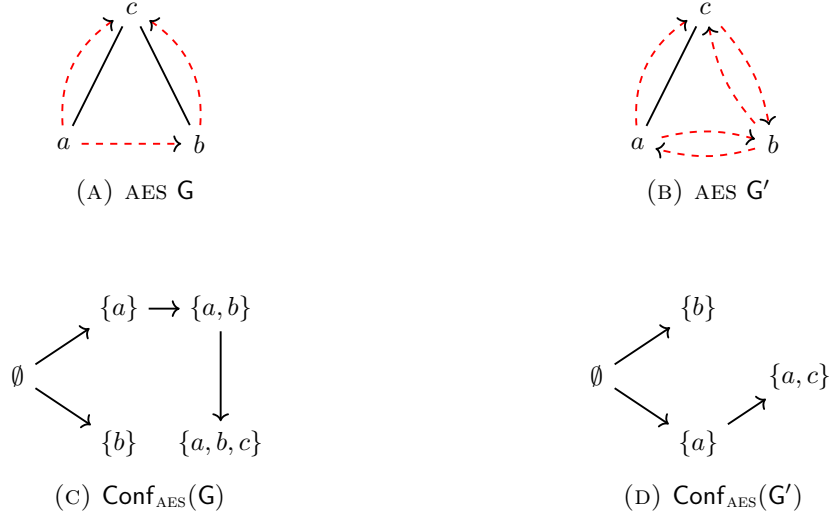
\begin{figure}[t]
   \begin{subfigure}{.45\textwidth}
      \centerline{\scalebox{1}{\input{figures/exaes-aes.tex}}}
    \caption{\aes $\mathsf{G}$}\label{fig:aesuno}
  \end{subfigure}\qquad
   \begin{subfigure}{.45\textwidth}
      \centerline{\scalebox{1}{\input{figures/exaes-aesbis.tex}}}
    \caption{\aes $\mathsf{G}'$}\label{fig:aesdue}
  \end{subfigure}
  \vspace*{1cm}
  
  \begin{subfigure}{.45\textwidth}
      \centerline{\scalebox{1}{\input{figures/exaes-confaes.tex}}}
      \caption{$\Conf{\mathsf{G}}{\aes}$}
    \label{fig:aesconfuno}
  \end{subfigure}\qquad
  \begin{subfigure}{.45\textwidth}
      \centerline{\scalebox{1}{\input{figures/exaes-confaesbis.tex}}}
          \caption{$\Conf{\mathsf{G}'}{\aes}$}
    \label{fig:aesconfdue}
  \end{subfigure}
 \caption{Two \aes{}es and their associated configurations.}
\end{figure}

Consider the \aes $\mathsf{G} = (E, <, \nearrow)$ depicted in
\Cref{fig:aesuno}, where the set of events is $E = \setenum{a, b, c}$, the
causality relation is defined such that $a < c$ and $b < c$, and the weak
causality is such that $a \nearrow b$, $a \nearrow c$, and $b \nearrow c$. The
set $E$ is finite, and hence countable~(Condition~\ref{def:aes-countable}). It
is straightforward to check that $<$ is an irreflexive partial order and also
that each event possesses a finite set of
causes~(Condition~\ref{def:aes-finite-causes}), being
$\hist{a} = \setenum{a}$, $\hist{b} = \setenum{b}$, and
$\hist{c} = \{a, b, c\}$. Additionally, we have that $a < c$ and
$a \nearrow c$; and also $b < c$ and $b \nearrow c$, in accordance with
Condition~\ref{def:aes-reflect-causality}.
It is immediate to check that $\nearrow$ is acyclic on $\hist{a}, \hist{b}$
and $\hist{c}$ 
(Condition~\ref{def:aes-acyclic-causality}). Moreover, the conflict relation
$\#$ is empty, and hence Condition~\ref{def:aes-conflict-inh} is trivially
satisfied.

In the \aes{} $\mathsf{G}' = (E', <', \nearrow')$ of Fig.~\ref{fig:aesdue},
the causality relation contains just $a <' c$ and the weak causality is
$a \nearrow' b, b \nearrow' a, a \nearrow' c, b\nearrow' c, c\nearrow' b$. In
this case $\nearrow'$ induces a symmetric conflict among $a$ and $b$ and one
among $b$ and $c$, hence $b \#' a$ and $b \#' c$, and also the inheritance of
conflicts along the causality relation is verified (Condition
\ref{def:aes-conflict-inh}).

In the \aes{} $\mathsf{G}' = (E', <', \nearrow')$ illustrated in
Fig.~\ref{fig:aesdue}, the causality relation comprises only $a <' c$, while
weak causality is defined by $a \nearrow' b$, $b \nearrow' a$,
$a \nearrow' c$, $b \nearrow' c$, and $c \nearrow' b$. In this case,
$\nearrow'$ results in a symmetric conflict between $a$ and $b$, as well as
one between $b$ and $c$. As a result, $b \#' a$ and $b \#' c$, satisfying the
inheritance of conflicts along causality (Condition
\ref{def:aes-conflict-inh}).
\end{exa}

\begin{defi}\label{de:aes-conf}
  A \emph{configuration} of an \aes{} $\mathsf{G} = (E, <, \nearrow)$ is a set
  $X\subseteq E$ of events such that
 \begin{enumerate}
 \item \label{def:conf-aes-well-founded}
   $\nearrow$ is well-founded on $X$;
 \item \label{def:conf-aes-left-close}
   $\forall e\in X.\ \hist{e}\subseteq X$; and
 \item \label{def:conf-aes-finite-prec}
   $\forall e\in X$ the set $\setcomp{e'\in X}{e'\nearrow e}$ is finite.
 \end{enumerate}
 The set of configurations of $\mathsf{G}$ is denoted by
 $\Conf{\mathsf{G}}{\aes}$.
\end{defi}

A configuration comprises a set of events representing a potential partial
execution of $\mathsf{G}$. Condition~\ref{def:conf-aes-well-founded} ensures
that events in $X$ are not in conflict, as circular weak-causal dependencies
are forbidden.
In accordance with Condition~\ref{def:conf-aes-left-close}, $X$ contains all
the causes of its constituent events: an event may occur only if its causes
have already occurred.
While a configuration may be infinite, as in a non-terminating execution, each
event within a configuration has a finite set of weak causes
(Condition~\ref{def:conf-aes-finite-prec}).

For any pair of sets $X, Y \subseteq E$ where $X \subseteq Y$, we say that $Y$ \emph{extends} $X$ if, for all $e \in X$ and $e' \in Y \setminus X$, there is no weak-causal dependency ($\neg (e' \nearrow e)$).
In the context of configurations, if $X$ and $Y$ are configurations, then $Y$
is considered reachable from $X$.

\begin{exa}\label{ex:conf-aes}
  The sets of configurations of the \aeses $\mathsf{G}$ and $\mathsf{G}'$
  introduced in \Cref{ex:aes} are shown in Figs.~\ref{fig:aesconfuno}
  and~\ref{fig:aesconfdue} respectively. The arrows represent the
  \emph{extends} relation.
  Straightforwardly configurations satisfy the conditions in
  Definition~\ref{de:aes-conf}.
\end{exa}

A (finite) configuration $X = \setenum{e_1, \dots, e_n, \dots}$ of an \aes{} $\mathsf{G}$ is \emph{reachable} whenever the events in $X$ can be ordered in such a way that for each $j\geq 1$ 
$\setenum{e_1, \dots, e_j}$ extends $\setenum{e_1, \dots, e_{j-1}}$  
and each $\setenum{e_1, \dots, e_j}\in \Conf{\mathsf{G}}{\aes}$.

\begin{defi}\label{de:aes-morphisms}
  Let $\mathsf{G}_0 = (E_0, <_0, \nearrow_0)$ and
  $\mathsf{G}_1 = (E_1, <_1, \nearrow_1)$ be \aeses. An \aes-morphism, denoted
  as $f : \mathsf{G}_0\rightarrow \mathsf{G}_1$, is defined as a partial
  function $f : E_0 \rightarrow E_1$ such that for all $e, e' \in E_0$

 \begin{enumerate}
 \item \label{def:aes-morphism-preserve} if $f(e)\neq \bot$ then
   $\hist{f(e)}\subseteq f(\hist{e})$; and
 \item if $f(e)\neq \bot \neq f(e')$ then
   \begin{enumerate}
   \item \label{def:aes-morphism-reflect}
     $f(e)\nearrow_1 f(e')$ implies $e\nearrow_0 e'$; and
   \item \label{def:aes-morphism-identify}
     $f(e) = f(e')$ and $e\neq e'$ imply $e \#_0 e'$.
   \end{enumerate}
 \end{enumerate}
\end{defi}
\noindent 
An \aes{}-morphism preserves the causes of each mapped event
(Condition~\ref{def:aes-morphism-preserve}) and reflects its weak causes
(Condition~\ref{def:aes-morphism-reflect}). Additionally, it allows the
mapping of two distinct events to the same event only when those events are in
conflict (Condition~\ref{def:aes-morphism-identify}). The aforementioned
conditions collectively guarantee that morphisms preserve computations, or
configurations, as stated below.

\begin{prop}\label{pr:aesmorph-conf}
  Let $f : \mathsf{G}_0\rightarrow \mathsf{G}_1$ be an \aes{}-morphism and $X$
  a configuration of $\mathsf{G}_0$, i.e., $X\in \Conf{\mathsf{G}_0}{\aes}$.
  Then, $f(X)\in \Conf{\mathsf{G}_1}{\aes}$.
\end{prop}
\begin{proof}
 See \cite[Lemma 3.6]{BCM01IC}.
\end{proof}
Since \aes{}-morphisms compose~\cite{BCM01IC}, \aeses and \aes-morphisms turn
into a category, which we denote by $\mathbf{AES}$.

%% file: figures/exaes-aes.tex
\scalebox{0.9}{\begin{tikzpicture}
\usetikzlibrary{decorations.pathmorphing}

\tikzset{snake it/.style={decorate, decoration=snake}}
\tikzstyle{cau}=[-,thick]
\tikzstyle{wcau}=[dashed,->,thick,draw=red]
\tikzstyle{wcaub}=[dashed,<-,thick,draw=red]
\tikzstyle{transition}=[rectangle, draw=none,thick,minimum size=5mm]

\node[transition] (b) at (2,0)  {$b$}
;

\node[transition] (a) at (0,0) {$a$}
edge[wcau] (b)
;
\node[transition] (c) at (1,2)  {$c$}
edge[cau] (a)
edge[cau] (b)
edge[wcaub, bend right = 30] (a)
edge[wcaub, bend left = 30] (b)
;
\end{tikzpicture}}

%% file: figures/exaes-aesbis.tex
\scalebox{0.9}{\begin{tikzpicture}
\usetikzlibrary{decorations.pathmorphing}

\tikzset{snake it/.style={decorate, decoration=snake}}
\tikzstyle{cau}=[-,thick]
\tikzstyle{wcau}=[dashed,->,thick,draw=red]
\tikzstyle{wcaub}=[dashed,<-,thick,draw=red]
\tikzstyle{transition}=[rectangle, draw=none,thick,minimum size=5mm]
\node[transition] (a) at (0,0) {$a$}
;
\node[transition] (b) at (2,0)  {$b$}
edge[wcau, bend left = 15] (a)
edge[wcaub, bend right = 15] (a)
;
\node[transition] (c) at (1,2)  {$c$}
edge[cau] (a)
edge[wcaub, bend right = 30] (a)
edge[wcau, bend left = 15] (b)
edge[wcaub, bend right = 15] (b)
;
\end{tikzpicture}}

%% file: figures/exaes-confaes.tex
\scalebox{0.9}{\begin{tikzpicture}
\usetikzlibrary{decorations.pathmorphing}

\tikzset{snake it/.style={decorate, decoration=snake}}
\tikzstyle{tr}=[<-,thick]
\tikzstyle{transition}=[rectangle, draw=none,thick,minimum size=5mm]
\node[transition] (empty) at (0,1)  {$\emptyset$}
;
\node[transition] (a) at (1.5,0)  {$\{b\}$}
edge[tr] (empty)
;
\node[transition] (b) at (1.5,2){$\{a\}$}
edge[tr] (empty)
;
\node[transition] (ba) at (3,2) {$\{a,b\}$}
edge[tr] (b)
;
\node[transition] (bac) at (3,0) {$\{a,b,c\}$}
edge[tr] (ba)
;

\end{tikzpicture}}

%% file: figures/exaes-confaesbis.tex
\scalebox{0.9}{\begin{tikzpicture}
\usetikzlibrary{decorations.pathmorphing}

\tikzset{snake it/.style={decorate, decoration=snake}}
\tikzstyle{tr}=[<-,thick]
\tikzstyle{transition}=[rectangle, draw=none,thick,minimum size=5mm]
\node[transition] (empty) at (0,1)  {$\emptyset$}
;
\node[transition] (a) at (1.5,0)  {$\{a\}$}
edge[tr] (empty)
;
\node[transition] (b) at (1.5,2){$\{b\}$}
edge[tr] (empty)
;
\node[transition] (ac) at (3,1) {$\{a,c\}$}
edge[tr] (a)
;

\end{tikzpicture}}

%% file: revasymmes.tex

\subsection{Reversible {\aeses}}
We now provide a summary of \emph{reversible} \aeses,  introduced
in~\cite{rpes,GPY:categories} as the reversible counterparts of\aeses.
Given a set $\anR$ of events and $u\in \anR$, we write $\un{\anr}$ for the
undoing of $\anr$, and $\un{\anR} = \setcomp{\un{\anr}}{\anr\in\anR}$ for the
set of \emph{undoings} of $\anR$.
\begin{defi}\label{de:raes}
  A \emph{Reversible Asymmetric Event Structure} (\raes) is a sextuple
  $\mathsf{H} \! = \! (E, \anR, <, \\ \nearrow, \prec, \lhd)$ where $E$ is the set of
  events, $\anR \subseteq E$ is the set of \emph{reversible} events, and
  \begin{enumerate}
  \item\label{cond:2} $\nearrow\ \subseteq E \times E$, called {\em weak
      causality};
  \item\label{cond:2bis} $\lhd \subseteq \un{\anR} \times E$, called
    \emph{prevention};
  \item\label{cond:3} $<\ \subseteq E \times E$, called {\em causation}, is an
    irreflexive relation defined such that for all $e\in E$,
    $\histtwo{e}{<} = \setcomp{e'\in E}{e' \leq e}$ is finite and
    ($\nearrow\cup <$) is acyclic on $\histtwo{e}{<}$; 
  \item $\prec\ \subseteq E \times \un{\anR}$, called {\em reverse causation},
    is defined such that
    \begin{enumerate}
    \item \label{cond:4} for all $\anr\in \anR.\ \anr \prec \un{\anr}$;
    \item \label{cond:3bis} for all $\anr\in \anR$,
      $\histtwo{\un{\anr}}{\prec} = \setcomp{e\in E}{e \prec \un{\anr}}$ is finite
      and ($\nearrow\cup <$) is acyclic on $\histtwo{\un{u}}{\prec}$;
    \end{enumerate}
  \item\label{cond:5b} for all $e\in E, \un{u}\in \un{\anR}.$
    $e\prec \un{u}\ \Rightarrow \neg(\un{u}\lhd e$); and
  \item\label{cond:6} $(E,\pprec, \nearrow)$ with
    $ {\pprec} = {<} \cap {\{(e,e')\ |\ e\not\in\anR \textit{ or } \un{e}\lhd
      e'\}}$ is an \aes.
  \end{enumerate}
\end{defi}
\noindent 
An \raes is defined in terms of a set of events $E$, with those in $\anR$
being reversible.

Causation ($<$) and weak causality ($\nearrow$) delineate the forward flow,
while reverse causation ($\prec$, depicted as solid blue arrows) and
prevention ($\lhd$, represented by dashed blue arrows) describe the backward
flow. Weak causality serves the same role as in \aeses, wherein $e\nearrow e'$
asserts that event $e$ cannot occur after event $e'$. Prevention, on the other
hand, governs the undoing of events: $\un{e}\lhd e'$ indicates that event $e$
cannot be undone if event $e'$ has occurred. Similar to causality in \aeses,
causation indicates causal dependencies.
Consistent with \aeses, each event has the finite set of causes
$\histtwo{e}{<}$, and these sets do not contain cyclic dependencies induced by
causation and weak causality (Condition~\ref{cond:3}). The acyclicity of
$\histtwo{e}{<}$ implies that for all $e, e'\in E$, if $e < e'$, then
$\neg(e'\nearrow e)$.
Reverse causation specifies the causes for the undoing of each event:
$e \prec \un{e'}$ states that $e'$ can be undone only if $e$ has occurred.
Hence, condition~\ref{cond:4} simply states that an event $u$ can be undone
only if it has occurred. Condition~\ref{cond:3bis}, which is analogous to
Condition~\ref{cond:3}, establishes that each undoing has a finite set of
causes.
Reverse causation precisely states the causes for undoing each event:
$e \prec \un{e'}$ asserts that event $e'$ can be undone only if event $e$ has
occurred. Therefore, Condition~\ref{cond:4} straightforwardly defines that an
event $u$ can be undone only if it has occurred. Condition~\ref{cond:3bis},
akin to Condition~\ref{cond:3}, establishes that each undoing has a finite set
of causes.
Just as causation and weak causality demand consistency, wherein an event
cannot have precedence over some of its causes (as defined in
Condition~\ref{cond:3}), a similar requirement is imposed on the backward flow
by Condition~\ref{cond:5b}. Specifically, reverse causation must align with
prevention: a cause $e$ of the undoing of $u$ (i.e., $e\prec \un{u}$) cannot
prevent the undoing, i.e., $\neg(\un{u}\lhd e)$.
The final condition is the most intricate. It's noteworthy that the definition
of \raes does not mandate $(E, <, \nearrow)$ to be an \aes, particularly
because conflicts may not be inherited along causation. This flexibility is
crucial for accommodating out-of-causal-order reversibility (see the example
below).
Condition~\ref{cond:5b} considers instead the relation $\pprec$, dubbed
\emph{sustained causation}. The sustained causation is derived by excluding
from causation all pairs $e < e'$ where event $e$ can be undone even when $e'$
has occurred (i.e., $\un{e}\lhd e'$ does not hold).
It's worth noting that $\pprec$ coincides with $<$ when $\anR = \emptyset$;
thus, $(E, \emptyset, <, \nearrow, \emptyset, \emptyset)$ indeed constitutes
an \aes.

\begin{exa}\label{ex:raes}
  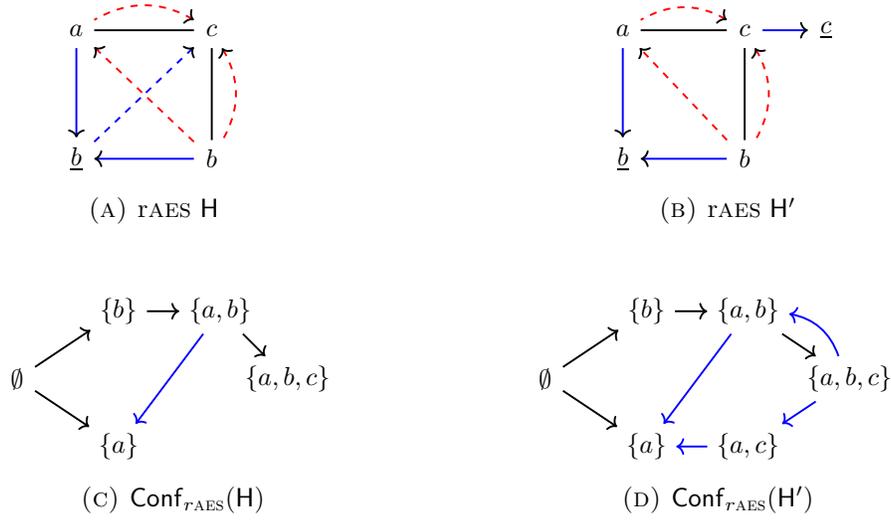
\begin{figure}[t]
    \begin{subfigure}{.45\textwidth}
      \centerline{\scalebox{1}{\input{figures/exaes-raes.tex}}}
      \caption{\raes{} $\mathsf{H}$}\label{fig:raesuno}
    \end{subfigure}\qquad
    \begin{subfigure}{.45\textwidth}
      \centerline{\scalebox{1}{\input{figures/exaes-raesbis.tex}}}
      \caption{\raes{} $\mathsf{H}'$}\label{fig:raesdue}
    \end{subfigure}
    \vspace*{.8cm}
    
    \begin{subfigure}{.45\textwidth}
      \centerline{\scalebox{1}{\input{figures/exaes-confraes.tex}}}
      \caption{$\Conf{\mathsf{H}}{\raes}$}
      \label{fig:raesconfuno}
    \end{subfigure}\quad
    \begin{subfigure}{.45\textwidth}
      \centerline{\scalebox{1}{\input{figures/exaes-confraesbis.tex}}}
      \caption{$\Conf{\mathsf{H}'}{\raes}$}
      \label{fig:raesconfdue}
    \end{subfigure}
    \caption{Two \raes{}es and their associated configurations.}
  \end{figure}

  Consider $\mathsf{H} = (E, \anR, <, \nearrow, \prec, \lhd)$ as depicted in
  \Cref{fig:raesuno}. The (forward) events are $\setenum{a, b, c}$, where $b$
  is the only reversible one ($\anR = \setenum{b}$). \emph{Causation} is
  defined such that $a < c$ and $b < c$, and weak causality states
  $a \nearrow c, b \nearrow c, b \nearrow a$. \emph{Reverse causation} is such
  that $a \prec \un{b}$ and $b \prec \un{b}$, and prevention is
  $\un{b}\lhd c$. Note that $c$ is caused by $a$ and $b$, with $b$ weakly
  causing $a$ (or $a$ being in asymmetric conflict with $b$). Moreover, $b$
  can be reversed only when $a$ is present and $c$ has not been executed. In
  this case, \emph{sustained causation} coincides with causation because the
  only reversible event $b$ cannot be reversed if $c$ (which causally depends
  on $b$) is present. It is routine to check that $(E, \pprec, \nearrow)$ is
  an \aes, and $\mathsf{H}$ is an \raes.

  Consider $\mathsf{H}' = (E, \anR', <, \nearrow, \prec', \lhd')$ shown in
  Fig.~\ref{fig:raesdue}, which has the same set of events, causation, and
  weak causality as $\mathsf{H}$ but also takes $c$ as reversible (i.e.,
  $\anR' = \setenum{b, c}$). Reverse causation is extended with the pairs
  $c \prec' \un{c}$, $a \prec' \un{b}$, and $b \prec' \un{b}$. Observe that
  $b$ can be reversed even if the event $c$ (which depends on $b$) has
  occurred. This is known as out-of-causal-order reversibility. In this case,
  sustained causation consists only of $a \pprec' c$, i.e., the pair $b < c$
  is removed because $b$ can be reversed despite $c$ (which causally depends
  on $b$) having occurred. It can be checked that $(E, \pprec', \nearrow)$ is
  an \aes, and $\mathsf{H'}$ is an \raes.
\end{exa}

\begin{rems}
  For the sake of the presentation, \Cref{de:raes} deviates in style from the
  original definition in~\cite{rpes}, where causation and reverse causation
  are merged, and weak causality and prevention are combined in a single
  relation. Additionally, we explicitly stipulate that $(E, \pprec, \nearrow)$
  must constitute an \aes, eliminating the need to restate conditions.
  Further discussion is available in the Appendix.
\end{rems}

The definition of configurations in \raeses has an operational flavor grounded
in the concept of enabling.
Consider $\mathsf{H} = (E, \anR, <, \nearrow, \prec, \lhd)$ as an \raes, and
let $X \subseteq E$ be a set of events such that $\nearrow$ is acyclic on $X$.
For $A\subseteq E$ and $B\subseteq \anR$, we say that $A\cup\un{B}$ is enabled
at $X$ if $A\cap X = \emptyset$, $B\subseteq X$, $\nearrow$ is acyclic on
$A \cup X$, and
\begin{enumerate}
\item for every $e\in A$, if $e' < e$ then $e'\in X\setminus B$ and if
  $e \nearrow e'$ then $e'\not\in X\cup A$; and
\item for every $\anr\in B$, if $e'\prec \un{\anr}$ then
  $e'\in X\setminus(B\setminus\setenum{\anr})$ and if $\un{\anr}\lhd e'$ then
  $e'\not\in X\cup A$.
\end{enumerate}
The first condition ensures that $X$ contains all the causes of the events
to be added (i.e., those in $A$) but none of their preventing events. The
second condition asserts that $X$ includes the reverse causes of the
events to be undone (i.e., those in $B$) but none of the preventing ones. If
$A\cup\un{B}$ is \emph{enabled} at $X$, then $X' = (X\setminus B)\cup A$ can be
reached from $X$, denoted as $X\ \xlongrightarrow{A\cup\underline{B}}\ X'$.
The initial condition above can be restated as follows:
$\forall e\in A. \histtwo{e}{<}\subseteq X$ and $X\cup A$ extends $X$.
\begin{defi}\label{de:rpes-conf}
  Let $\mathsf{H} = (E, \anR, <, \nearrow, \prec, \lhd)$ be an \raes{} and
  $X\subseteq E$ a set of events that is well-founded with respect to
  $(\nearrow \cup <)$. Then, $X$ is a \emph{(reachable) configuration}
  if there exist two sequences of sets $A_i$ and $B_i$, for $i=1,\ldots,n$,
  such that
 \begin{enumerate}
  \item $A_i\subseteq E$ and $B_i\subseteq \anR$ for all $i$, and
  \item $X_i\ \xlongrightarrow{A_i\cup\underline{B_i}}\ X_{i+1}$ for all $i$,
    and $X_1 = \emptyset$ and $X_{n+1}=X$.
 \end{enumerate}
 The set of configurations of $\mathsf{H}$ is denoted by
 $\Conf{\mathsf{H}}{\raes}$.
\end{defi}

\begin{exa}\label{ex:raes-conf}
  The configurations of the \raeses in Example~\ref{ex:raes} (along with the
  steps to reach them) are illustrated in
  \Cref{fig:raesconfuno,fig:raesconfdue}. Notice that $\setenum{a, c}$---a
  configuration of $\mathsf{H}'$ but not of $\mathsf{H}$---is reached from the
  configuration $\setenum{a, b, c}$ by undoing $b$.
\end{exa}

\begin{defi}\label{de:raes-morphisms}
  Let $\mathsf{H}_i = (E_i, \anR_i, <_i, \nearrow_i, \prec_i, \lhd_i)$ with
  $i=0,1$ be two \raeses. An \raes{}-morphism
  $f : \mathsf{H}_0\rightarrow \mathsf{H}_1$ satisfies the conditions of an \aes{}-morphism
  $f : (E_0, <_0, \nearrow_0) \rightarrow (E_1, <_1, \nearrow_1)$ such that
  \begin{enumerate}
  \item\label{de:raes-morphisms-1} $f(\anR_0)\subseteq \anR_1$;
  \item\label{de:raes-morphisms-2} for all $\anr \in \anR_0$, if $f(\anr)\neq \bot$ then
    $\histtwo{\un{f({\anr})}}{\prec_1}\subseteq
    f(\histtwo{\un{\anr}}{\prec_0})$; and
  \item\label{de:raes-morphisms-3} for all $e \in E_0$ and ${\anr}\in{\anR_0}$, if
    $f(e)\neq \bot \neq f(\anr)$ then
    $\un{f({\anr})}\lhd_1 f(e)\ \Rightarrow \un{\anr}\lhd_0 e$.
  \end{enumerate}
\end{defi}
\noindent 
Observe that $(E_0, <_0, \nearrow_0)$ and $(E_1, <_1, \nearrow_1)$ are not
necessarily \aeses, but here it is enough that $f$ works on causation and weak
causality as an \aes{}-morphism.
Notice that \raes{}-morphisms preserve the causes (and the reverse causes) of
each event (resp., reversing event), and reflect preventions.

\begin{prop}\label{pr:raesmorph-conf}
  Let $f : \mathsf{H}_0\rightarrow \mathsf{H}_1$ be an \raes{}-morphism and let
  $X\in \Conf{\mathsf{H}_0}{\raes}$.
  Then $f(X)\in  \Conf{\mathsf{H}_1}{\raes}$.
\end{prop}

\ifreport
\begin{proof}
  It suffices to demonstrate that, given $X\subseteq E_0$ such that
  $\nearrow_0$ is acyclic on $X$, if $A\cup\un{B}$ is enabled at $X$ (with
  $A\subseteq E_0$ and $B\subseteq \anR_0$), and
  $X\ \xlongrightarrow{A\cup\un{B}}\ X'$ (with $X' = (X\setminus B)\cup A$),
  then $f(A)\cup\un{f(B)}$ is enabled at $f(X)$.
 Observe that $f(A)\cap f(X) = \emptyset$ since $A \cap X = \emptyset$ and
 $f(B) \subseteq f(X)$ due to $B \subseteq X$.
 We now prove that $\nearrow_1$ is acyclic on $f(X\cup A) = f(X)\cup f(A)$ by
 contradiction. Suppose there exists a sequence of events
 $f(e_0), \dots, f(e_n)$ in $f(X\cup A)$ such that
 $f(e_i) \nearrow_1 f(e_{i+1})$, with $0\leq i < n$, and
 $f(e_n) \nearrow_1 f(e_0)$. However, as
 $f\!:\! (E_0, <_0,\\ \nearrow_0)\rightarrow (E_1, <1, \nearrow_1)$ is an
 \aes-morphism, we have $e_i\nearrow_0 e{i+1}$ for all $i\in\setenum{0,n-1}$,
 and $e_n\nearrow_0 e_0$, contradicting the acyclicity of $\nearrow_0$ on
 $X\cup A$.

 Now consider $e\in A$ such that $f(e)$ is defined and take $e' <_0 e$ such
 that $f(e')$ is defined. As $f$ is an \aes-morphism, if
 $e' \in X\setminus B$, then $f(e') \in f(X)\setminus f(B)$, and if
 $f(e) \nearrow_1 f(e')$, we have $e\nearrow_0 e'$, implying that
 $e'\not\in X\cup A$. Therefore, $f(e')\not\in f(X)\cup f(A)$.

 Finally, consider $\anr\in B$ and assume $f(\anr)$ is defined. Take
 $e' \prec_0 \anr$. If $f(e')$ is defined, and since
 $\histtwo{\un{f({\anr})}}{\prec_1}\subseteq f(\histtwo{\un{\anr}}{\prec_0})$,
 we have that $f(e')\in f(X)\setminus (f(B)\setminus\setenum{f(\anr)})$
 because $e'\in X\setminus (B\setminus\setenum{\anr})$. Now, consider
 $\un{f(\anr)} \lhd_1 f(e')$, which implies $f(e')\not\in f(X)\cup f(A)$
 because $\un{f(\anr)} \lhd_1 f(e')$ implies $\anr \lhd_0 e'$ and
 $e'\not\in X\cup A$. Hence
 $f(X)\ \xlongrightarrow{f(A)\cup\un{f(B)}}\ f(X')$. Observing that
 configurations are subsets of events reachable from the empty one, we
 establish the thesis.
\end{proof}
\fi

As shown in \cite{GPY:categories}, \raes{}-morphisms compose; hence \raeses
and \raes{}-morphisms are a category, denoted with $\mathbf{RAES}$.
Moreover, $\mathbf{AES}$ is a full and faithful subcategory of
$\mathbf{RAES}$.

%% file: figures/exaes-raes.tex
\scalebox{0.9}{\begin{tikzpicture}
\usetikzlibrary{decorations.pathmorphing}

\tikzset{snake it/.style={decorate, decoration=snake}}
\tikzstyle{cau}=[-,thick]
\tikzstyle{wcau}=[dashed,->,thick,draw=red]
\tikzstyle{wcaub}=[dashed,<-,thick,draw=red]
\tikzstyle{revcau}=[->,thick,draw=blue]
\tikzstyle{prev}=[dashed,<-,thick,draw=blue]
\tikzstyle{transition}=[rectangle, draw=none,thick,minimum size=5mm]
\node[transition] (unb) at (0,0) {$\underline{b}$}
;
\node[transition] (a) at (0,1.9) {$a$}
edge[revcau] (unb)
;
\node[transition] (b) at (2,0)  {$b$}
edge[revcau] (unb)
edge[wcau] (a)
;
\node[transition] (c) at (2,1.9)  {$c$}
edge[cau] (a)
edge[wcaub, bend right = 30] (a)
edge[cau] (b)
edge[wcaub, bend left = 30] (b)
edge[prev] (unb)
;
\end{tikzpicture}}

%% file: figures/exaes-raesbis.tex
\scalebox{0.9}{\begin{tikzpicture}
\usetikzlibrary{decorations.pathmorphing}

\tikzset{snake it/.style={decorate, decoration=snake}}
\tikzstyle{cau}=[-,thick]
\tikzstyle{wcau}=[dashed,->,thick,draw=red]
\tikzstyle{wcaub}=[dashed,<-,thick,draw=red]
\tikzstyle{revcau}=[->,thick,draw=blue]
\tikzstyle{prev}=[dashed,<-,thick,draw=blue]
\tikzstyle{transition}=[rectangle, draw=none,thick,minimum size=5mm]
\node[transition] (unb) at (0,0) {$\underline{b}$}
;
\node[transition] (unc) at (3,1.9) {$\underline{c}$}
;
\node[transition] (a) at (0,1.9) {$a$}
edge[revcau] (unb)
;
\node[transition] (b) at (1.8,0)  {$b$}
edge[revcau] (unb)
edge[wcau] (a)
;
\node[transition] (c) at (1.8,1.9)  {$c$}
edge[cau] (a)
edge[wcaub, bend right = 30] (a)
edge[cau] (b)
edge[wcaub, bend left = 30] (b)
edge[revcau] (unc)
;
\end{tikzpicture}}

%% file: figures/exaes-confraes.tex
\scalebox{0.9}{\begin{tikzpicture}
\usetikzlibrary{decorations.pathmorphing}

\tikzset{snake it/.style={decorate, decoration=snake}}
\tikzstyle{tr}=[<-,thick]
\tikzstyle{rtr}=[<-,thick,blue]
\tikzstyle{rtrb}=[->,thick,blue]
\tikzstyle{transition}=[rectangle, draw=none,thick,minimum size=5mm]
\node[transition] (empty) at (0,1)  {$\emptyset$}
;
\node[transition] (a) at (1.5,0)  {$\{a\}$}
edge[tr] (empty)
;
\node[transition] (b) at (1.5,2){$\{b\}$}
edge[tr] (empty)
;
\node[transition] (ba) at (3,2) {$\{a,b\}$}
edge[tr] (b)
edge[rtrb] (a)
;
\node[transition] (bac) at (4.0,1) {$\{a,b,c\}$}
edge[tr] (ba)
;

\end{tikzpicture}}

%% file: figures/exaes-confraesbis.tex
\scalebox{0.9}{\begin{tikzpicture}
\usetikzlibrary{decorations.pathmorphing}

\tikzset{snake it/.style={decorate, decoration=snake}}
\tikzstyle{tr}=[<-,thick]
\tikzstyle{rtr}=[<-,thick,blue]
\tikzstyle{rtrb}=[->,thick,blue]
\tikzstyle{transition}=[rectangle, draw=none,thick,minimum size=5mm]
\node[transition] (empty) at (0,1)  {$\emptyset$}
;
\node[transition] (a) at (1.5,0)  {$\{a\}$}
edge[tr] (empty)
;
\node[transition] (b) at (1.5,2){$\{b\}$}
edge[tr] (empty)
;
\node[transition] (ba) at (3,2) {$\{a,b\}$}
edge[tr] (b)
edge[rtrb] (a)
;
\node[transition] (ac) at (3,0) {$\{a,c\}$}
edge[rtrb] (a)
;
\node[transition] (bac) at (4.5,1) {$\{a,b,c\}$}
edge[tr] (ba)
edge[rtrb] (ac)
edge[rtrb, bend right] (ba)
;

\end{tikzpicture}}

%% file: coprodraes.tex

\subsection{Constructions}\label{sec:constr-raes}
We conclude the section showing a categorical construction for \raes.
The category of \raes{s} has coproduct, as shown in \cite{GPY:categories}. We recall here the construction
adapting it to our definition of \raes which is equivalent to the one in \cite{rpes} and
\cite{GPY:categories}, as we discuss in the appendix.

\begin{prop}\label{de:raes-coprod}
 Let $\mathsf{H}_0 = (E_0, \anR_0, <_0, \nearrow_0, \prec_0, \lhd_0)$ and
 $\mathsf{H}_1 = (E_1, \anR_1, <_1, \nearrow_1, \prec_1, \lhd_1)$ be two \raes{s}. Then
 $\mathsf{H}_0 + \mathsf{H}_1 = (E, \anR, <, \nearrow, \prec, \lhd)$ where
 \begin{itemize}
  \item $E = \setenum{0}\times E_0 \cup \setenum{1}\times E_1$;
  \item $\anR = \setenum{0}\times\anR_0 \cup \setenum{1}\times\anR_1$;
  \item $(i,e) < (j,e')$ whenever $i = j$ and $e <_i e'$;
  \item $(i,e) \nearrow (j,e')$ whenever either $i = j$ and $e \nearrow_i e'$ or $i\neq j$;
  \item $(i,e) \prec (j,\anr)$ whenever $i = j$ and $e \prec_i \anr$; and
  \item $(i,\anr) \lhd (j,e)$ whenever either $i = j$ and $\anr \lhd_i e$ or $i\neq j$,
 \end{itemize}
 is their \emph{coproduct} and $\iota_i : \mathsf{H}_i \to \mathsf{H}_0 + \mathsf{H}_1$ defined as
 $\iota_i(e) = (i,e)$ are the injections.
\end{prop}
\begin{proof}
 $\mathsf{H}_0 + \mathsf{H}_1$ is clearly a \raes. To show that it is indeed a coproduct, we consider any other
 \raes $\mathsf{H}' = (E', \anR', <', \nearrow', \prec', \lhd')$ together with two
 morphisms $f_i : \mathsf{H}_i \to \mathsf{H}'$. We now show that the mapping
 $g : \mathsf{H}_0 + \mathsf{H}_1 \to \mathsf{H}'$ defined as $g(i,e) = f_i(e)$ is a \raes-morphism and it is
 unique.
 Clearly $\histtwo{g(i,e)}{<'} = \histtwo{f_i(e)}{<'} \subseteq f_i(\histtwo{e}{<_i})$ as $f_i$ is a
 \raes-morphism, moreover if $g(i,e)$ and $g(j,e')$ are defined and $g(i,e) \nearrow' g(j,e')$ we have two
 cases: either $i = j$ and then $g(i,e) = f_i(e)$, $g(j,e') = f_i(e')$ and $f_i(e)\nearrow f_i(e')$ is because
 $e\nearrow_i e'$, or $i\neq j$, but in this case by construction we have $g(i,e) \nearrow g(j,e')$.
 Assume now that $g(i,e)$ and $g(j,e')$ are defined and equal. There are two cases: if $i = j$ then
 $f_i(e) \# f_i(e')$ and $e \#_i e'$, and if $i \neq j$ then we have again
 $g(i,e) \nearrow g(j,e')$ and $g(j,e') \nearrow g(i,e)$ which implies $g(i,e) \# g(j,e')$.
 Turning to the reverse part, clearly $g(\setenum{0}\times\anR_0 \cup \setenum{1}\times\anR_1) =
 g(\setenum{0}\times\anR_0) \cup g(\setenum{1}\times\anR_1) = f_0(\anR_0) \cup
 f_1(\anR_1) \subseteq \anR'$, and
 $\histtwo{\un{g(i,\anr)}}{\prec'} = \histtwo{\un{f_i(\anr)}}{\prec'} \subseteq f_i(\histtwo{\un{\anr}}{\prec})$
 being $f_i$ a \raes-morphism, and for
 $g(i,\anr) \lhd' g(j,e')$ the reasoning goes as for $\nearrow'$, hence $(i,\anr) \lhd (j,e')$, as required.
 Uniqueness follows from the fact that the events are a coproduct in the category of sets and partial mapping.
\end{proof}

%% file: racn.tex

\section{Reversible Asymmetric Causal Nets}\label{sec:rcn}
We now introduce Asymmetric Causal Nets as a subclass of nets with inhibitor
arcs, wherein standard notions of causality and conflicts among transitions
are modelled via inhibitor arcs.

Firstly, we revisit the notion of nets with inhibitor arcs and subsequently
introduce asymmetric causal nets along with their reversible versions. We
develop appropriate notions of morphisms, turning asymmetric causal nets and
reversible asymmetric causal nets into categories.

\input{ipt}
\input{revised-causal-nets-defs}
\ifreport
\input{acn-configuration}
\fi
\input{revised-morphisms}
\ifreport
\input{acn-morph-configuration}
\fi
\input{revised-rev-causal-nets-defs}
\ifreport
\input{racn-configurations}
\fi
\input{revised-rev-morphisms}
\ifreport
\input{racn-morph-configuration}
\fi
\ifreport
\input{coprodracn}

\fi

%% file: ipt.tex

\subsection{Nets with inhibitor arcs}\label{sec:inet}
We summarise the basics of Petri net with inhibitor arcs along the
lines of \cite{MR:CN,BBCP:rivista}.
We write $\nat$ for the set of natural numbers.
 A \emph{multiset} over a set $A$ is a function $m : A
 \rightarrow \nat$.
 We assume the usual operations of union ($+$) and difference ($-$) on
 multisets, and write $m \subseteq m'$ if $m(a) \leq m'(a)$ for all $a \in A$.
 We will use $\multisetenum{\ldots}$ when enumerating the elements of
   a multiset.
The multiset $\flt{m}$ is defined such that $\flt{m}(a) = 1$ if $m(a) > 0$ and
$\flt{m}(a) = 0$ otherwise.
We often confuse a multiset $m$ with the set $\setcomp{a\in A}{m(a) \neq 0}$
when $m = \flt{m}$. In such cases, $a\in m$ denotes $m(a) \neq 0$, and
$m\subseteq A$ signifies that $m(a) = 1$ implies $a\in A$ for all $a$.
The underlying set of a multiset $m$, namely the one formed by the elements
$a$ with $m(a)$, is precisely $\flt{m}$.
Additionally, we will employ standard set operations like $\cap$, $\cup$, or
$\setminus$.
The set of all multisets over $A$ is denoted as $\mu A$; the symbol $\zero$
stands for the unique multiset such that $\flt{\zero} = \emptyset$.

\begin{defi}\label{de:contextualnet}
  A \emph{Petri net with inhibitor arcs} (\inet for short) is a tuple
  $N = \langle S, T, F, I, \mathsf{m}\rangle$, where $S$ is a set of
  \emph{places}, $T$ is a set of \emph{transitions} such that
  $S \cap T = \emptyset$, $F \subseteq (S\times T)\cup (T\times S)$ is the
  \emph{flow} relation, $I \subseteq S\times T$ is the \emph{inhibiting}
  relation, and $\mathsf{m}\in \mu S$ is the \emph{initial marking}.
\end{defi}

Given an \inet $N = \langle S, T, F, I, \mathsf{m}\rangle$ and $x\in S\cup T$,
the {\em pre-} and {\em postset} of $x$ are respectively defined as the
(multi)sets $\pre{x} = \setcomp{y}{(y,x)\in F}$ and
$\post{x} = \setcomp{y}{(x,y)\in F}$.
If $x\in S$ then $\pre{x} \in \mu T$ and $\post{x} \in \mu T$; analogously, if
$x\in T$ then $\pre{x}\in\mu S$ and $\post{x} \in \mu S$.
The {\em inhibitor set} of a transition $t$ is the (multi)set
$\inib{t} = \setcomp{s}{(s,t)\in I}$.
The definition of $\pre{\cdot}, \post{\cdot}, \inib{\cdot}$ generalise
straightforwardly to multisets of transitions.

\begin{exa}\label{ex:cont-net1}

  \begin{figure}[t]
    \begin{subfigure}{.45\textwidth}
      \centerline{\scalebox{1}{\input{examples/example-cn-a}}}
      \caption{$N_1$}
      \label{fig:pcn-a}
    \end{subfigure}
    \qquad
    \begin{subfigure}{.45\textwidth}
      \centerline{\scalebox{1}{\input{examples/example-cn-b}}}
      \caption{$N_2$}
      \label{fig:pcn-b}
    \end{subfigure}
    \vspace*{.8cm}
    
    \begin{subfigure}{.45\textwidth}
      \centerline{\scalebox{1}{\input{examples/example-cn-c}}}
      \caption{$N_3$}
      \label{fig:pcn-c}
    \end{subfigure}
    \quad
    \begin{subfigure}{.45\textwidth}
      \centerline{\scalebox{1}{\input{examples/example-cn-d}}}
      \caption{$N_4$}
      \label{fig:pcn-d}
  \end{subfigure}
  \caption{Some \inet{}s.}
  \label{fig:pcn}
\end{figure}

  \Cref{fig:pcn} introduces some \inets.
  The \inet $N_1$ in \Cref{fig:pcn-a} has six places (named $s_i$) and three
  transitions $a$, $b$, and $c$. The initial marking is
  $ \mathsf{m} = \setenum{s_1, s_2, s_3}$. For instance, the transition $b$
  consumes a token from $s_2$ and produces a token in $s_5$ and it is
  inhibited by $s_1$, meaning its pre-, post-, and inhibiting sets are
  $\pre b = \setenum{s_2}$, $\post b = \setenum{s_5}$, and
  $\inib b = \setenum{s_1}$, respectively.
\end{exa}

A (multiset of) transition(s) $A\in \mu T$ is {\em enabled at a marking}
$m\in \mu S$, written $m\trans{A}$, if ${\pre{A}} \subseteq m$,
$\inib{A} \cap\ \flt{m} = \emptyset$ and
$\forall t\in \flt{A}.\ \inib{t}\cap \post{(A-\setenum{t})} = \emptyset$.
Intuitively, $A$ is enabled at $m$ if $m$ contains the tokens to be consumed
by $A$ (${\pre{A}} \subseteq m$) and none of the transitions in $A$ is
inhibited in $m$ ($\inib{A} \cap\ \flt{m} = \emptyset$). The last condition
avoids cases in which a transition in $A$ produces tokens that inhibit other
transition in $A$. Observe that the multiset $\zero$ is enabled at every
marking.
If $A$ is enabled at $m$, then it can \emph{fire} and its firing produces the
marking $m' = m - \pre{A} + \post{A}$, which is written $m\trans{A}m'$.
Hereafter, we assume that each transition $t$ is defined such that
$\pre{t}\neq\emptyset$, i.e., it cannot fire \emph{spontaneously} without
consuming tokens.

A marking $m$ is \emph{reachable} if there exists a sequence of firings
$m_i\trans{A_i}m_{i+1}$ originated in the initial marking and leading to $m$;
$\reachMark{N}$ stands for the set of reachable markings of $N$. An \inet $N$
is {\em safe} if every reachable marking is a set, i.e.,
$\forall m\in \reachMark{N}.m = \flt{m}$.
From now on, we will only consider safe \inets.

\begin{exa}
  Consider the \inet{} $N_4$ in \Cref{fig:pcn-d}. Both $b$ and $c$ are enabled
  at marking $m = \setenum{s_1, s_2, s_3}$. On the contrary, $a$ is not
  enabled because it is inhibited by the token in the place $s_2$. The firing
  of $b$ on $m$ produces the marking $m' = \setenum{s_1, s_3, s_5}$, i.e.
  $m\trans{b} m'$. The transition $a$ is enabled at $m'$ while $c$ is disabled
  and cannot be fired because of the token in $s_5$. The firing of $a$ on $m'$
  produces $m'' = \setenum{s_3, s_4, s_5}$. The reachable markings of $N_4$
  are $\setenum{s_1, s_2, s_3}, \setenum{s_1, s_3, s_5}$
  $\setenum{s_1, s_2, s_6}$ and $\setenum{s_3, s_4, s_5}$.
\end{exa}

%% file: examples/example-cn-a.tex
\scalebox{0.9}{\begin{tikzpicture}
\tikzstyle{inhibitorred}=[o-, draw=red,thick]
\tikzstyle{inhibitorblu}=[o-, draw=blue,thick]
\tikzstyle{pre}=[<-,thick]
\tikzstyle{post}=[->,thick]
\tikzstyle{readblue}=[-, draw=blue,thick]
\tikzstyle{transition}=[rectangle, draw=black,thick,minimum size=5mm]
\tikzstyle{place}=[circle, draw=black,thick,minimum size=5mm]
\tikzstyle{invplace}=[circle, draw=black!0,thick,minimum size=5mm]
\node[place,tokens=1] (p1) at (0,2.5) [label=left:$s_1$] {};
\node[place,tokens=1] (p3) at (2,2.5) [label=right:$s_2$] {};
\node[place,tokens=1] (p5) at (4,2.5) [label=right:$s_3$] {};

\node[place] (p2) at (0,0) [label=left:$s_4$] {};
\node[place] (p4) at (2,0) [label=right:$s_5$] {};
\node[place] (p6) at (4,0) [label=right:$s_6$] {};
\node[invplace] (ii) at (5,0) {};

\node[transition] (a) at (0,1.25)  {$a$}
edge[pre] (p1)
edge[inhibitorred] (p6)
edge[inhibitorred] (p4)
edge[post](p2);

\node[transition] (b) at (2,1.25) {$b$}
edge[pre] (p3)
edge[post] (p4)
edge[inhibitorred] (p1)
edge[inhibitorred] (p6)
;

\node[transition] (c) at (4,1.25) {$c$}
edge[pre] (p5)
edge[post] (p6)
edge[inhibitorred] (p3)
edge[inhibitorred] (p1)
;

\end{tikzpicture}}

%% file: examples/example-cn-b.tex
\scalebox{0.9}{\begin{tikzpicture}
\tikzstyle{inhibitorred}=[o-, draw=red,thick]
\tikzstyle{inhibitorblu}=[o-, draw=blue,thick]
\tikzstyle{pre}=[<-,thick]
\tikzstyle{post}=[->,thick]
\tikzstyle{readblue}=[-, draw=blue,thick]
\tikzstyle{transition}=[rectangle, draw=black,thick,minimum size=5mm]
\tikzstyle{place}=[circle, draw=black,thick,minimum size=5mm]
\tikzstyle{invplace}=[circle, draw=black!0,thick,minimum size=5mm]
\node[place,tokens=1] (p1) at (0,2.5) [label=left:$s_1$] {};
\node[place,tokens=1] (p3) at (2,2.5) [label=right:$s_2$] {};
\node[place] (p2) at (0,0) [label=left:$s_3$] {};
\node[place] (p4) at (2,0) [label=right:$s_4$] {};
\node[transition] (a) at (0,1.25)  {$a$}
edge[pre] (p1)
edge[post](p2);

\node[transition] (b) at (2,1.25) {$b$}
edge[pre] (p3)
edge[post] (p4)
edge[inhibitorred] (p2)
;
\end{tikzpicture}}

%% file: examples/example-cn-c.tex
\scalebox{0.9}{\begin{tikzpicture}
\tikzstyle{inhibitorred}=[o-, draw=red,thick]
\tikzstyle{inhibitorblu}=[o-, draw=blue,thick]
\tikzstyle{pre}=[<-,thick]
\tikzstyle{post}=[->,thick]
\tikzstyle{readblue}=[-, draw=blue,thick]
\tikzstyle{transition}=[rectangle, draw=black,thick,minimum size=5mm]
\tikzstyle{place}=[circle, draw=black,thick,minimum size=5mm]
\node[place,tokens=1] (p1) at (0,2.5) [label=left:$s_1$] {};
\node[place,tokens=1] (p3) at (2,2.5) [label=right:$s_2$] {};
\node[place] (p2) at (0,0) [label=left:$s_3$] {};
\node[place] (p4) at (2,0) [label=right:$s_4$] {};
\node[transition] (a) at (0,1.25)  {$a$}
edge[pre] (p1)
edge[inhibitorred] (p4)
edge[post](p2);

\node[transition] (b) at (2,1.25) {$b$}
edge[pre] (p3)
edge[post] (p4)
edge[inhibitorred] (p2)
;
\end{tikzpicture}}

%% file: examples/example-cn-d.tex
\scalebox{0.9}{\begin{tikzpicture}
\tikzstyle{inhibitorred}=[o-, draw=red,thick]
\tikzstyle{inhibitorblu}=[o-, draw=blue,thick]
\tikzstyle{pre}=[<-,thick]
\tikzstyle{post}=[->,thick]
\tikzstyle{readblue}=[-, draw=blue,thick]
\tikzstyle{transition}=[rectangle, draw=black,thick,minimum size=5mm]
\tikzstyle{place}=[circle, draw=black,thick,minimum size=5mm]
\node[place,tokens=1] (p0) at (-2,2.5) [label=left:$s_1$] {};
\node[place] (p6) at (-2,0) [label=left:$s_4$] {};

\node[place,tokens=1] (p1) at (0,2.5) [label=left:$s_2$] {};
\node[place,tokens=1] (p3) at (2,2.5) [label=right:$s_3$] {};
\node[place] (p2) at (0,0) [label=left:$s_5$] {};
\node[place] (p4) at (2,0) [label=right:$s_6$] {};
\node[transition] (a) at (-2,1.25)  {$a$}
edge[pre] (p0)
edge[inhibitorred] (p1)
edge[inhibitorred] (p4)
edge[inhibitorred] (p2)
edge[post](p6)
;

\node[transition] (b) at (0,1.25)  {$b$}
edge[pre] (p1)
edge[inhibitorred] (p4)
edge[post](p2);

\node[transition] (c) at (2,1.25) {$c$}
edge[pre] (p3)
edge[post] (p4)
edge[inhibitorred] (p2)
edge[inhibitorred] (p6)
;
\end{tikzpicture}}

%% file: revised-causal-nets-defs.tex

\subsection{Asymmetric Causal Nets}
In this section, we introduce Asymmetric Causal Nets, a class of \inets that
generalises the concept of Causal Nets introduced in \cite{lics} to
accommodate asymmetric conflicts. Broadly speaking, we focus on \inets where
dependencies between transitions arise solely from \emph{inhibitor} arcs.
Similar to causal nets, the property $\post{t} \cap \pre{t'} = \emptyset$
holds for all transitions $t$ and $t'$. This means that if a place appears in
the preset of a transition, it does not appear in the postset of any
transition, and vice versa. Consequently, the flow relation induces an empty
causal relation.
However, causality can be recovered through inhibitor arcs. Intuitively, a
transition $t$ connected via an inhibitor arc to some place in the preset of
another transition $t'$ cannot be fired before $t'$ if we assume that the
preset of $t'$ is marked. This is exemplified by transitions $a$ and $b$ in
\Cref{fig:pcn-d}, where $a$ can only be fired after $b$. The induced
(immediate) causality relation $\lessdot$ is defined by $t \lessdot t'$ if and
only if $\pre{t}\cap\inib{t'}\neq\emptyset$, signifying that the firing of $t$
consumes (at least) one of the tokens inhibiting the firing of $t'$.

Additionally, asymmetric causal nets impose that places should not be shared
between the presets and postsets of transitions. Formally,
$\post{t} \cap \post{t'} \neq \emptyset\ \lor\ \pre{t} \cap \pre{t'} \neq
\emptyset$ implies $t = t'$ for all transitions $t$ and $t'$. This ensures
that the flow relation does not introduce forward or backward conflicts, which
need to be recovered from inhibitor arcs.
Note that a transition $t$ inhibited by some place in the postset of another
transition $t'$ cannot be fired if $t'$ has been fired, i.e., $t'$ prevents
$t$. This is exemplified by transitions $b$ and $a$ in \Cref{fig:pcn-b}, where
$b$ cannot be fired after $a$ (i.e., $b$ is prevented by $a$).
Therefore, the induced prevention relation $\preventedby$ is defined by
$t \preventedby t'$ if and only if $\post{t}\cap\inib{t'}\neq\emptyset$. We
use $\rcnprevent$ to denote the inverse of $\preventedby$. It is important to
note that $\rcnprevent$ is analogous to the weak causality of \aes{}es: if
$t'\ \rcnprevent\ t$, then $\post{t}\cap\inib{t'}\neq\emptyset$, indicating
that $t'$ cannot be fired if $t$ has been fired; however, $t$ can be fired
after $t'$.
Similar to AES, \emph{symmetric} conflicts are retrieved from prevention. That
is, transitions $t$ and $t'$ are in \emph{symmetric} conflict, denoted as
$t \cnconf t'$, whenever $t \preventedby t'$ and $t\ \rcnprevent\ t'$.

\begin{defi}\label{de:pre-acausal-net}
  Let $C = \langle S, T, F, I, \mathsf{m}\rangle$ be an \inet. $C$ is a
  \emph{pre asymmetric causal net} (p\paca) if the following conditions hold:
  \begin{enumerate}
  \item\label{pcn:cond1}
    $\forall t, t'\in T.\ \post{t}\cap\pre{t'} = \emptyset$;

  \item\label{pcn:cond2} $\forall t\in T$.
    $\card{\pre{t}} = \card{\post{t}} = 1$;
    $\mathsf{m} = S \setminus \post{T}$; and
  $\inib T \subseteq \pre T \cup \post T$;

  \item\label{pcn:cond3} $\lessdot^{+}$ is a partial order;

  \item\label{pcn:cond4}
    $\forall t\in T.\ \histtwo{t}{\lessdot} = \setcomp{t'\in
      T}{t'\lessdot^{\ast} t}$ is finite and $(\rcnprevent\cup\lessdot)$ is
    acyclic on $\histtwo{t}{\lessdot}$; and
  \item\label{pcn:cond5} $\forall t, t'\in T.$ if
      $\pre{t}\cap\inib{t'} \neq \emptyset$ then
      $\post{t}\cap\inib{t'}= \emptyset$ and if
      $\post{t}\cap\inib{t'} \neq \emptyset$ then
      $\pre{t}\cap\inib{t'}= \emptyset$.
  \end{enumerate}
\end{defi}
\noindent 
The first condition asserts that causal dependencies in a p\paca{} are not
established through the flow relation.
The second condition stipulates that each transition has just one place in its preset and one place in its postset, and places that are either isolated
  or part of the presets of all transitions are initially marked. The condition on inhibitor arcs, denoted as $\inib T \subseteq \pre T \cup \post T$, specifies that these arcs should not link transitions to isolated places. This is because their purpose is to represent dependencies between transitions that are associated via the flow relation to those places.
Since $\lessdot$ is meant to model causal dependencies, the third condition
requires its transitive closure $\lessdot^{+}$ to form a partial order.
The fourth condition mandates that each transition $t$ has a finite set of
causes, ensuring that $\lessdot^{+}$ is a well-founded partial order. By
enforcing the acyclicity of $(\rcnprevent\cup\lessdot)$ on the causes of every
transition $t$, we ensure that causes of each transition can be ordered to
satisfy both causality and prevention. This constraint excludes situations
where (i) prevention contradicts causality (e.g., $t \lessdot t'$ and
$t' \rcnprevent t$), (ii) circular chains of prevention exist (i.e.,
$t_0 \rcnprevent t_1\rcnprevent \ldots \rcnprevent t_n\rcnprevent t_0$), where
symmetric conflicts are a particular case, and (iii) self-blocked transitions
occur (i.e., $\lessdot$ needs to be irreflexive, implying
${\pre{t}} \cap{\inib{t}}=\emptyset$ for all transitions $t$). The
last condition simply says that a transition cannot depends on one that
prevents it.
It is worth to stress that, though in general markings are multisets, in our case they are indeed sets.

\begin{defi}\label{de:acausal-net}
  A p\paca{} $C = \langle S, T, F, I, \mathsf{m}\rangle$ is an
  \emph{asymmetric causal net} (\paca) if
  \begin{enumerate}
   \item\label{def:causality-saturated} for all $t,t'\in T$.
    $t \lessdot^{+} t'$ implies $t \lessdot t'$;
   \item\label{cond:lessdotimpliesnearrow}
    for all $t,t'\in T$. $t \lessdot t'$ implies $t\ \rcnprevent\ t'$; and
   \item\label{cond:confl-ereditati} for all $t, t', t''\in T$,
  $t\cnconf t' \land\ t'\lessdot t''$ imply $t \cnconf t''$.
  \end{enumerate}
\end{defi}

\noindent 
The additional conditions imposed on \paca{}s entails
that all dependencies among transitions must be
explicitly represented in the structure of the net; in other words, causality
is saturated (Condition~\ref{def:causality-saturated}) and
that all conflicts must
be explicitly represented in the net's structure, ensuring the presence of
inhibitor arcs for all inherited symmetric conflicts (Condition~\ref{cond:confl-ereditati}).
Condition~\ref{cond:lessdotimpliesnearrow} simply states that it
$\pre{t}\cap\inib{t'} \neq \emptyset$ then $\post{t'}\cap \inib{t}\neq \emptyset$. The
inhibitor arc $(s,t)$ with $\{s\} = \post{t'}$ is somehow superflous, but it follow
the intuition that in an \aes{} causality implies weak causality.

\begin{exa}\label{ex:relations-explained}
  The \inets depicted in \Cref{fig:pcn} satisfy the conditions of \pacas. The
  first two conditions of \cref{de:pre-acausal-net} are met by all four nets,
  as transitions do not share places in their pre and postsets. Additionally,
  the places in the presets of all transitions are the only ones initially
  marked.
  For $N_1$ (\Cref{fig:pcn-a}), we observe that $a \lessdot b$ and
  $b \lessdot c$, making $\lessdot^{+}$ a total order, and causality is
  saturated with the inclusion of $a \lessdot c$. Furthermore, $\preventedby$
  is empty (as no transitions have inhibitor arcs connected to the postset of
  other transitions), thus satisfying the fourth condition.
  In the case of the net $N_2$ (\Cref{fig:pcn-b}), the causality relation is
  empty, while prevention contains the sole pair $a \preventedby b$: $b$
  cannot be fired after $a$, but $a$ can be executed after $b$.
  \Cref{fig:pcn-c} presents a similar net where $a$ and $b$ are in symmetric
  conflict ($a \cnconf b$), meaning the execution of one prevents the other.
  Finally, \Cref{fig:pcn-d} illustrates a net where $b \lessdot a$,
  $b \cnconf c$, and $a \cnconf c$, with conflicts being inherited along the
  causality relation $\lessdot$.
\end{exa}

%% file: acn-configuration.tex

\subsection{Configurations of a (pre) Asymmetric Causal Nets}
We introduce additional concepts and findings related to (pre) \paca{s}, in
line with our previous work in \cite{lics}.

\input{aux-ipt}

\begin{defi}\label{de:ca-conf}
  Let $C = \langle S, T, F, I, \mathsf{m}\rangle$ be a p\paca. A set of
  transitions $X\subseteq T$ is a \emph{configuration} of $C$ if
  \begin{enumerate}
  \item $\forall t\in X$. $\histtwo{t}{\lessdot}\subseteq X$ (i.e., $X$ is
    \emph{left closed} with respect to $\lessdot$); and
  \item $\rcnprevent\cup\lessdot$ is acyclic on $X$.
  \end{enumerate}
  The set of configurations of a p\paca{} $C$ is denoted by
  $\Conf{C}{p\paca}$.
\end{defi}

The conditions imposed in the definition of \paca ensure that
$\rcnprevent = \preventedby$. In such case, the second condition can be
alternatively expressed as $\forall t, t' \in X.\ \neg (t \cnconf t')$.

If $X \in \Conf{C}{p\paca}$, then the transitions of $X$ can be partially
ordered with respect to $\lessdot \cup \ \preventedby$. This means that there
exists a sequence $t_1, \dots, t_n$ of the transitions in $X$ such that
$t_i \lessdot t_j$ or $t_i \ \rcnprevent\ t_j$ imply $i < j$.

\begin{prop}\label{lm:ordering-configuration}
 Let $C = \langle S, T, F, I, \mathsf{m}\rangle$ be an p\paca{} and let $X\subseteq T$ a
 \emph{configuration} of $C$. Then $X$ can be partially ordered with respect
 to $\rcnprevent\cup\lessdot$.
\end{prop}
\ifreport
\begin{proof}
 Take $X\in \Conf{C}{p\paca}$.
 As $X$ is acyclic with respect to $\rcnprevent\cup\lessdot$, then
 we have that 
 $(\rcnprevent\cup\lessdot)^{\ast} \cap (X \times X)$ is a partial
 order and then we have the thesis.
\end{proof}
\fi

It is important to note a close correspondence between the configurations of a
\paca\ and its reachable markings: any reachable marking corresponds to a
configuration of the net, and conversely.
\begin{prop}\label{pr:ca-reach-markings-are-conf}
  Let $C = \langle S, T, F, I, \mathsf{m}\rangle$ be an p\paca{}. Then,
  \begin{enumerate}
  \item if $m'\in\reachMark{C}$ then $\pre{m'}\in\Conf{C}{p\paca}$; and
  \item if $X\in\Conf{C}{p\paca}$ is a reachable configuration then
    $\mathsf{m}-\pre{X}+\post{X}\in\reachMark{C}$.
  \end{enumerate}
\end{prop}
\ifreport
\begin{proof}
  \
  \begin{enumerate}
  \item If $m'\in\reachMark{C}$, then there exists a firing sequence
    $\sigma$ such that $m' = \lead{\sigma}$.
    We prove by induction on the length of the firing sequence that
    $\pre{m'}\in\Conf{C}{p\paca}$.
    \begin{itemize}
    \item  \emph{Base case} ($\len{\sigma} = 0)$. Then,
      $\lead{\sigma} = \mathsf{m}$ and $\pre{\mathsf{m}} = \emptyset$, which
      is a configuration in $\Conf{C}{p\paca}$.

    \item \emph{Inductive step} ($\len{\sigma} = n+1$). Then,
      $\sigma = \sigma'\trans{A} m'$ with $\len {\sigma'} = n$.
      By inductive hypothesis on $\sigma'$,
      $\pre{\sigma(n)} \in \Conf{C}{p\paca}$. Define $X' = \pre{\sigma(n)}$.
      From $\sigma(n)\trans{A}$, we conclude that $A \cap X' = \emptyset$,
      because transitions cannot be fired twice in a p\paca.
      Also, for all $t \in A$ and $t'\lessdot t$, we have that $t'\in X'$
      because $\inib{t}\cap\pre{t'}\neq\emptyset$ (otherwise $t$ would not be
      enabled at $\sigma(n)$).
      This implies that $X'\cup A$ is left-closed with respect to
      $\lessdot$, i.e., $\histtwo{A}{\lessdot}\subseteq X'\cup A$.

      It remains to show that $\rcnprevent\cup\lessdot$ is acyclic on
      $X'\cup A$. Since $X'$ is a configuration,  its transitions
      can be partially ordered according to $\rcnprevent\cup\lessdot$.
      Moreover, $\forall t'\in X'$, $\post{t'}\subseteq \sigma(n)$. Hence,
      $t'\in X'$ implies $\neg (t'\rcnprevent t)$ for all $t \in A$
      (otherwise, $t$ would not be enabled at $\sigma(n)$).
      Since $A$ is enabled, for all $t, t' \in A$, neither $t \rcnprevent t'$
      nor $t \rcnprevent t'$, because $A$ enabled ensures that
      $\forall t\in \flt{A}.\ \inib{t}\cap \post{(A-\setenum{t})} =
      \emptyset$.
      Consequently,
      $\rcnprevent\cup\lessdot$ is also acyclic on
      $X'\cup A = \pre{m'}$.

    \end{itemize}
    Therefore, we conclude that $m'\in\reachMark{C}$ implies
    $\pre{m'}\in\Conf{C}{p\paca}$.

  \item Let $X$ be a reachable configuration. By
    \cref{lm:ordering-configuration}, the elements of $X$ can be partially
    ordered. That is, there exists a sequence $t_1, \dots, t_n$ of the
    elements of $X$ such that $t_i \lessdot t_j$ or $t_i \ \rcnprevent\ t_j$
    imply $i < j$.
    We prove by induction on the size $n$ of $X$ that
    $m_0\trans{t_1}m_1\trans{t_2}\cdots m_{n-1}\trans{t_n}m_n$ with
    $\mathsf{m} = m_0$ is a firing sequence and
    $\forall i \in \setenum{1, \dots n}$,
    $\pre{m_i} = \setenum{t_1, \dots, t_i}$ and
    $\mathsf{m}-\pre{X}+\post{X}\in\reachMark{C}$.
    %
    \begin{itemize}
    \item \emph{Base case} ($n= 0$). Hence, $X = \emptyset$. Then, the firing
      sequence is empty, with $m_0 = \mathsf{m}$. Moreover,
      $\pre {m_0} = \emptyset = X$ and also
      $m_0 = \mathsf{m} = \mathsf{m} - \pre \emptyset + \post \emptyset =
      \mathsf{m} - \pre X + \post X$.
    \item \emph{Inductive step} ($n = k + 1$). By inductive hypothesis, there
      exists a firing sequence
      $m_0\trans{t_1}m_1\trans{t_2}\cdots m_{k-1}\trans{t_{k}}m_k$ such that
      $\pre{m_{k}} = X\setminus \setenum{t_{k+1}}$ and
      $m_{k} = \mathsf{m} - \pre{(X\setminus\{t_{k+1}\})} +
      \post{(X\setminus\{t_{k+1}\})}$.
      Since $X$ is a configuration, $\histtwo{t_{k+1}}{\lessdot}\subseteq X$.
      This implies that $\inib{t_{k+1}}\cap m_{k} = \emptyset$. Additionally,
      if $t'\rcnprevent t_{k+1}$ then $t'\not\in X\setminus \setenum{t_{k+1}}$
      (otherwise, $X$ would not be a configuration). Hence,
      $\pre{m_{k}}\trans{t_{k+1}}$.
      Then, we can conclude that
      $m_0\trans{t_1}m_1\trans{t_2}\cdots
      m_{k-1}\trans{t_{k}}m_k\trans{t_{k+1}}m_{k+1}$ is a reachable marking.
      Moreover,
      $m_{k+1}= m_k - \pre {t_{k+1}} + \post {t_{k+1}} = \mathsf{m} -
      \pre{(X\setminus\{t_{k+1}\})} + \post{(X\setminus\{t_{k+1}\})} - \pre
      {t_{k+1}} + \post {t_{k+1}} = \mathsf{m} - \pre {X} + \post {X}$ (where
      the last equality holds because transitions do not share places). Also,
      $\pre{m_{k}} = X\setminus \setenum{t_{k+1}}$ implies that
      $\pre{m_{k+1}} = X$ because $\pre{(\pre {t_{k+1}})} = \emptyset$ and
      $\pre{(\post {t_{k+1}})} = t_{k+1}$.
      \qedhere
    \end{itemize}
  \end{enumerate}
\end{proof}
\fi
\noindent 
We emphasize that the previous results remain valid even if the nets are
\paca, as the inheritance of conflicts along the $\lessdot$ relation does not
play any role in the proofs above.

%% file: aux-ipt.tex
Given a function $f : A \to B$, the domain of $f$ is defined as
$\mathit{dom}(f) = \setcomp{a\in A}{\exists b\in B.\ f(a) = b}$.
A sequence of elements in $A$ is a (possibly partial) mapping $\rho:
\nat\rightarrow A$ defined such that $n\in \mathit{dom}(\rho)$ implies $n'\in\mathit{dom}(\rho)$ for all $n' < n$.
Its  length, denoted by $\len{\rho}$, is defined as the cardinality
of its domain, i.e., $\len{\rho} = |\mathit{dom}(\rho)|$. We say that $\rho$ is
finite when its length is so (i.e., $\len{\sigma} < \infty$).
We often write a sequence $\rho$ as $a_1a_2\cdots$ where $a_i = \rho(i)$. Note
that all elements in a sequence are distinct if the the mapping is injective.

A \emph{firing sequence} $\sigma$ (abbreviated as \fs{} $\sigma$) of an \inet $N
= \langle S, T, F, I, \mathsf{m}\rangle$ is a sequence of markings defined
such that for each $i\in dom(\sigma)$ there is a multiset of
transitions $A_i$ enabled at $\sigma(i)$ and
$\sigma(i)\trans{A_i}\sigma(i+1)$.
A \fs\ $\sigma$ is written as $m_0\trans{A_0}m_1$ $\cdots$ $m_{n}\trans{A_n}
m_{n+1}$ $\cdots$. Additionally, $\start{\sigma}$ indicates its initial
marking $\sigma(0) = m_0$. If $\sigma$ is finite, $\lead{\sigma}$ designates
its final marking $\sigma(\len{\sigma})$. We say that $\sigma$ starts at a
marking $m$ if $\start{\sigma} = m$, and let $\firseq{N}{m}$ denote the set of
all firing sequences of the \inet $N$ that start at $m$.
A marking $m$ is \emph{reachable} in $N$ if there exists an \fs\ $\sigma \in
\firseq{N}{\mathsf{m}}$ such that $m = \lead{\sigma}$.
The set of all reachable markings of $N$ is $\reachMark{N} =
 \setcomp{\lead{\sigma}}{\sigma\in\firseq{N}{\mathsf{m}}}$.

%% file: revised-morphisms.tex

\subsection{Morphisms for (pre) Asymmetric Causal Nets}

We introduce a suitable notion of morphisms for p\pacas, which takes into
account that inhibitor arcs correspond to two distinct types of dependencies:
causality and prevention.
In particular, inhibitor arcs representing prevention demand a specific
treatment of markings when compared to classical notions of morphisms for
nets~\cite{Win:ES,Win:PNAMC}.

\begin{defi}\label{de:pca-morphism}
  Let $C_0 = \langle S_0, T_0, F_0, I_0, \mathsf{m}_0\rangle$ and
  $C_1 = \langle S_1, T_1, F_1, I_1, \mathsf{m}_1\rangle$ be p\pacas. An
  \paca-morphism is a pair $(f_S,f_T)$ consisting of a relation
  $f_S \subseteq S_0\times S_1$ and a partial function
  $f_T : T_0 \rightarrow T_1$ defined such that
  \begin{enumerate}
  \item\label{cond:c} for all $t\in T_0$ if $f_T(t) \neq \bot$ then
    \begin{enumerate}
    \item\label{cond:preset-preserved} $\pre{f_T(t)} = f_S(\pre{t})$ and
      $\post{f_T(t)} = f_S(\post{t})$;
    \item\label{cond:inhibitor-reflected} $\forall (s,f_T(t))\in I_1$.
     \begin{enumerate}
     \item\label{cond:inhibitor-reflected-1}
       if $\pre{s} = \emptyset$
         then $\exists s'\in f_S^{-1}(s)$. $(s',t)\in I_0$;
     \item\label{cond:inhibitor-reflected-2}
       if $\pre{s} \neq \emptyset$ then
       $\forall s'\in f_S^{-1}(s)$. $(s',t)\in I_0$.
     \end{enumerate}
    \end{enumerate}
  \item\label{cond:e} $\forall t, t'\in T_0$ if $f_T(t) \neq \bot\neq f_T(t')$
    then $f_T(t) = f_T(t')\ \Rightarrow\ t\ \cnconf_0\ t'$.

  \item\label{cond:b}
    $\forall s_1\in S_1.\ \forall s_0, s_0'\in f_S^{-1}(s_1).$ $s_0\neq s_0'$
    implies $\post{s_0}\ \cnconf_0\ \post{s_0'}$ or
    $\pre{s_0}\ \cnconf_0\ \pre{s_0'}$;
  \item\label{cond:a} $\mrflt{f_S(\mathsf{m}_0)} = \mathsf{m}_1$.
  \end{enumerate}
\end{defi}
\noindent 
The initial two conditions adhere to the standard criteria for morphisms between safe nets: the preservation of presets and postsets (Condition~\ref{cond:preset-preserved}), and the reflection of inhibitor arcs (Conditions~\ref{cond:inhibitor-reflected-1} and \ref{cond:inhibitor-reflected-2}). The inhibitor arcs modeling causality  have to be reflected by finding a witness of the causality, whereas those representing preventions have to be reflected in every possible instance. Notably, Condition~\ref{cond:inhibitor-reflected} implies that $\inib{f_T(t)} \subseteq \mrflt{f_S(\inib{t})}$. Furthermore, only conflicting transitions can be identified, as stipulated by Condition~\ref{cond:e}.

In contrast to conventional requirements, Condition~\ref{cond:b} permits $f_S$
to actively \emph{identify} places in the preset of different transitions.
This might raise concerns, as a place in the target of the morphism could
represent different tokens in the source, potentially evolving independently.
However, under the constraints of Condition~\ref{cond:b}, $f_S$ is only
allowed to identify places connected to transitions that are in (symmetric)
conflict. Moreover, the initial markings corresponds via the relation $f_S$
(Condition~\ref{cond:a}).
It is worth noting that the notion of morphisms remains unaffected by the inheritance of symmetric conflicts along the causality relation, making these conditions applicable to \paca{s} as well.

\begin{exa}\label{ex:arnmorph}
Examine the \pacas $C_0$ and $C_1$ illustrated in \Cref{fig:acn-morph}. The morphism $(f_S, f_T): C_0\rightarrow C_1$ is defined as follows.
For transitions, the mapping $f_T$ is defined by $f_T(a) = a'$, $f_T(b) = b'$, and $f_T(c) = c' = f_T(d)$, thereby identifying the conflicting transitions $c$ and $d$. The relationship on places is established as expected, specifically $f_S(s^{0}_{i},s^{1}_i)$ for $1 \leq i \leq 6$, $f_S(s^{0}_{7},s^{1}_3)$, and $f_S(s^{0}_{8},s^{1}_6)$.
The inhibitor arc $(s^{1}_{3},b')$ in $C_1$ is reflected by the arc
$(s^{0}_{3},b)$ in $C_0$. It is noteworthy that the mapping does not preserve
the remaining arcs of $C_0$. We also stress that if flattening is not applied we would have two tokens in $s^1_3$, one \emph{originated} from $s^0_3$ and the other from $s^0_7$.

\begin{figure}[t]
  \qquad\quad
  \begin{subfigure}{.40\textwidth}
    \centerline{\scalebox{1}{\input{examples/example-acn-morph1}}}
    \caption{$C_0$}
    \label{fig:acn-morph1}
  \end{subfigure}
  \qquad
  \begin{subfigure}{.40\textwidth}
    \centerline{\scalebox{1}{\input{examples/example-acn-morph2}}}
    \caption{$C_1$}
    \label{fig:acn-morph2}
  \end{subfigure}
  \vspace*{.8cm}

  \begin{subfigure}{.80\textwidth}
    \centerline{\scalebox{1}{\input{examples/example-arcn-c}}}
    \caption{$V$}
    \label{fig:racn-c}
  \end{subfigure}
  \caption{}\label{fig:acn-morph}
\end{figure}
\end{exa}

The following result is instrumental for the following developments, and
ensures that any place in the preset of a transition (in the target of an
\paca-morphisms) possesses a corresponding pre-image within the morphism.

\begin{prop}\label{lem:mapping-minimal-morphism} Let
$C_0 = \langle S_0, T_0, F_0, I_0, \mathsf{m}_0\rangle$ and
$C_1 = \langle S_1, T_1, F_1, I_1, \mathsf{m}_1\rangle$ be p\pacas, and
$(f_S,f_T) : C_0 \rightarrow C_1$  an \paca-morphism. Then, for all $s_1\in S_1$, if  $\pre {s_1} = \emptyset$, then $f_S^{-1}(s_1) \neq \emptyset$.
\end{prop}

\begin{proof}
  By contradiction. Assume $f_S^{-1}(s_1) = \emptyset$. Hence, there does does
  not exists $s_0 \in S_0$ such that $f_S(s_0) = s_1$. Hence,
  $s_1\not\in f_S(\mathsf{m}_0)$; and consequently,
  $\mrflt{f_S(\mathsf{m}_0)} \neq \mathsf{m}_1$, which contradicts the
  hypothesis that $(f_S,f_T)$ is an \paca-morphism
  (\Cref{de:pca-morphism}(\ref{cond:a})).
\end{proof}

In connection with a previous property, it is noteworthy that this property does
not hold for transitions, which do not have a guaranteed pre-image.
Additionally, places in the postset of transitions that are outside of the
image of the morphism, may not necessarily be within the image of the
morphism.

Now, our focus is on demonstrating that \paca-morphisms preserve the token
game; specifically, a firing ($m\trans{A}m'$) in the source net is
correspondingly translated to a firing in the target net.
It is noteworthy that not every marking is meaningful in a \paca. For
instance, we anticipate that a marking should not simultaneously contain
tokens for both the pre- and postset of the same transition. Similarly, a
valid marking is not expected to place tokens in the postset of conflicting
transitions. The upcoming definition introduces the concept of markings that
 are coherent for \paca{s}.

\begin{defi}\label{de:coherentmarking}
  Given a \paca{} $C$, we say $m$ is a \emph{coherent marking} for $C$ if
  $\mrflt{m} = m$,
  \begin{enumerate}
  \item for all $t$, $\pre{t} \in m$ iff $\post{t}\not\in m$.
  \item for all $t, t'$, $t\ \cnconf\ t'$ and $\post{t} \in m$ implies
    $\post{t'}\not\in m$.
  \end{enumerate}
\end{defi}

\begin{prop}\label{pr:initialmarkingiscohenrent}
 Let  $C = \langle S, T, F, I, \mathsf{m}\rangle$ be a \paca{}. Then $\mathsf{m}$ is coherent
\end{prop}
\begin{proof}
  Since $C$ is a \paca{}, $\flt{\mathsf{m}} = \mathsf{m}$. Moreover,
  $\mathsf{m} = S\setminus\post{T}$. Hence, the two conditions are trivially
  satisfied.
\end{proof}

Another important consideration is that an \paca-morphism may collapse  different conflicting transitions in the source into a single transition in the target. In other words, two conflicting transitions $t$ and $t'$ may be mapped to the same transition in the target, denoted as $f_T(t) = f_T(t')$. Consequently, $f_S(\pre t) = f_S(\pre{t'})$ and $f_S(\post t) = f_S(\post {t'})$. Therefore, a coherent marking in the source that includes $\post t$ and $\pre {t'}$ corresponds to an execution where the conflict has been resolved--—$t$ has been fired, and subsequently, $t'$ is precluded.
When mapping such a marking to the target, the conventional approach in
morphisms for Petri nets, applying $f_S$ to the marking, would include both
$f_S(\post t)$ and $f_S(\pre {t'})$ in the target marking. Since
$f_T(t) = f_T(t')$, this would result in a marking that is not coherent for
the net in the target of the morphism. Therefore, when mapping a marking, it
is imperative to account for conflicts that have already been resolved.
Specifically, tokens in the preset of transitions that were in conflict with
others that have been fired must be disregarded. This is formally addressed by
the next definition, which defines the relevant information of a marking.

  The next results establish the correspondence among firing between \paca{s} related via a morphism.

\begin{defi}
  Let $(f_S,f_T) : C_0 \rightarrow C_1$ be an \paca-morphism, and $m$ a
   coherent marking of $C_0$. The \emph{relevant information of} $m$ is
   $\underline{m} = m - \multisetcomp{ s }{s,s' \in m \land s\in\pre{t} \land  s'\in \post{t'}  \land   t\ \cnconf\ t' } $
\end{defi}

\begin{prop}\label{pr:morph-preserve-token-game}
  Let $(f_S,f_T) : C_0 \rightarrow C_1$ be an \paca-morphism, and $m$
    a coherent marking of $C_0$. Then, $m\trans{A}m'$ implies $m'$ coherent
    and
    $\mrflt{f_S (\underline m)} \trans{f_{T}(A)}\mrflt{f_S(\underline{m'})}$.
\end{prop}
\ifreport
\begin{proof}
  Assume $m \trans{A}$. Since $A$ is enabled at $m$, we have (i)
  ${\pre{A}} \subseteq m$, (ii) $\inib{A} \cap\ \flt{m} = \emptyset$, and
  (iii)
  $\forall t\in \flt{A}.\ \inib{t}\cap \post{(A-\setenum{t})} = \emptyset$.
  Firstly, we show that $f_S(\pre{A}) = \mrflt{f_S(\pre{A})}$. Suppose the
  contrary, i.e., $f_S(\pre{A}) \neq \mrflt{f_S(\pre{A})}$. Since, $m$ is a
  set, then all transitions in $A$ are different, otherwise $A$ would not be
  enabled at $A$. Since $C_0$ is an \paca{}s, each of its transitions has just
  one place in its preset; i.e., for all $t\in C_0$, $\pre{t}$ is a singleton.
  Consequently, there should be two different transitions $t \neq t' \in A$
  whose preset is mapped to the same place, i.e.,
  $f_S(\pre{t}) = f_S(\pre{t'})$.
  Since $(f_S,f_T)$ is a p\paca{} morphism, $t\ \cnconf_0\ t'$ by
  \Cref{de:pca-morphism}(\ref{cond:e}), which contradicts the assumption that
  $A$ is enabled at $m$. Hence, $f_S(\pre{A}) = \mrflt{f_S(\pre{A})}$. From
  (i), we conclude that $\mrflt{f_S(\pre{A})} \subseteq \mrflt{f_S(m)}$.
   By contradiction, we show that
    $\mrflt{f_S(\pre{A})} \subseteq \mrflt{f_S(m)}$ implies
    $\mrflt{f_S(\pre{A})} \subseteq \mrflt{f_S(\underline{m})}$. Suppose there
    is $s \in {f_S(\pre{A})}$, $s \in f_S(m)$, and
    $s \not\in f_S(\underline{m})$. Hence, there exist $s'$ and $t'$ such that
    $s' \in m$, $s' \in \post{t'}$ and $t'\ \cnconf t$. But this implies $t$ is
    not enabled at $m$, contradicting (ii). Hence,
    $\mrflt{f_S(\pre{A})} \subseteq \mrflt{f_S(\underline{m})}$. Since
  $(f_S,f_T)$ is a p\paca{} morphism,
  $\pre{f_T(A)} = f_S(\pre{A}) = \mrflt{f_S(\pre{A})}$; and consequently,
  $\pre{f_T(A)} \subseteq \mrflt{f_S(\underline m)}$.

  Secondly, we show that
  $\inib{f_T (A)} \cap\ \mrflt{f_S(\underline m)} = \emptyset$. We proceed by
  contradiction. Assume that there exist $t\in A$ and
  $s\in\mrflt{f_S(\underline m)}$ such that $s$ inhibits $f_T(t)$, i.e.,
  $s \in \inib{f_T(t)}$. We have two cases:
  \begin{itemize}
  \item $\pre s = \emptyset$: Since $s\in\mrflt{f_S(\underline m)}$, we
    conclude $ f_S^{-1}(s) \neq \emptyset$. Hence, there exists $s' \in S_0$
    such that $(s',s)\in f_S$ and $s'\in \inib t$, because inhibitor arcs are
    reflected by \Cref{de:pca-morphism}(\ref{cond:inhibitor-reflected-1}).
    Since $C_0$ is an \paca, there exists exactly one
    transition $t'\in T_0$ such that $s' \in \pre{t'}$. There are two cases:
    \begin{itemize}
    \item $s'\in \underline m$: Hence, we have $s'\in m$ and $s'\in \inib t$,
      $t\in A$; which is in contradiction with (ii).

    \item $s'\not\in\underline m$. Since
      $f_S(s')\in\mrflt{f_S(\underline m)}$, there should exist
      $s''\in \underline m$ such that $f(s'') = f(s)' = s$. Let $t''$ be the
      transition such that $s''\in\pre t''$. Since $(f_S,f_T)$ is an
      \paca-morphism, $\post {s'}\ \cnconf\ \post{s''}$
      (\Cref{de:pca-morphism}(\ref{cond:b})). Hence, $t'\ \cnconf\ t''$. Since
      $m$ is a coherent marking, $s'\not\in m$ implies $\post{t'}\in m$. Since
      $\post{t'}\in m$ and $t'\ \cnconf\ t''$, by definition of relevant
      information of a marking, we conclude that
      $\pre t'' \not\in \underline m$, which contradicts the assumption
      $s''\in \underline m$.
    \end{itemize}
  \item $\pre s \neq \emptyset$. Since $s\in\mrflt{f_S(\underline m)}$, then
    there exist $s' \in \underline m \subseteq m$ such that $f_S(s') = s$. By
    \Cref{de:pca-morphism}(\ref{cond:inhibitor-reflected-2}), inhibitor arcs
    are reflected, and hence $s'\in \inib t$, which contradicts (ii).
  \end{itemize}
  Finally, we show that
  $\forall t\in \flt{f_T (A)}.\ \inib{t}\cap \post{(
    f_T(A)-\setenum{t})} = \emptyset$. If this were not the case, then there
  would exist $t_1 \in  f_T (A)$ such that $t_1 \neq t$ and
  $\inib t \cap \post{t_1} \neq \emptyset$. Since $(f_S,f_T)$ is a p\paca{}
  morphism, there would exists $t', t_1' \in A$ such that $f_S(t') = t$ and
  $f_S(t_1') = t_1$ and $\inib t' \cap \post{t_1'}$, which contradicts (iii).
  Consequently, ${f_T(A)}$ is enabled at $\mrflt{f_S(\underline m)}$.
  Moreover,
  $\mrflt{f_S(\underline m)}\trans{f_T(A)}\mrflt{f_S(\underline m')}$ as
  $\mrflt{f_S(\underline m')} = \mrflt{f_S(\underline m)} -
  \mrflt{f_S(\pre{A})} + \mrflt{f_S(\post{A})}$.
\end{proof}
\fi

The next results shows that \paca-morphisms are closed under composition.

\begin{prop}\label{lm:morph-compose}
  Let $(f_S,f_T) : C_0 \rightarrow C_1$ and $(g_S,g_T) : C_1 \rightarrow C_2$
  be two \paca-morphisms. Then
  $(\composizione{f_S}{g_S},\composizione{f_T}{g_T}) : C_0 \rightarrow C_2$ is
  a \paca-morphism as well.
\end{prop}

\ifreport
  \begin{proof}
   We check the conditions of \Cref{de:pca-morphism}.
  \begin{enumerate}
  \item Take $t\in T_0$ if $f_T;g_T(t) \neq \bot$ then
    \begin{enumerate}
    \item Since $f_T$ and $g_T$ define morphisms, we have that
      $\pre (f_T;g_T)(t) = \pre g_T(f_T(\pre t)) =  g_S(f_S(\pre{t})) =
      (f_S;g_S) (\pre t)$.
      By analogous reasoning, we deduce that
      $\post {(f_T;g_T)(t)} = (f_S;g_S) (\post t)$.
    \item Consider $(s,f_T;g_T(t))\in I_2$.
      \begin{enumerate}
      \item Assume $\post{s}\neq \emptyset$, and
        $f_S^{-1}(g_S^{-1}(s)) \neq \emptyset$. By the definition of
        $(g_S,g_T)$, there exists $\tilde{s}\in g_S^{-1}(s)$.
        $(\tilde{s},f_T(t))\in I_1$. Furthermore
        $\post{\tilde{s}}\neq \emptyset$. Again by definition of $(f_S,f_T)$,
        we have again that there is an $s'\in g_T^{-1}(g_S^{-1}(s))$ such that
        $(s',t)\in I_0$. We can conclude that if $(s,f_T;g_T(t))\in I_2$ and
        if $f_S^{-1}(g_S^{-1}(s)) \neq \emptyset$ then we have that there
        exists $s'\in (f_T;g_T)^{-1}(s))$, $(s',t)\in I_0$.
      \item Assume $\post{s} = \emptyset$. By the definition of $(g_S,g_T)$,
        for all $\tilde{s}\in g_S^{-1}(s)$. $(\tilde{s},f_T(t))\in I_1$.
        Furthermore $\post{\tilde{s}} = \emptyset$. Again by definition of
        $(f_S,f_T)$, we have that for all $s'\in g_T^{-1}(g_S^{-1}(s))$ we
        have $(s',t)\in I_0$. We can conclude that if $(s,f_T;g_T(t))\in I_2$,
        then for all $s'\in (f_T;g_T)^{-1}(s))$ we have $(s',t)\in I_0$.
      \end{enumerate}
    \end{enumerate}
  \item Assume $f_T;g_T(t) = f_T;g_T(t')$. There are two cases:
    \begin{itemize}
    \item $f_T(t) = f_T(t')$. Then, $t\ \cnconf_0\ t'$ because $(f_S,f_T)$ is
      an \paca morphism.

    \item $f_T(t) \neq f_T(t')$ and $g_T(t) = g_T(t')$. Since, $(g_S,g_T)$ is
      an \paca morphism, $f_S(t)\ \cnconf_1\ f_s(t')$. By definition of
      symmetric conflicts,
      $\post{f_S(t)}\ \cap \ \inib{f_S(t')} \neq \emptyset$ and
      $\post{f_S(t')}\ \cap \ \inib{f_S(t)} \neq \emptyset$. Since p\paca are
      defined such that the postset of every transition is a singleton, we
      have that $\post{f_S(t)}\ \subseteq \ \inib{f_S(t')}$ and
      $\post{f_S(t')}\ \subseteq \ \inib{f_S(t)}$. Since $(f_S,f_T)$ is a
      p\paca-morphism, we have that $\post {t'}\ \subseteq \ \inib{t}$ and
      $\post{t}\ \subseteq \ \inib{t'}$, by
      \Cref{de:pca-morphism}(\ref{cond:inhibitor-reflected}). Consequently,
      $t \ \cnconf_0\ t'$.
    \end{itemize}

  \item We check that $\forall s_2\in S_2$,
    $\forall s_0, s_0'\in f_S^{-1}(g_S^{-1}(s_2))$ either
    $\post{s_0}\ \cnconf_0\ \post{s_0'}$ or
    $\pre{s_0}\ \cnconf_0\ \pre{s_0'}$. Therefore, either there exists a place
    $s_1\in g_S^{-1}(s_2)$ such that both $s_0$ and $s_0'$ belong to
    $f_S^{-1}(s_1)$, or there are two places $s_1$ and $s_1'$ in
    $ g_S^{-1}(s_2)$ and $s_0$ and $s_0'$ are both in
    $f_S^{-1}(\setenum{s_1, s_1'})$. In the first case
    $\post{s_0}\ \cnconf_0\ \post{s_0'}$ or $\pre{s_0}\ \cnconf_0\ \pre{s_0'}$
    holds as $(f_S,f_T)$ is an \paca-morphism; in the second case we have
    $\post{s_1}\ \cnconf_1\ \post{s_1'}$ or $\pre{s_1}\ \cnconf_1\ \pre{s_1'}$
    as $(g_S,g_T)$ is a \paca-morphism. Since, conflicts are reflected,
    also $\post{s_0}\ \cnconf_0\ \post{s_0'}$ or
    $\pre{s_0}\ \cnconf_0\ \pre{s_0'}$ holds.
  \item Condition $\mrflt{\composizione{f_S}{g_S}(m_0)} = m_2$
    straightforwardly follows by the definitions of $f_S$ and $g_S$ that
    ensure $\mrflt{f_S(m_0)} = m_1$ and $\mrflt{g_S(m_1)} = m_2$.
    \qedhere
  \end{enumerate}
\end{proof}
\fi
\noindent 
We designate the category of p\pacas and \paca-morphisms as $\mathbf{pACN}$.
Within this category, there exists a full and faithful subcategory denoted as
$\mathbf{ACN}$, wherein the objects are \pacas. This subcategory,
$\mathbf{ACN}$, is the specific category of interest.

%% file: examples/example-acn-morph1.tex
\scalebox{0.9}{\begin{tikzpicture}
\tikzstyle{inhibitorred}=[o-, draw=red,thick]
\tikzstyle{inhibitorblu}=[o-, draw=blue,thick]
\tikzstyle{pre}=[<-,thick]
\tikzstyle{post}=[->,thick]
\tikzstyle{readblue}=[-, draw=blue,thick]
\tikzstyle{transition}=[rectangle, draw=black,thick,minimum size=5mm]
\tikzstyle{place}=[circle, draw=black,thick,minimum size=5mm]
\node[place,tokens=1] (p1) at (1,2.5) [label=above:$s_1^{0}$] {};
\node[place,tokens=1] (p3) at (2.5,2.5) [label=above:$s_2^{0}$] {};
\node[place,tokens=1] (p5) at (4,2.5) [label=above:$s_3^{0}$] {};
\node[place,tokens=1] (p7) at (5.5,2.5) [label=above:$s_7^{0}$] {};

\node[place] (p2) at (1,0) [label=below:$s_4^{0}$] {};
\node[place] (p4) at (2.5,0) [label=below:$s_5^{0}$] {};
\node[place] (p6) at (4,0) [label=below:$s_6^{0}$] {};
\node[place] (p8) at (5.5,0) [label=below:$s_8^{0}$] {};

\node[transition] (a) at (1,1.25)  {$a$}
edge[pre] (p1)
edge[post](p2)
edge[inhibitorred] (p3)
edge[inhibitorred] (p5)
edge[inhibitorred] (p4)
edge[inhibitorred] (p6)
;

\node[transition] (b) at (2.5,1.25) {$b$}
edge[pre] (p3)
edge[post] (p4)
edge[inhibitorred] (p5)
edge[inhibitorred] (p6)
;

\node[transition] (c) at (4,1.25) {$c$}
edge[pre] (p5)
edge[post] (p6)
edge[inhibitorred] (p8)
;

\node[transition] (d) at (5.5,1.25) {$d$}
edge[pre] (p7)
edge[post] (p8)
edge[inhibitorred] (p6)
;
\end{tikzpicture}}

%% file: examples/example-acn-morph2.tex
\scalebox{0.9}{\begin{tikzpicture}
\tikzstyle{inhibitorred}=[o-, draw=red,thick]
\tikzstyle{inhibitorblu}=[o-, draw=blue,thick]
\tikzstyle{pre}=[<-,thick]
\tikzstyle{post}=[->,thick]
\tikzstyle{readblue}=[-, draw=blue,thick]
\tikzstyle{transition}=[rectangle, draw=black,thick,minimum size=5mm]
\tikzstyle{place}=[circle, draw=black,thick,minimum size=5mm]
\node[place,tokens=1] (p1) at (0,2.5) [label=above:$s_1^{1}$] {};
\node[place,tokens=1] (p3) at (1.5,2.5) [label=above:$s_2^{1}$] {};
\node[place,tokens=1] (p5) at (3,2.5) [label=above:$s_3^{1}$] {};

\node[place] (p2) at (0,0) [label=below:$s_4^{1}$] {};
\node[place] (p4) at (1.5,0) [label=below:$s_5^{1}$] {};
\node[place] (p6) at (3,0) [label=below:$s_6^{1}$] {};

\node[transition] (a) at (0,1.25)  {$a'$}
edge[pre] (p1)
edge[post](p2)
;

\node[transition] (b) at (1.5,1.25) {$b'$}
edge[pre] (p3)
edge[post] (p4)
edge[inhibitorred] (p5)
;

\node[transition] (c) at (3,1.25) {$c'$}
edge[pre] (p5)
edge[post] (p6)
edge[inhibitorred] (p4)
;
\end{tikzpicture}}

%% file: examples/example-arcn-c.tex
\begin{tikzpicture}[scale=.9]
\tikzstyle{inhibitorred}=[o-, draw=red,thick]
\tikzstyle{pre}=[<-,thick]
\tikzstyle{post}=[->,thick]
\tikzstyle{readblue}=[-, draw=blue,thick]
\tikzstyle{rev}=[-, draw=gray,thick]
\tikzstyle{prerev}=[<-,thick,draw=gray]
\tikzstyle{postrev}=[->,thick,draw=gray]

\tikzstyle{transition}=[rectangle, draw=black,thick,minimum size=5mm]
\tikzstyle{place}=[circle, draw=black,thick,minimum size=5mm]
\node[place,tokens=1] (p1) at (1,2.5) [label=above:$s_1$] {};
\node[place,tokens=1] (p2) at (2.5,2.5) [label=above:$s_2$] {};
\node[place,tokens=1] (p3) at (4.5,2.5) [label=above:$s_3$] {};
\node[place] (p4) at (1,0) [label=below:${s_4}$] {};
\node[place] (p5) at (2.5,0) [label=below:${s_5}$] {};
\node[place] (p6) at (4.5,0) [label=below:${s_6}$] {};

\node[transition] (t1) at (1,1.25)  {$a$}
edge[pre] (p1)
edge[post](p4)
edge[inhibitorred] (p5)
;
\node[transition] (t2) at (2.5,1.25)  {$b$}
edge[pre] (p2)
edge[post](p5)
edge[inhibitorred] (p4)
edge[inhibitorred] (p6)
;

\node[transition] (t3) at (4.5,1.25)  {$c$}
edge[inhibitorred, bend right](p2)
edge[pre](p3)
edge[post](p6)
;

\node[rev] (t5) at (3.5,1.25)  {$\underline{b}$}
edge[prerev, bend left](p5)
edge[inhibitorred](p2)
edge[postrev, bend right](p2)

;

\end{tikzpicture}

%% file: acn-morph-configuration.tex

\subsubsection{Morphisms and configurations:}

We conclude this part by noticing that morphisms preserve configurations, a
fact established through the following proposition.

\begin{prop}\label{lm:cause-preservation}
  Let $(f_S,f_T) : C_0 \rightarrow C_1$ be an \paca-morphism. Then
  \begin{enumerate}
  \item for all $t\in T_0$, if $f_T(t) \neq \bot$ then
    $\histtwo{f_T(t)}{\lessdot_1} \subseteq f_T(\histtwo{t}{\lessdot_0})$; and
  \item for all $t_0, t_0'\in T_0$ such that
    $f_T(t_0) \neq \bot\neq f_T(t_0')$, if
    $f_T(t_0) \preventedby_1\ f_T(t_0')$ then $t_0 \preventedby_0\ t_0'$.
  \end{enumerate}
\end{prop}

\ifreport
\begin{proof}
 \begin{enumerate}
  \item Take $t\in T_0$ such that $f_T(t) \neq \bot$. For each
    $t_0\in \histtwo{f_T(t)}{\lessdot_1}$ we have that either
    $\pre{t_0}\cap\inib{f_T(t)} \neq\emptyset$ or there exists
    transitions $t_1,\dots, t_n \in \histtwo{f_T(t_0)}{\lessdot_1}$ such that
    $(\pre{t_0},t_1), (\pre{t_1},t_2), \dots,$ $(\pre{t_n},f_T(t)) \in I_1$.
    Since $(f_S,f_T)$ is a \paca-morphism, this leads to either a place
    $s_0\in S_0$ with $\post{s_0} = t_0'$, $f_T(t) = t_0'$ and
    $(s_0, t_0')\in I_0$, or the existence of $s_0, s_1, \dots, s_n$ such that
    $\post{s_i} = t_i'$, $f_T(t_i') = t_i$ and $(s_i, t_i')\in I_0$. This
    implies $f_T(t_0') = t_0\in f_T(\histtwo{t_0}{\lessdot_1})$. Thus, the
    inclusion is established.
  \item Assume $f_T(t_0) \preventedby_1\ f_T(t_0')$, then $(s,f_T(t_0))\in I_1$ and
    $s\in \post{f_T(t_0')}$, but then for all $s'\in f_S^{-1}(s)$ we have 
    $(s',t_0)\in I_0$ and there must be an $s'\in \post{t_0'}$. 
    But then $t_0 \preventedby_0\ t_0'$.\qedhere
 \end{enumerate}
\end{proof}
\fi

\begin{prop}
  Let $(f_S,f_T) : C_0 \rightarrow C_1$ be an \paca-morphism. If
  $X\in\Conf{C_0}{p\paca}$, then $f_T(X)\in\Conf{C_1}{p\paca}$.
\end{prop}

\ifreport
\begin{proof}
  Take $X\in\Conf{C_0}{p\paca}$. Hence, for every $t_1\in f_T(X)$, there
  exists $t_0\in X$ such that $f_T(t_0) = t_1$. By
  \Cref{lm:cause-preservation}
  $\histtwo{f_T(t_0)}{\lessdot_1} \subseteq f_T(\histtwo{t_0}{\lessdot_0})$.
  Then, it follows that $\histtwo{t_1}{\lessdot_1} \subseteq f_T(X)$. It
  remains to show that $\rcnprevent_1\cup\lessdot_1$ is acyclic on $f_T(X)$.
  We proceed by contradiction. Assume that $\rcnprevent_1\cup\lessdot_1$ has a
  cycle on $f_T(X)$. Since both relations are induced by inhibitor arcs, which
  are reflected by \paca-morphisms, this implies that
  $\rcnprevent_0\cup\lessdot_0$ has a cycle on $X$, which contradicts the
  assumption that $X$ is a configuration. Therefore,
  $\rcnprevent_1\cup\lessdot_1$ is acyclic on $f_T(X)$ and
  $f_T(X)\in\Conf{C_1}{p\paca}$.
\end{proof}
\fi

\begin{cor}
  Let $(f_S,f_T) : C_0 \rightarrow C_1$ be an \paca-morphism and $C_0, C_1$ be
  two \paca{s}. If $X\in\Conf{C_0}{\paca}$ then $f_T(X)\in\Conf{C_1}{\paca}$.
\end{cor}

%% file: revised-rev-causal-nets-defs.tex

\subsection{Reversible Asymmetric Causal Nets}
In this section, we introduce the concept of Reversible Asymmetric Causal
Nets, following the approach outlined in \cite{lics}. We extend \pacas by
incorporating \emph{backward} transitions, responsible for undoing or
reversing the effects of previously executed \emph{forward} transitions, i.e.,
ordinary transitions. We assume that the set $T$ of transitions in a net is
divided into two sets: $\fwdset{}$ for forward transitions and $\bwdset{}$ for
backward transitions. Furthermore, each backward transition
$\abwd\in\bwdset{}$ is designed to undo the effect of precisely one forward
transition $\afwd\in\fwdset{}$. Nonetheless, there may be forward transitions
that are irreversible. For simplicity, we will use $\afwd$ to refer to the
forward transition and $\abwd$ to denote its associated reversing transition,
when applicable.

\begin{defi}\label{de:reversible-causal-net}
  An \inet $V = \langle S, T, F, I, \mathsf{m}\rangle$ is a \emph{reversible
    Asymmetric Causal Net} (\parcn) if there exists a partition
  $\setenum{\fwdset{} , \bwdset{}}$ of $T$ (where $\fwdset{}$ represents the
  forward transitions, and $\bwdset{}$ denotes the backward transitions)
  satisfying the following conditions:
  \begin{enumerate}
  \item\label{rcn:cond1}
    $\resarcn{V}{}{\fwdset{}} = \langle S, \fwdset{},
    F_{|\fwdset{}\times\fwdset{}}, I_{|\fwdset{}\times\fwdset{}},
    \mathsf{m}\rangle$ is a p\paca\ net;
  \item\label{rcn:cond2}
    $\forall \abwd\in\bwdset{}.\ \exists!\ \afwd\in \fwdset{}$ such that
    $\post{\afwd} = \pre{\abwd}$, $\pre{\afwd} = \post{\abwd}$, and
    $\pre{\afwd}\subseteq\inib{\abwd}$;
  \item\label{rcn:cond3} $\forall \abwd\in\bwdset{}$.
    $K_{\abwd} = \setcomp{\afwd' \in \fwdset{}}{\inib{\abwd} \cap
      \pre{\afwd'{}} \neq \emptyset}$ is finite and $\rcnprevent$ acyclic on
    $K_{\abwd}$;
  \item\label{rcn:cond4}
    $\forall \abwd\in\bwdset{}.\ \forall \afwd\in\fwdset{}$ if
    $\pre\afwd\cap \inib{\abwd}{} \neq \emptyset$ then
    $\post\afwd\cap \inib{\abwd}{} = \emptyset$;
  \item\label{rcn:cond6}
    $\forall \afwd, \afwd', \afwd''\in \fwdset{} .\ \afwd\ \cnconf\ \afwd'\
    \land\ \afwd'\lll \afwd''\ \Rightarrow\ \afwd\ \cnconf\ \afwd''$ with
    $\lll$ being the transitive closure of
    $\lessdot \cap \{(\afwd, \afwd')\ |\ \abwd\not\in\bwdset{}\ \textit{or}\
    \inib{\abwd}\cap\post{\afwd'}\neq\emptyset \}$.
  \end{enumerate}
\end{defi}
\noindent 
Sometimes we use  $\arcn{V}{\bwdset{}}$ to represent a \parcn{}
$V$ with the set $\bwdset{}$ of backward transitions.

In accordance with Condition~\ref{rcn:cond1}, the underlying net
$\resarcn{V}{}{\fwdset{}}$, encompassing solely the forward transitions of
$V$, is a p\paca. The insistence on it being a p\paca rather than an \paca
stems from the fact that the conflict relation is not always inherited along
$\lessdot$, which serves as the causation relationship. This deviation is due
to the fact that reversing transitions may allow potentially conflicting
transitions to be executed.
Condition~\ref{rcn:cond2} stipulates that each backward transition $\abwd$
unequivocally reverses one and only one forward transition $\afwd$.
Consequently, $\abwd$ consumes the tokens produced by $\afwd$
($\post{\afwd} = \pre{\abwd}$) and generates the tokens consumed by $\afwd$
($\post{\abwd} = \pre{\afwd}$).
The requirement $\pre{\afwd}\subseteq\inib{\abwd}$ in
Condition~\ref{rcn:cond2} signifies that the reversal of $\afwd$ (i.e.,
$\abwd$) can only occur if $\afwd$ has been executed. In other words, a
transition can only be reversed if it has been fired.
Condition~\ref{rcn:cond3} requires a finite set of causes for undoing each
transition; in essence, $\pre{K_{\abwd}}$ encompasses all the forward
transitions $\afwd'$ that enable the execution of $\abwd$.
Condition~\ref{rcn:cond4} asserts that if a backward transition $\abwd'$
causally depends on the forward transition $\afwd$ (i.e.,
$\pre\afwd\cap \inib{\abwd'}{} \neq \emptyset$), then $\abwd'$ cannot be
prevented by the same transition $\afwd$
($\post\afwd\cap \inib{\abwd'}{} = \emptyset$), as otherwise, it would be
blocked.
Condition~\ref{rcn:cond6} introduces the relation $\lll$, which is analogous
to the \emph{sustained causation} in \raes. This relation coincides with
causality, except in cases in which a cause can be reversed even after a
causally-dependent transition has been fired. Note that conflicts should be
inherited along $\lll$ rather than along the $\lessdot$ relation.

The inhibitor arcs of an \parcn $V$ induce four distinct relations. Two of
these pertain to the forward flow, defined on $\fwdset{}\times\fwdset{}$, and
correspond to those found in p\paca{}s, namely $\lessdot$ and $\rcnprevent$.
Additionally, two relations concern the backward flow. These are \emph{reverse
  causation} ${\prec} \subseteq {\fwdset{}\times\bwdset{}}$, characterized by
$t \prec \un{t'}$ if and only if $\pre{t}\cap\inib{\un t'}\neq \emptyset$, and
\emph{prevention} ${\lhd} \subseteq {\bwdset{}\times\fwdset{}}$, defined by
$\un{t'} \lhd t$ if and only if $\post{t}\cap\inib{\un t'}\neq \emptyset$.

\begin{exa}

  \begin{figure}[bt]
    \begin{subfigure}{.45\textwidth}
      \centerline{\scalebox{1}{\input{examples/example-arcn-a.tex}}}
      \caption{$V$}
      \label{fig:rcn-a}
    \end{subfigure}
    \quad
    \begin{subfigure}{.45\textwidth}
      \centerline{\scalebox{1}{\input{examples/example-arcn-a-forw.tex}}}
      \caption{$V_{a,b,c}$}
      \label{fig:rcn-a-forw}
    \end{subfigure}
    \vspace*{.8cm}
    
    \begin{subfigure}{.80\textwidth}
      \centerline{\scalebox{1}{\input{examples/example-arcn-b.tex}}}
      \caption{$V'$}
      \label{fig:rcn-b}
    \end{subfigure}
    \caption{}
    \label{fig:rcn}
  \end{figure}

  Verifying that the \inet{} $V$ in \Cref{fig:rcn-a} is an \racn is
  straightforward. Note that $c$ causally depends on $a$ and $b$ due to the
  inhibitor arcs connecting $c$ with the presets of $a$ and $b$.
  Moreover, $a$ and $b$ are in asymmetric conflict: the inhibitor arc linking
  $s_4$ and $b$ indicates that $b$ is prevented by $a$. The sole reversible
  transition is $b$, and its associated undoing $\un{b}$ causally depends on
  both $a$ and $b$ and is prevented by $c$. In other words, the reversal of
  $b$ can only occur after the firing of $a$ and $b$, but only if $c$ has not
  taken place.

  The net $V_{a,b,c}$ in \Cref{fig:rcn-a-forw} is obtained by eliminating
  the transition $\un{b}$ along with the connected inhibitor arcs from $V$.
  It is immediate that $V_{a,b,c}$ is a \paca{}.

  The net $V'$ in \Cref{fig:rcn-b} is obtained by extending $V_1$ with the reversal of
  $c$, i.e., $\un{c}$. If $\un{b}$ and $\un{c}$ along with their associated
  inhibitor arcs are removed from $V'$, we obtain the \paca{} depicted in
  Fig.~\ref{fig:rcn-a-forw}.
\end{exa}

\begin{exa}
  Consider the net $V$ in Fig.~\ref{fig:acn-morph}. Note that the removal of
  the reversing transition $\un{b}$ does not yield a \paca, as the conflict
  between $a$ and $b$ is not inherited along $b\lessdot c$.
  It is noteworthy that the following represents a valid execution of $V$:
  initiating with $b$, followed by $c$, and then reversing $b$ (performing
  $\un{b}$), enabling the subsequent execution of $a$. The ability to execute
  $a$ after the firing and reversal of $b$ would be forbidden if the conflict
  between $a$ and $b$ were inherited along $b\lessdot c$.
\end{exa}

%% file: examples/example-arcn-a.tex
\begin{tikzpicture}[scale=.9]
\tikzstyle{inhibitorred}=[o-, draw=red,thick]
\tikzstyle{pre}=[<-,thick]
\tikzstyle{post}=[->,thick]
\tikzstyle{readblue}=[-, draw=blue,thick]
\tikzstyle{rev}=[-, draw=gray,thick]
\tikzstyle{prerev}=[<-,thick,draw=gray]
\tikzstyle{postrev}=[->,thick,draw=gray]
\tikzstyle{transition}=[rectangle, draw=black,thick,minimum size=5mm]
\tikzstyle{place}=[circle, draw=black,thick,minimum size=5mm]
\node[place,tokens=1] (p1) at (1,2.5) [label=above:$s_1$] {};
\node[place,tokens=1] (p2) at (3,2.5) [label=above:$s_2$] {};
\node[place,tokens=1] (p3) at (5.4,2.5) [label=above:$s_3$] {};
\node[place] (p4) at (1,0) [label=below:${s_4}$] {};
\node[place] (p5) at (3,0) [label=below:${s_5}$] {};
\node[place] (p6) at (5.4,0) [label=below:${s_6}$] {};

\node[transition] (t1) at (1,1.25)  {$a$}
edge[pre] (p1)
edge[post](p4)
;
\node[transition] (t2) at (3,1.25)  {$b$}
edge[pre] (p2)
edge[post](p5)
edge[inhibitorred] (p4)
;
\node[rev] (t5) at (4,1.25)  {$\underline{b}$}
edge[prerev](p5)
edge[postrev](p2)
edge[inhibitorred, bend right] (p2)
edge[inhibitorred] (p6)
edge[inhibitorred, bend right = 10] (p1)
;

\node[transition] (t3) at (5.4,1.25)  {$c$}
edge[inhibitorred, bend right = 20](p2)
edge[inhibitorred, bend right = 8] (p1)
edge[pre](p3)
edge[post](p6)
edge[inhibitorred, bend left = 8] (p4)
edge[inhibitorred, bend left = 8] (p5)
;

%
%
%

\end{tikzpicture}

%% file: examples/example-arcn-a-forw.tex
\begin{tikzpicture}[scale=.9]
\tikzstyle{inhibitorred}=[o-, draw=red,thick]
\tikzstyle{pre}=[<-,thick]
\tikzstyle{post}=[->,thick]
\tikzstyle{readblue}=[-, draw=blue,thick]
\tikzstyle{rev}=[-, draw=gray,thick]
\tikzstyle{prerev}=[<-,thick,draw=gray]
\tikzstyle{postrev}=[->,thick,draw=gray]

\tikzstyle{transition}=[rectangle, draw=black,thick,minimum size=5mm]
\tikzstyle{place}=[circle, draw=black,thick,minimum size=5mm]
\node[place,tokens=1] (p1) at (1,2.5) [label=above:$s_1$] {};
\node[place,tokens=1] (p2) at (3,2.5) [label=above:$s_2$] {};
\node[place,tokens=1] (p3) at (5,2.5) [label=above:$s_3$] {};
\node[place] (p4) at (1,0) [label=below:${s_4}$] {};
\node[place] (p5) at (3,0) [label=below:${s_5}$] {};
\node[place] (p6) at (5,0) [label=below:${s_6}$] {};

\node[transition] (t1) at (1,1.25)  {$a$}
edge[pre] (p1)
edge[post](p4)
;
\node[transition] (t2) at (3,1.25)  {$b$}
edge[pre] (p2)
edge[post](p5)
edge[inhibitorred] (p4)
;

\node[transition] (t3) at (5,1.25)  {$c$}
edge[inhibitorred, bend right = 20](p2)
edge[inhibitorred, bend right = 0.8] (p1)
edge[pre](p3)
edge[post](p6)
edge[inhibitorred, bend left = 8] (p4)
edge[inhibitorred, bend left = 8] (p5)
;

\end{tikzpicture}

%% file: examples/example-arcn-b.tex
\begin{tikzpicture}[scale=.9]
\tikzstyle{inhibitorred}=[o-, draw=red,thick]
\tikzstyle{pre}=[<-,thick]
\tikzstyle{post}=[->,thick]
\tikzstyle{readblue}=[-, draw=blue,thick]
\tikzstyle{rev}=[-, draw=gray,thick]
\tikzstyle{prerev}=[<-,thick,draw=gray]
\tikzstyle{postrev}=[->,thick,draw=gray]

\tikzstyle{transition}=[rectangle, draw=black,thick,minimum size=5mm]
\tikzstyle{place}=[circle, draw=black,thick,minimum size=5mm]
\node[place,tokens=1] (p1) at (1,2.5) [label=above:$s_1$] {};
\node[place,tokens=1] (p2) at (3,2.5) [label=above:$s_2$] {};
\node[place,tokens=1] (p3) at (5.4,2.5) [label=above:$s_3$] {};
\node[place] (p4) at (1,0) [label=below:${s_4}$] {};
\node[place] (p5) at (3,0) [label=below:${s_5}$] {};
\node[place] (p6) at (5.4,0) [label=below:${s_6}$] {};

\node[transition] (t1) at (1,1.25)  {$a$}
edge[pre] (p1)
edge[post](p4)
;
\node[transition] (t2) at (3,1.25)  {$b$}
edge[pre] (p2)
edge[post](p5)
edge[inhibitorred] (p4)
;
\node[rev] (t5) at (4.2,1.25)  {$\underline{b}$}
edge[prerev](p5)
edge[postrev](p2)
edge[inhibitorred, bend right] (p2)
edge[inhibitorred, bend right = 10] (p1)
;

\node[transition] (t3) at (5.4,1.25)  {$c$}
edge[inhibitorred, bend right = 20](p2)
edge[inhibitorred, bend right = 8] (p1)
edge[pre](p3)
edge[post](p6)
edge[inhibitorred, bend left = 8] (p4)
edge[inhibitorred, bend left = 8] (p5)
;

\node[rev] (t5) at (6.6,1.25)  {$\underline{c}$}
edge[prerev, bend left](p6)
edge[inhibitorred](p3)
edge[postrev, bend right](p3)

;

\end{tikzpicture}

%% file: racn-configurations.tex
\subsection{Configurations of \parcn{s}}

We now introduce the concept of configuration for \parcn{s}. In contrast to
the configuration for standard nets, which is typically defined in terms of
causality, a configuration of \parcn{} is simply  required reachable via a
firing sequence.

\begin{defi}\label{de:prcn-configuration}
  Let $\arcn{V}{\bwdset{}} = \langle S, T, F, I, \mathsf{m}\rangle$ be a
  \parcn{}. A \emph{configuration} of $\arcn{V}{\bwdset{}}$ is any subset
  $X\subseteq \fwdset{}$ of forward transitions such that there exists a
  firing sequence $\mathsf{m}\trans{A_1}\dots m'$ with
  $X = \pre{m'}\cap \fwdset{}$.
\end{defi}

\begin{exa}
  Consider the net $V_2$ in Fig.~\ref{fig:rcn-b}; one of its configuration is
  $\setenum{a, c}$, obtained by executing $b$ first, followed by $a$, then $c$
  and subsequently undoing $b$ (executing $\un{b}$).
  This configuration is only reachable by undoing $b$ since, given that $b$ is
  a cause of $c$, the presence of $b$ is necessary to execute $c$.
\end{exa}

%% file: revised-rev-morphisms.tex

\subsection{Morphisms for Reversible Asymmetric Causal Nets}
The notion of morphisms for \parcns is given below.

\begin{defi}\label{de:rev-morphism}
  Let $\arcn{V}{\bwdset{0}} = \langle S_0, T_0, F_0, I_0, \mathsf{m}_0\rangle$
  and $\arcn{V}{\bwdset{1}} = \langle S_1, T_1, F_1, I_1, \mathsf{m}_1\rangle$
  be two \parcns. A \parcn-morphism is a pair $(f_S,f_T)$ consisting of a
  relation $f_S\subseteq S_0\times S_1$ and a partial function
  $f_T : T_0 \rightarrow T_1$ satisfying the following conditions
  \begin{enumerate}
  \item $f_T(\fwdset{0})\subseteq \fwdset{1}$ and
    $f_T(\bwdset{0})\subseteq \bwdset{1}$;
  \item \label{pacamorf-in-restriction}
    $(f_S,f_T|_{\fwdset{0}}) : \resarcn{V}{}{\fwdset{0}} \rightarrow
    \resarcn{V}{}{\fwdset{1}}$ is an \paca-morphism,
  \item $\forall \abwd\in \bwdset{0}.$ if $f_T(t) \neq \bot$ then
    \begin{enumerate}
    \item \label{mapping-rev} $f_T(\abwd) \neq \bot$ and
      $f_T(\abwd) = \un{f_T(\afwd)}$; and
    \item \label{non-collapsing-inhib}
      $\forall (s,f_T({\un{t}}))\in I_1$.
       \begin{enumerate}
       \item\label{non-collapsing-inhib-1} if $\post{s}\cap \fwdset{1}\neq \emptyset$,
             {$ f_S^{-1}(s)\neq \emptyset$ implies $\exists$} $s'\in f_S^{-1}(s)$.
             $(s',{\un t})\in I_0$; and
       \item\label{non-collapsing-inhib-2} if $\post{s}\cap \fwdset{1} = \emptyset$,
             {$\forall$} $s'\in f_S^{-1}(s)$.
             $(s',{\un t})\in I_0$.
       \end{enumerate}
    \end{enumerate}
  \end{enumerate}
\end{defi}
\noindent 
The first condition stipulates that forward and backward transitions are
respectively mapped to forward and backward transitions. According to
Condition~\ref{pacamorf-in-restriction}, when considering forward transitions
only (i.e., the restriction of $f_T$ to ${\fwdset{0}}$) we have an
\paca{}-morphism on the underlying $\resarcn{V}{}{\fwdset{0}}$ and
$\resarcn{V}{}{\fwdset{1}}$ (i.e., the nets consisting only of forward
transitions). Condition~\ref{mapping-rev} ensures that the transition $\abwd$,
responsible for reversing $t$, must be mapped to the transition
$\un{f_T(\afwd)}$, which, in turn, reverses $f_T(\afwd)$.
{Finally, Condition~\ref{non-collapsing-inhib-1} prevents the merging of
inhibitor arcs, meaning that the distinction between various causes that might
prevent the reversal of a transition is preserved, whereas Condition~\ref{non-collapsing-inhib-2}
assures that is a prevention is present in the target then all the reverse images of this prevention
must be present in the origin.
We have to restrict our attention to $\post{s}\cap \fwdset{1}$ as we have to consider the
relations induced by the inhibitor arcs in this case that are either reverse causation or reverse
prevention, and these are defined among forward transitions and backward ones.}

\begin{exa}\label{ex:revmorph}
  \begin{figure}[tb]
    \begin{subfigure}{.45\textwidth}
      \centerline{\scalebox{1}{\input{examples/example-racn-morph1}}}
      \caption{$V_0$}
      \label{fig:racn-morph1}
    \end{subfigure}
    \qquad
    \begin{subfigure}{.45\textwidth}
      \centerline{\scalebox{1}{\input{examples/example-racn-morph2}}}
      \caption{$V_1$}
      \label{fig:racn-morph2}
    \end{subfigure}
    \caption{Two \racn}\label{fig:racn-morph}
  \end{figure}

  An \racn{}-morphism $(f_S,f_T) : V_0 \rightarrow V_1$ for the \parcns $V_0$
  and $V_1$ in Fig.~\ref{fig:racn-morph} is as follows. Consider the
  multirelation $f_S$ on places, as detailed in \Cref{ex:arnmorph}, and define
  the mapping on transitions $f_T$ such that it mirrors the one in
  \Cref{ex:arnmorph} for forward transitions. On the reversing transition
  $\un{a}$, set $f_T(\un{a}) = \un{a}'$. It is evident that
  $\un{f_T(a)} = f_T(\un{a})$.
  The other conditions are straightforwardly satisfied.
\end{exa}

The following two results state that \parcn-morphisms preserve behaviours and
that they are closed under composition.

We first need to specialize the notion of coherent marking for \racn. The idea
is that we have to focus on forward transitions only.
In fact, a coherent marking
does not place tokens simultaneously in both the preset and postset of a transition, and does not place tokens in the postsets of two conflicting transitions. Note that this can
guaranteed by just looking at forward transitions.
Therefore given a
marking $m$ of the \racn{} $V = \arcn{V}{\bwdset{}}$, $m$ is coherent if it is
coherent in $V|_{\fwdset{}}$ (recall that the places in $V$ and
$V|_{\fwdset{}}$ are the same). Similarly let
$(f_S,f_T) : \arcn{V}{\bwdset{0}} \rightarrow \arcn{V}{\bwdset{1}}$ be an
\racn-morphism, and let $m$ be a coherent marking of $\arcn{V}{\bwdset{0}}$.
The relevant information of $m$ are the relevant information of the marking
$m$ in $V|_{\fwdset{0}}$.

\begin{prop}\label{pr:rev-morph-preserve-token-game}
  Let $(f_S,f_T) : \arcn{V}{\bwdset{0}} \rightarrow \arcn{V}{\bwdset{1}}$ be a
  \parcn{}-morphism. {Let $m$ be a coherent marking of $\arcn{V}{\bwdset{0}}$.
    Then, $m\trans{A}m'$ implies $m'$ coherent and
    $\mrflt{f_S (\underline m)} \trans{f_{T}(A)}\mrflt{f_S(\underline{m'})}$}.
\end{prop}

\ifreport
\begin{proof}
  We notice that conditions \ref{mapping-rev} and \ref{non-collapsing-inhib}
  of \Cref{de:rev-morphism} are essentially the same of a \paca{}-morphism,
  thus without distinguish among reverse transitions and forward ones, we have
  that a \parcn-morphism is a \paca-morphism. Consequently, the tokens game is
  preserved.
\end{proof}
\fi

\begin{prop}\label{lm:rcnmorph-compose}
  Let $(f_S,f_T) : V_0 \rightarrow V_1$ and $(g_S,g_T) : V_1 \rightarrow V_2$
  be two \parcn{}-morphisms. Then
  $(\composizione{f_S}{g_S},\composizione{f_T}{g_T}) : V_0 \rightarrow V_2$ is
  a \parcn{}-morphism as well.
\end{prop}

\ifreport
\begin{proof}
  The proof follows the same lines as the one in \Cref{lm:morph-compose}.
\end{proof}
\fi

Therefore, \parcns and \parcn-morphisms constitute a category, which is
denoted by $\mathbf{RACN}$.

%% file: examples/example-racn-morph1.tex
\scalebox{0.9}{\begin{tikzpicture}
\tikzstyle{inhibitorred}=[o-, draw=red,thick]
\tikzstyle{pre}=[<-,thick]
\tikzstyle{post}=[->,thick]
\tikzstyle{readblue}=[-, draw=blue,thick]
\tikzstyle{rev}=[-, draw=gray,thick]
\tikzstyle{prerev}=[<-,thick,draw=gray]
\tikzstyle{postrev}=[->,thick,draw=gray]
\tikzstyle{transition}=[rectangle, draw=black,thick,minimum size=5mm]
\tikzstyle{place}=[circle, draw=black,thick,minimum size=5mm]
\node[place,tokens=1] (p1) at (0,2.5) [label=above:$s_1^{0}$] {};
\node[place,tokens=1] (p3) at (2,2.5) [label=above:$s_2^{0}$] {};
\node[place,tokens=1] (p5) at (4,2.5) [label=above:$s_3^{0}$] {};
\node[place,tokens=1] (p7) at (6,2.5) [label=above:$s_7^{0}$] {};
\node[place] (p2) at (0,0) [label=below:$s_4^{0}$] {};
\node[place] (p4) at (2,0) [label=below:$s_5^{0}$] {};
\node[place] (p6) at (4,0) [label=below:$s_6^{0}$] {};
\node[place] (p8) at (6,0) [label=below:$s_8^{0}$] {};

\node[transition] (a) at (0,1.25)  {$a$}
edge[pre] (p1)
edge[post](p2)
edge[inhibitorred] (p3)
edge[inhibitorred] (p5)
;

\node[rev] (reva) at (1,0.7) {$\underline{a}$}
edge[prerev] (p2)
edge[postrev] (p1)
edge[inhibitorred, bend right] (p1)
edge[inhibitorred, bend left = 10] (p8)
;

\node[transition] (b) at (2,1.25) {$b$}
edge[pre] (p3)
edge[post] (p4)
edge[inhibitorred] (p5)
edge[inhibitorred, bend left=35] (p2)
;

\node[transition] (c) at (4,1.25) {$c$}
edge[pre] (p5)
edge[post] (p6)
edge[inhibitorred] (p8)
edge[inhibitorred, bend left=5] (p2)
edge[inhibitorred] (p4)

;

\node[transition] (d) at (6,1.25) {$d$}
edge[pre] (p7)
edge[post] (p8)
edge[inhibitorred] (p6)

;
\end{tikzpicture}}

%% file: examples/example-racn-morph2.tex
\scalebox{0.9}{\begin{tikzpicture}
\tikzstyle{inhibitorred}=[o-, draw=red,thick]
\tikzstyle{pre}=[<-,thick]
\tikzstyle{post}=[->,thick]
\tikzstyle{readblue}=[-, draw=blue,thick]
\tikzstyle{rev}=[-, draw=gray,thick]
\tikzstyle{prerev}=[<-,thick,draw=gray]
\tikzstyle{postrev}=[->,thick,draw=gray]
\tikzstyle{transition}=[rectangle, draw=black,thick,minimum size=5mm]
\tikzstyle{place}=[circle, draw=black,thick,minimum size=5mm]
\node[place,tokens=1] (p1) at (0,2.5) [label=above:$s_1^{1}$] {};
\node[place,tokens=1] (p3) at (2,2.5) [label=above:$s_2^{1}$] {};
\node[place,tokens=1] (p5) at (4,2.5) [label=above:$s_3^{1}$] {};
\node[place] (p2) at (0,0) [label=below:$s_4^{1}$] {};
\node[place] (p4) at (2,0) [label=below:$s_5^{1}$] {};
\node[place] (p6) at (4,0) [label=below:$s_6^{1}$] {};
\node[transition] (a) at (0,1.25)  {$a'$}
edge[pre] (p1)
edge[post](p2)
edge[inhibitorred] (p3)
;

\node[rev] (reva) at (1,0.8) {$\underline{a'}$}
edge[prerev] (p2)
edge[postrev] (p1)
edge[inhibitorred, bend right] (p1)
;

\node[transition] (b) at (2,1.25) {$b'$}
edge[pre] (p3)
edge[post] (p4)
edge[inhibitorred] (p5)
edge[inhibitorred, bend left=30] (p2)
;

\node[transition] (c) at (4,1.25) {$c'$}
edge[pre] (p5)
edge[post] (p6)
edge[inhibitorred] (p4)
;

\end{tikzpicture}}

%% file: racn-morph-configuration.tex
\subsubsection{Configurations and morphisms:}
We show that morphisms preserve configurations.

\begin{prop}
  Let $(f_S,f_T) : \arcn{V}{\bwdset{0}} \rightarrow \arcn{V}{\bwdset{1}}$ be a
  \parcn-morphism and $X$ a configuration, i.e.,
  $X\in\Conf{\arcn{V}{\bwdset{0}}}{\parcn}$. Then,
  $f_T(X)\in\Conf{\arcn{V}{\bwdset{1}}}{\parcn}$.
\end{prop}

\ifreport
\begin{proof}
  It is sufficient to observe that configurations in \parcn{} are
  characterized by reachable markings, and \parcn-morphisms preserve them.
\end{proof}
\fi

%% file: coprodracn.tex

\subsection{Constructions}
Similarly to what we have done for \raes, we have a coproduct also in the category of \racn.

\begin{prop}\label{de:racn-coprod}
  Let $\resarcn{V}{\bwdset{0}}{0} = (S_0, T_0, F_0, I_0, \mathsf{m}_0)$ and
  $\resarcn{V}{\bwdset{1}}{1} = (S_1, T_1, F_1, I_1, \mathsf{m}_1)$ be two
  \racn{s}. Then
  $\resarcn{V}{\bwdset{0}}{0} + \resarcn{V}{\bwdset{1}}{1} = (S, T, F, I,
  \mathsf{m})$ where
 \begin{itemize}
 \item $S = \setenum{0}\times S_0 \cup \setenum{1}\times S_1$;
 \item $T = \setenum{0}\times T_0 \cup \setenum{1}\times T_1$;
 \item
   $\forall (i,a)\in S\cup T,\ \forall (j,b)\in S\cup T.\ ((i,a),(j,b))\in F$
   whenever $i = j$ and $(a,b)\in F_i$;
 \item $\forall (i,s)\in S,\ \forall (j,t)\in T.\ ((i,s),(j,t))\in I$ whenever
   either $i = j$ and $(s,t)\in I_i$ or $i\neq j$; and
 \item $\forall (i,s)\in S.\ \mathsf{m}(i,s) = \mathsf{m}_i(s)$
 \end{itemize}
 is their \emph{coproduct} and
 $(\mathit{in}^i_S, \mathit{in}^i_T) : \resarcn{V}{\bwdset{i}}{i} \to
 \resarcn{V}{\bwdset{0}}{0} + \resarcn{V}{\bwdset{1}}{1}$ defined as
 $(s,(i,s))\in \mathit{in}^i_S$ and $\mathit{in}^i_T(t) = (i,t)$ are the
 injections.
\end{prop}
\begin{proof}
 $\resarcn{V}{\bwdset{0}}{0} + \resarcn{V}{\bwdset{1}}{1}$ is clearly an \racn. It remains to prove that is indeed
 a coproduct in the category $\mathbf{RACN}$. Consider an \racn{} $\resarcn{V}{\bwdset{2}}{2}$ and two
 morphisms $(f_S,f_T) : \resarcn{V}{\bwdset{0}}{0} \to \resarcn{V}{\bwdset{2}}{2}$ and
 $(g_S,g_T) : \resarcn{V}{\bwdset{1}}{1} \to \resarcn{V}{\bwdset{2}}{2}$, we show that there exists a unique
 morphisms $(h_S,h_T) : \resarcn{V}{\bwdset{0}}{0} + \resarcn{V}{\bwdset{1}}{1} \to \resarcn{V}{\bwdset{2}}{2}$.
 Define $h_S$ as the relation comprising the pairs $((i,s),s')$ if either $(s,s')\in f_S$ or $(s,s')\in g_S$
 and the mapping on transitions $h_T(i,t)$ equal to $f_T(t)$ if $i=0$ and to $g_T(t)$ if $i = 1$.
 We check first that this is indeed an \racn-morphism.
 Consider $(i,t) \in T$ such that $h_T(i,t)$ is defined. Then we have that
 $\pre{h_T(i,t)} = \pre{f_T(t)}$ if $i=0$ and $\pre{h_T(i,t)} = \pre{g_T(t)}$ if $i=1$ and
 as $(f_S,f_T)$ and $(g_S,g_T)$ are morphisms we have that either $\pre{h_T(i,t)} = \mu f_S(\pre{t})$ or
 $\pre{h_T(i,t)} = \mu g_S(\pre{t})$ and similarly for $\post{h_T(i,t)} = \mu f_S(\post{t})$ or
 $\post{h_T(i,t)} = \mu g_S(\post{t})$.
 Consider now $((i,s),h_T(j,t))\in I_2$ and $s'\in f_S^{-1}(i,s) \cup g_S^{-1}(i,s)$. If $s'\in f_S^{-1}(i,s)$
 and $j=0=i$ then $((i,s'),(i,t))\in I$ and if $i\neq j$ then $((i,s'),(j,t))\in I$ by construction.
 The case $s'\in g_S^{-1}(i,s)$ is the same.
 Assume now $h_T(i,t)$ and $h_T(j,t')$ are both defined and equal. If $i=j$ then
 $(i,t)\ \cnconf\ (i,t')$ and if $i\neq j$ we have $(i,t)\ \cnconf\ (j,t')$ as well as
 each transition of the first one is prevented by the happening of a transition of the second one and vice versa.
 Condition~\ref{cond:b} of \Cref{de:pca-morphism} is proven similarly to the previous one, and
 the same argument is used to show that fact that $\flt{h_S}(\mathsf{m}) = \mathsf{m}_2$.
 Observing that forward transitions are mapped to forward ones and backward transitions are mapped to
 backward ones, and if $(i,\un t)\in \bwdset{}$ and $h_T(i,t)$ is defined we have
 that $h_T(i, \un t)$ is defined and $h_T(i, \un t) = \un{h_T(i,t)}$ as $(f_S,f_T)$ and $(g_S,g_T)$
 are morphisms.
 The \racn-morphism $(h_S,h_T)$ is unique as $h_T$ is unique and there is just one $h_S$ satisfying
 the requirements.
\end{proof}
Also the category $\mathbf{RACN}$ does not have products.

%% file: coreflection.tex
\section{Relating models}\label{sec:adjunctions}
In this section, we explore the interplay between the categories of
(reversible) asymmetric causal nets and (reversible) asymmetric event
structures. We establish functors and demonstrate the emergence of an
adjunction.

\input{relatingmodels}

%% file: relatingmodels.tex

We delve into the connection between \raeses and \racns. The following
definition outlines the process of recovering an \raes from the flow and
inhibitor arcs of an \racn.

As expected, the relations induced by an \raes—--namely $\lessdot$
(causality), $\rcnprevent$ (weak causality), $\prec$ (reverse causality),
$\lhd$ (prevention), and $\lll$ (sustained causation)---play a crucial role in
defining the relations of the corresponding \raes.

\begin{defi}\label{def:rpcntoraes}
  Let $\arcn{V}{\bwdset{}} = \langle S, T, F, I, \mathsf{m}\rangle$ be an
  \racn, and let $\lessdot$, $\rcnprevent$, $\prec$, $\lhd$ and $\lll$ denote
  the causation, weak causality, reverse causality, prevention and sustained
  causation induced by $F$ and $I$. Then, the corresponding structure
  $\acntoaes{r}(\arcn{V}{\bwdset{}})$ is the tuple
  $(\fwdset{}, \anR, \lessdot, {\lll}\cup {\rcnprevent}$, $\prec, \lhd)$,
  where $\anR = \{ t \in \fwdset{} \ |\ \abwd\in\bwdset{}\}$ represents the
  reversible events.
\end{defi}

The following result ensures that the above construction generates an \raes.

\begin{restatable}{thm}{rpcntoraes}
\label{th:rpcntoraes}
Let $\arcn{V}{\bwdset{}}$ be an \racn{}. Then
$\acntoaes{r}(\arcn{V}{\bwdset{}})$ is an \raes{}.
\end{restatable}

\input{racnraes-proof}

\begin{exa}
  Consider the \racn{} $V_1$ \Cref{fig:rcn}. The corresponding \raes
  $\acntoaes{r}(V_1)$ is defined as $\mathsf{H}_1$ in \Cref{ex:raes}. In fact,
  within $V_1$, causality is characterized by $b \lessdot c$ and
  $a \lessdot c$ due to the conditions $\pre{a}\cap\inib{c} \neq \emptyset$
  and $\pre{b}\cap\inib{c} \neq \emptyset$. Moreover, weak causality includes
  $b \nearrow a$ as a consequence of $b\ \rcnprevent\ a$, and also encompasses
  $a \nearrow c$ and $b\nearrow c$ induced by the sustained causation, which
  coincides with $\lessdot$.
  On the other hand, reverse causality encompasses $b\prec \un{b}$ and
  $a \prec \un{b}$, while prevention is limited to $\un{b}\lhd c$.
\end{exa}

Recall that an \racn{} lacking reversing transitions (i.e.,
$\arcn{V}{\emptyset}$) is essentially an \acn{}. Consequently, the
construction outlined in \Cref{def:rpcntoraes} is applicable to \acns as well.

\begin{cor}
  Let $C = \langle S, T, F, I, \mathsf{m}\rangle$ be an \acn. Then
  $\acntoaes{r}(C)$ is an \aes.
\end{cor}

$\acntoaes{r}$ extends to a functor by observing that an \racn-morphism
$(f_S,f_T) : V_0 \rightarrow V_1$ induces an \raes-morphism
$\acntoaes{r}(f_S,f_T) = f_T$.

\begin{prop}\label{pr:racntoraes-functor}
  $\acntoaes{r} : \mathbf{RACN} \rightarrow \mathbf{RAES}$ is a well-defined
  functor.
\end{prop}

The construction linking a net to an event structure adheres to the
conventional intuition, where places represent appropriate subsets of events,
and transitions are connected to places based on the relations specified in
the event structure. In our context, the event subsets have a cardinality of
at most one, meaning they are either the empty set or a singleton. To
streamline notation, we will use the symbol $e$ to represent the singleton set
$\{e\}$ and, consequently, avoid the use of braces.

\begin{defi}\label{de:raestoracn}
  Let $\mathsf{H} = (E, \anR, <, \nearrow, \prec, \lhd)$ be an \raes. Then,
  the associated net $\aestoacn{r}(\mathsf{H})$ is defined as
  $\langle S, E\cup \setcomp{\un{\anr}}{\anr\in\anR}, F, I,
  \mathsf{m}\rangle$ where

  \[
    \begin{array}{r@{\ }c@{\ }l@{\ }c@{\ }l@{\ }c@{\ }l}
      S
      & =
      &
        \setcomp{\ (\emptyset,e)\ }{\ e\in E}
      & \cup\
      & \setcomp{\ (e, e)\ }{\ e\in E}
      & \cup\
      & \setenum{\ (\emptyset,\emptyset)\ }
      \\[6pt]
      F
      & =
      & \setcomp{\ ((\emptyset, {e}), e)\hspace{.23cm} }{\ e\in E}
      & \cup\
      & \setcomp{\ (e, ({e}, {e}))\ }{\ e\in E}
      & \cup
      \\
      &
      &
        \setcomp{\ (({u}, {u}), \un{\anr})\ }{\, \anr\in U}
      & \cup\
      &
        \setcomp{\ (\un{\anr}, (\emptyset, {u}))}{\ \anr\in U}
      \\[6pt]
      I
      & =
      & \setcomp{\ ((\emptyset, {e'}), e)\ }{\ e' < e}
      & \cup\
      &         \setcomp{\ (({e'}, {e'}), e)\ }{\ e \nearrow e'
        }
      & \cup
      \\
      &
      & \setcomp{\ ((\emptyset, {e}), \un{\anr})\hspace{.18cm}}{\ e\prec\un{\anr}\,}
      & \cup\
      & \setcomp{\ (({e}, {e}), \un{\anr}) }{\ \un{\anr}\lhd e}
      \\[6pt]
      \mathsf{m}
      & =
      & \setcomp{\ (\emptyset, {e})\ }{\ e\in E}\
      & \cup
      & \setenum{\ (\emptyset,\emptyset)\ }
    \end{array}
  \]

\end{defi}

Events in $E$ are mapped to the forward transitions of the net. For each event
$e$, two associated places are considered: $(\emptyset, e)$ signifies that the
event $e$ has not occurred, while $(e, e)$ indicates its execution. As a
result, the preset of the forward transition $e$ is ${(\emptyset, e)}$, and
its postset is ${(e, e)}$.
A transition $\un{\anr}$ corresponds to the undoing of the forward
transition $u$. This transition consumes tokens from $(u, u)$ and produces
them in $(\emptyset, u)$.
The inclusion of the special place $(\emptyset, \emptyset)$ serves no
operational purpose; its motivation is purely technical. Its significance will
become clearer when demonstrating that the mapping extends to a functor, as
explained in the subsequent discussion.
The inhibitor arcs in $I$ model both causality (forward or backward) and
precedence (forward or backward). Additionally, all places not appearing in
the postset of forward transitions are initially marked.

\begin{restatable}{thm}{raesisracn}
  \label{th:raesisracn}
  If $\mathsf{H} = (E, \anR, <, \nearrow, \prec, \lhd)$ is a \raes{}, then
  $\aestoacn{r}(\mathsf{H})$ is an \racn{}.
\end{restatable}

\ifreport
\input{raesracn-proof}
\fi

\begin{exa}\label{ex:raestoracn}
  Consider the \raes $\mathsf{H'}$ from \Cref{ex:raes}. The places are
  $(\emptyset,\emptyset)$, $(\emptyset,{a})$, $(\emptyset,{b})$,
  $(\emptyset,{c})$, $({a},{a})$, $({b},{b})$ and $({c},{c})$. The
  corresponding transitions are $a, b, c$ (forward ones) and $\un{b}, \un{c}$
  (reversing ones). The flow arcs adhere to the specifications outlined in
  \Cref{de:raestoracn}, while the inhibitor arcs are determined by the
  relations within $\mathsf{H'}$. The resulting net bears resemblance to the
  one depicted in \Cref{fig:rcn-b}, with distinct place labels and the
  exclusion of the isolated place $(\emptyset,\emptyset)$.
\end{exa}

\begin{cor}
  If $\mathsf{G} = (E, <, \nearrow)$ is an \aes{}, then
  $\aestoacn{r}(E, \emptyset, <, \nearrow, \emptyset, \emptyset)$ is an
  \acn{}.%
\end{cor}

We establish the extension of the mapping $\aestoacn{r}$ to a functor by
demonstrating that any \raes{}-morphism
$f: \mathsf{H}_0 \rightarrow \mathsf{H}_1$ induces a \racn-morphism
$\aestoacn{r}(f): \aestoacn{r}(\mathsf{H}_0) \rightarrow
\aestoacn{r}(\mathsf{H}_1)$.

\begin{defi}\label{def:induced-morphism}
  Let $f: \mathsf{H}_0 \rightarrow \mathsf{H}_1$ be \raes{}-morphism. Assume
  $\aestoacn{r}(\mathsf{H}_i) = \langle S_i, E_i\cup
  \setcomp{\un{\anr}}{\anr\in\anR_i} , F_i, I_i, \mathsf{m}_i \rangle$ for
  $i=0,1$. We define $\aestoacn{r}(f) = (f_S, f_T)$ as follows:

  \begin{enumerate}
  \item\label{def:induced-morphism-1}
    $f_S \subseteq S_0 \times S_1$ where
    \[
      (s_0, s_1) \in f_S \iff
      \begin{cases}
        s_0 = s_1 = (\emptyset,\emptyset);\ or
        \\
        s_0 = (\emptyset,\emptyset) \ \land\
        s_1 = (\emptyset, e) \ \land\ f^{-1}(e) = \bot;\ or
        \\
        s_0 = (\emptyset, e)  \ \land\  f(e) \neq \bot  \ \land\ s_1 = (\emptyset,f(e));\ or
        \\
        s_0 = (e, e) \ \land\  f(e) \neq \bot \ \land\ s_1 = (f(e), f(e)).
      \end{cases}
    \]

  \item\label{def:induced-morphism-2}
    $f_T : E_0 \cup \setcomp{\un{\anr}}{\anr\in\anR_0} \rightarrow E_1
    \cup \setcomp{\un{\anr}}{\anr\in\anR_1}$ is defined as follows:
    \[
      f_T(t) =
      \begin{cases}
        f(t) & \text{if } t \in E_0, \ \land\  f(e) \neq \bot  \\
        \un{f(\anr)} & \text{if } t =  \un{\anr} \text{ and } f(\anr) \neq \bot.
      \end{cases}
    \]
  \end{enumerate}
\end{defi}

\noindent 
According to Condition (1), the relation $f_S$ is established to associate the
place $(\emptyset, \emptyset)$ in the source with its counterpart
$(\emptyset, \emptyset)$ in the target. Additionally, it is linked to all the
places $(\emptyset, e)$ in the target, where $e$ is not in the image of the
morphism $f$. This connection is crucial to guarantee that the definition
satisfies the condition regarding the initial markings of \acn-morphisms
(\Cref{de:pca-morphism}(\ref{cond:a})).
Each place in the source, denoted as $(\emptyset, e)$, which corresponds to
the preset of the event $e$, is paired with the place in the target
representing the preset of the corresponding event $f(e)$; denoted as
$(\emptyset, f(e))$.
Similarly, each place in the source, denoted as $(e, e)$, indicating the
postset of the event $e$, is associated with its counterpart in the target
representing the postset of the corresponding event $f(e)$; designated as
$(f(e), f(e))$.
The mapping for transitions is straightforward. A transition representing a
forward event $e$ is mapped to a transition representing the corresponding
event $f(e)$. Reversing transitions are similarly mapped to reversing
transitions.

The following result ensures that the construction above actually gives an
\racn-morphism.

\begin{prop}
  If $f : \mathsf{H}_0 \rightarrow \mathsf{H}_1$ is an \raes{}-morphism, then
  $\aestoacn{r}(f) : \aestoacn{r}(\mathsf{H}_0) \rightarrow
  \aestoacn{r}(\mathsf{H}_1)$ is a well-defined \racn-morphism.
\end{prop}

\input{functor-proof}

\begin{prop}\label{pr:raestoracn-functor}
  $\aestoacn{r} : \mathbf{RAES} \rightarrow \mathbf{RACN}$ is a well-defined
  functor.
\end{prop}

\ifreport
\begin{prop}\label{pr:raes-same}
  Let $\mathsf{H}$ be an \raes, then
  $\acntoaes{r}(\aestoacn{r}(\mathsf{H})) = \mathsf{H}$.
\end{prop}
\begin{proof}
  Consider the $\mathsf{H} = (E, \anR, <, \nearrow, \prec, \lhd)$ and the
  associated \racn{}
  $\aestoacn{r}(\mathsf{H}) = \langle S, T, F, I, \mathsf{m}\rangle$ as
  defined in \Cref{de:raestoracn}. Let
  $\acntoaes{r}(\aestoacn{r}(\mathsf{H}))= (E', \anR', <', \nearrow', \prec',
  \lhd')$. We now show that the elements coincides:
  \begin{itemize}
  \item $E'$: By \Cref{de:raestoracn},
    $T = E\cup \setcomp{\un{\anr}}{\anr\in\anR}$ with $\fwdset{} = E$.
    By \Cref{def:rpcntoraes}, $E' = \fwdset{} = E$.
  \item $\anR'$: By \Cref{de:raestoracn},
    $\bwdset{} = \setcomp{\un{\anr}}{\anr\in\anR} $. By
    \Cref{def:rpcntoraes},
    $\anR' = \{ t \in \fwdset{} \ |\ \abwd\in\bwdset{}\}$. By substituting
    $\fwdset{}$ and $\bwdset{}$,
    $\anR' = \{ t \in E \ |\ \abwd\in
    \setcomp{\un{\anr}}{\anr\in\anR}\}$. Hence, $\anR' = \anR$.
  \item $<'$: By \Cref{def:rpcntoraes}, $<' = \lessdot$. Recall that
    $t \lessdot t'$ iff $\pre{t}\cap\inib{t'}\neq\emptyset$. By inspecting the
    definition of $F$ and $I$, $\pre{t}\cap\inib{t'} \neq \emptyset$ implies
    $t = e$, $t = e'$ with $e, e'\in E$ and $e < e'$. Hence, $<' = <$.
  \item $\nearrow'$: By \Cref{def:rpcntoraes},
    ${\nearrow'} = {\lll \cup \leadsto}$. We now show that $e \nearrow e'$ iff
    $e \nearrow' e'$. Assume that $e \nearrow e'$. Then,
    $(({e'}, {e'}), e) \in I$, by \Cref{de:raestoracn}. Hence,
    $e \leadsto e'$. Therefore, $e \nearrow' e'$. On the contrary, assume
    $e \nearrow' e'$. Then, either $e \lll e'$ or $e \leadsto e'$. If
    $e \leadsto e'$, then $(({e'}, {e'}), e) \in I$. By \Cref{de:raestoracn},
    $e \nearrow e'$. In case $e \lll e'$, we recall that $\lll$ is the
    transitive closure of
    $\lessdot \cap \{(\afwd, \afwd')\ |\ \abwd\not\in\bwdset{}\ \textit{or}\
    \inib{\abwd}\cap\post{\afwd'}\neq\emptyset \}$. Hence, $e \lll e'$ implies
    $e \lessdot e'$. Therefore, $\pre e \cap \inib{e'} \neq \emptyset$. Then,
    $((\emptyset, e), e') \in I$. By inspecting \Cref{de:raestoracn},
    $e < e'$. Since $\mathsf{H}$ is an \raes, $e < e'$ implies $e \nearrow e'$
    (\Cref{de:aes}(\ref{def:aes-reflect-causality})).

  \item $\prec'$: Then, $\prec' = \prec_{\aestoacn{r}(\mathsf{H})}$.
    Therefore, $e \prec' \un{u}$ implies
    $\pre{e}\cap\inib{\un u}\neq \emptyset$. Then,
    $((\emptyset, e), \un u) \in I$. By inspecting \Cref{de:raestoracn}, we
    conclude that $e \prec \un {u}$.

  \item $\lhd'$:  Then, $\lhd' = \lhd_{\aestoacn{r}(\mathsf{H})}$. Therefore,
    $\un{u}\lhd' e$ implies $\post{e}\cap\inib{\un u}\neq \emptyset$. Then,
    $((e, e), \un u) \in I$. By inspecting  \Cref{de:raestoracn}, we conclude that
    $\un{u}\lhd e$.
    \qedhere
  \end{itemize}
\end{proof}
\fi

\noindent 
The main result is that there is a precise relation between $\mathbf{RACN}$
and $\mathbf{RAES}$, namely the functor $\aestoacn{r}$ is the left adjoint of
$\acntoaes{r}$.

\begin{restatable}{thm}{raesacncorefl}
  \label{th:raesrcn-corefl}
  The functor $\aestoacn{r} : \mathbf{RAES}\to \mathbf{RACN}$ is the left
  adjoint of the functor $\acntoaes{r} : \mathbf{RACN} \to \mathbf{RAES}$.
\end{restatable}
%
\input{coreflection-proof}

Denoting with $\acntoaes{}$ and $\aestoacn{}$ the functors defined as $\acntoaes{r}$ and
$\aestoacn{r}$ but acting on
objects that are either \acns or \aeses, i.e. if we restrict our attentions to the
two full and faithful subcategories of $\mathbf{ACN}$ and $\mathbf{AES}$, we have
that the same relation the  between them exists.

\begin{restatable}{thm}{aesacncorefl}
\label{th:aesacn-corefl}
  The functor $\aestoacn{} : \mathbf{AES}\to \mathbf{ACN}$
is the left adjoint of the functor $\acntoaes{} : \mathbf{ACN} \to \mathbf{AES}$.
\end{restatable}

\begin{cor}
  \label{cor:raesrcn-corefl}
  The adjunctions $\acntoaes{r} \vdash \aestoacn{r}$ and $\acntoaes{} \vdash \aestoacn{}$ are coreflections.
\end{cor}
\begin{proof}
It is suffices to observe that the units of the adjunctions are natural isomorphisms.
\end{proof}

%% file: racnraes-proof.tex
\ifreport
\begin{proof}
  We show that the conditions in \Cref{de:raes} are satisfied:
  \begin{enumerate}
  \item Weak causality is set to ${\lll}\cup {\rcnprevent}$. By definition,
    both ${\lll}$ and $\rcnprevent$ are relations over
    $\fwdset{} \times \fwdset{}$. Hence,
    $({\lll}\cup {\rcnprevent}) \subseteq \fwdset{} \times \fwdset{}$.

  \item Prevention corresponds to $\lhd$, which, by definition satisfies
    $\lhd \subseteq \bwdset{} \times \fwdset{}$.

  \item Firstly, causation is set to $\lessdot$, which, by definition,
    satisfies $\lessdot \subseteq \fwdset{} \times \fwdset{}$.
    Moreover, since $\arcn{V}{\bwdset{}}$ is an \racn{},
    $\arcn{V_{\fwdset{}}}{}$ is an \acn{} by
    \Cref{de:reversible-causal-net}(\ref{rcn:cond1}). Hence, $\lessdot$ is
    irreflexive because of \Cref{de:pre-acausal-net}(\ref{pcn:cond4}).
    By \Cref{de:pre-acausal-net}(\ref{pcn:cond4}),
    $\forall t\in \fwdset{}.\ \histtwo{t}{\lessdot} = \setcomp{t'\in
      \fwdset{}}{t'\lessdot^{\ast} t}$ is finite and
    $(\rcnprevent\cup\lessdot)$ is acyclic on $\histtwo{t}{\lessdot}$.
    By \Cref{de:reversible-causal-net}(\ref{rcn:cond6}),
    ${\lll} \subset {\lessdot}$, hence
    $ ({\lll} \cup {\rcnprevent} \cup {\lessdot}) = ({\rcnprevent} \cup
    {\lessdot})$. Therefore, $({\rcnprevent} \cup {\lessdot})$ acyclic on
    $\histtwo{t}{\lessdot}$ implies
    $({\lll} \cup {\rcnprevent} \cup {\lessdot})$ acyclic on
    $\histtwo{t}{\lessdot}$.

  \item Reverse causation is set to $\prec$, which is by definition
    $\prec\ \subseteq \fwdset{} \times \bwdset{}$.
    \begin{enumerate}
    \item By \Cref{de:reversible-causal-net}(\ref{rcn:cond2}),
      $\forall \abwd\in\bwdset{}$, $\pre{\afwd}\subseteq\inib{\abwd}$. Hence,
      $t \prec \un{t}$.
    \item By \Cref{de:reversible-causal-net}(\ref{rcn:cond3}), for all
      $\abwd\in\bwdset{}$,
      $K_{\abwd} = \setcomp{\afwd' \in \fwdset{}}{\inib{\abwd} \cap
        \pre{\afwd'{}} \neq \emptyset}$ is finite and $\rcnprevent$ acyclic on
      $K_{\abwd}$. Note that
      $K_{\abwd} = \setcomp{\afwd' \in \fwdset{}}{t \prec {\un{t'}}} =
      \histtwo{\un t}{\prec}$. Hence, $\histtwo{\un{t}}{\prec}$ is finite. It remains
      to show that ${\lll} \cup {\rcnprevent} \cup \lessdot$ is acyclic on
      $K_{\abwd}$. As in the previous item,
      $({\lll}\cup {\rcnprevent}\cup \lessdot) = ({\rcnprevent}\cup
      \lessdot)$. Hence, we need to show that $ ({\rcnprevent}\cup \lessdot)$
      is acyclic on $K_{\abwd}$.
      We know that $\rcnprevent$ is acyclic on $K_{\abwd}$, and $\lessdot^{+}$
      is a partial order since $\acntoaes{r}(\arcn{V}{\bwdset{}})$ is a \racn.
      Hence, acyclicity on $K_{\abwd}$ can only be violated if, for any two
      $t', t''\in \histtwo{\un t}{\prec}$, we have $t'\rcnprevent t''$ and
      $t''\lessdot t'$. However, this implies that
      $\post{t''}\cap\inib{t'}\neq \emptyset$ and
      $\pre{t''}\cap\inib{t'}\neq \emptyset$, which contradicts the fact that
      $\acntoaes{r}(\arcn{V}{\bwdset{}})$ is an \racn.

    \end{enumerate}
  \item for all $t\in \fwdset{}, \un{t'}\in\bwdset{}$.
    $t\prec \un{t'}\ \Rightarrow \neg(\un{t'}\lhd t$). Note that
    $t \prec \un{t'}$ implies that
    $\pre{t} \cap \inib{\un{t'}} \neq \emptyset$. Consequently,
    $\post{t} \cap \inib{\un{t'}} = \emptyset$, indicating that
    $\neg(\un{t'} \lhd t)$.

  \item We need to show that $(\fwdset{},\pprec, {\lll}\cup {\rcnprevent})$
    with
    $ {\pprec} = {\lessdot} \cap {\{(t,t')\ |\ t\not\in\anR \textit{ or }
      \un{t}\lhd t'\}}$ is an \aes. We now check the conditions of
    \Cref{de:aes}. First note that $\pprec\ = \lll$ by the definition of
    $\lll$.

    \begin{enumerate}

    \item We need to show that $\pprec\ \subseteq \fwdset{}\times\fwdset{}$ is
      an irreflexive partial order defined such that $\forall t\in\fwdset{}$.
      $\hist{t}_{\pprec}$ is finite. The fact that $\pprec$ irreflexive and
      antisymmetric because $\lll\subseteq\lessdot+$ and $\lessdot^+$ is an
      irreflexive partial order. Transitivity follows from the definition of
      $\lll$ (\Cref{de:reversible-causal-net}(\ref{rcn:cond6})).

      $\hist{t}_{\pprec}$ is finite because $\pprec = \lll \subseteq\lessdot$
      and $\hist{t}_{\lessdot}$ is finite
      (\Cref{de:pre-acausal-net}(\ref{pcn:cond4})).

    \item $({\lll}\cup {\rcnprevent}) \subseteq \fwdset{}\times\fwdset{}$ by
      definition. We need to show that for all $t, t'\in \fwdset{}$:
      \begin{enumerate}
      \item $t \lll t'\ \Rightarrow\ t\ ({\lll}\cup {\rcnprevent})\ t'$; which
        immediately follows because
        ${\lll} \subseteq {(\lll \cup \rcnprevent)}$.
      \item
        $({\lll}\cup {\rcnprevent}) \cap (\hist{t}_{\pprec}\times
        \hist{t}_{\pprec})$ is acyclic. Note that
        $({\lll}\cup {\rcnprevent}) \cap (\hist{t}_{\pprec}\times
        \hist{t}_{\pprec}) = ({\lll}\cup {\rcnprevent}) \cap
        (\hist{t}_{\lll}\times \hist{t}_{\lll}) = (\hist{t}_{\lll}\times
        \hist{t}_{\lll}) \subseteq (\hist{t}_{\lessdot}\times
        \hist{t}_{\lessdot})$, which is acyclic because
        $\arcn{V_{\fwdset{}}}{}$ is an \acn{}
        (\Cref{de:pre-acausal-net}(\ref{pcn:cond4})).
      \item if $t \# t'$ and $t'\pprec t''$ then $t \# t''$. It follows from
        \Cref{de:reversible-causal-net}(\ref{rcn:cond6}).
      \end{enumerate}
    \end{enumerate}
\end{enumerate}

  We can conclude that $\acntoaes{r}(V)$ is indeed an \raes{}.
\end{proof}
\fi

%% file: raesracn-proof.tex
\ifreport
\begin{proof}
  Consider the \raes{} $\mathsf{H} = (E, \anR, <, \nearrow, \prec, \lhd)$ and
  $\aestoacn{r}(\mathsf{H}) = \langle S, E\cup
  \setcomp{\un{\anr}}{\anr\in\anR}, F, I, \mathsf{m}\rangle$ as defined in
  \Cref{de:raestoracn}. Initially, we show that the net
  $\langle S, E, F', I', \mathsf{m}\rangle$, where $F'$ and $I'$ are the
  restrictions of $F$ and $I$ to the transitions in $E$, is a p\paca{}.
  The first three conditions of \Cref{de:pre-acausal-net} follow through a
  straightforward examination of the definition of $\aestoacn{r}(\mathsf{H})$.
  The fourth condition is a consequence of $\mathsf{H}$ being an \raes{},
  satisfying \Cref{de:raes}(\ref{cond:3}), which translates to
  Condition~\ref{pcn:cond4} in \Cref{de:pre-acausal-net}.
  The final condition is implied by the fact that $(E,\pprec,\nearrow)$
  constitutes an \aes with ${\pprec} \subseteq {<}$.

  For each transition in $\setcomp{\un{\anr}}{\anr\in\anR}$, there is
  precisely one corresponding transition in $E$, represented by $\anr$, as
  dictated by $\mathsf{H}$ being an \raes{}.
  Condition~\ref{rcn:cond3} in \Cref{de:reversible-causal-net}—--asserting
  that $K_{\un{\anr}}$ is finite---stems from $\histtwo{\un{\anr}}{\prec}$ being
  finite in an \raes{}. For the same reason, causation and weak causality are
  acyclic on such sets. Hence, Condition~\ref{rcn:cond4} is also satisfied by
  construction.
  The condition \ref{rcn:cond6} of \Cref{de:reversible-causal-net} is verified
  as it mimic the requirement of sustained causality of the \raes{}
  $\mathsf{H}$. The final condition depends again on the analogous one in
  \raes{}.
  We can conclude that indeed $\aestoacn{r}(\mathsf{H})$ is an \racn{}.
\end{proof}
\fi

%% file: functor-proof.tex

\ifreport
\begin{proof}
  We verify that $\aestoacn{r}(f)$ constitutes an \rpacn-morphism from
  $\aestoacn{r}(\mathsf{H}_0)$ to $\aestoacn{r}(\mathsf{H}_1)$ by checking the
  conditions of \Cref{de:rev-morphism}.
  Assume
  $\aestoacn{r}(\mathsf{H}_i) = \langle S_i, T_i , F_i, I_i, \mathsf{m}_i
  \rangle$ for $i=0,1$. By definition of $\aestoacn{r}(\mathsf{H}_i)$, note
  that $\fwdset{i} = E_i$ and
  $\bwdset{i} = \setcomp{\un{\anr}}{\anr\in\anR_i}$.

  \begin{enumerate}
  \item By definition of $\aestoacn{r}(f)$,
    $f_T(\fwdset{0}) = f_T(E_0) \subseteq f_T(E_1) = \fwdset{1}$. Similarly,
    $f_T(\bwdset{0})= f_T(\setcomp{\un{\anr}}{\anr\in\anR_0}) \subseteq
    f_T(\setcomp{\un{\anr}}{\anr\in\anR_1}) = \bwdset{1}$.
  \item We need to show that
    $(f_S,f_T|_{E_0}) : \resarcn{V}{}{E_0} \rightarrow \resarcn{V}{}{E_1}$ is
    an \paca-morphism. Hence, we check the conditions of
    \Cref{de:pca-morphism}.
    \begin{enumerate}
    \item for all $e\in E_0$ if $f_T(e) \neq \bot$ then
      \begin{enumerate}
      \item We show that $\pre{f_T(e)} = f_S(\pre{e})$ and
        $\post{f_T(e)} = f_S(\post{e})$
        \begin{align*}
          \pre{f_T(e)}
          & = \pre{f(e)} & \text{By def. of } f_T(e).\\
          & = \{(\emptyset, f(e))\} & \text{By def. of } F_0. \\
          & =  f_S(\{ (\emptyset, e)\}) & \text{By def. of } f_S. \\
          & =  f_S(\pre{e}) & \text{By def. of } F_0.
        \end{align*}

        Analogously,
        $\post{f_T(t)} =\post{f(e)} = \{ (f(e), f(e))\} =  f_S( \{ (e, e)\}
        ) =  f_S(\post{t})$.

      \item Assume now $(s,f_T(e))\in I_1$. We distinguish two cases, either
        $\post{s} \neq \emptyset$ or $\post{s} = \emptyset$.

        In the first case, if $f_S^{-1}(s) = \emptyset$ we have nothing to prove, thus
        assume that $f_S^{-1}(s) \neq \emptyset$
        By \Cref{de:raestoracn}, if $(s,f_T(e))\in I_1$, then
        there are two cases:
        \begin{itemize}
        \item $s = (\emptyset, e')$ and $e' <_1 f_T(e)$: Since $f$ is an
          \raes{}-morphism, it is also an \aes-morphism. Consequently, it
          preserves causes
          (\Cref{de:aes-morphisms}(\ref{def:aes-morphism-preserve})).
          Therefore, it should hold that $\hist{f(e)}\subseteq f(\hist{e})$.
          Since $e' \in \hist{f(e)}$, there must be $e_0 \in \hist{e} $ such
          that $f(e_0) = e'$. Then, it suffices to consider
          $s' = (\emptyset, e_0)$. It is immediate that $s' \in f_S^{-1}(s)$
          because $f(e_0) = e'$. Moreover, $((\emptyset, e_0), e)\in I_0$
          because $e_0 <_0 e$ (\Cref{de:raestoracn}).
        \item $s = (e', e')$ and $f_T(e) \nearrow e'$. As
          $f_S^{-1}(s) \neq \emptyset$, we can consider all $e_0\in E_0$ such
          that $f(e_0) = e'$. Since $f$ is an \raes{}-morphism, it reflects
          weak causality
          (\Cref{de:aes-morphisms}(\ref{def:aes-morphism-reflect})). Hence,
          for all $e_0$ such that $f(e_0) = e'$, we have that
          $e \nearrow e_0$. Hence, $((e_0, e_0), e)\in I_0$ by
          \Cref{de:raestoracn}.
        \end{itemize}
      \end{enumerate}
    \item We show that $\forall e, e'\in E_0$ if
      $f_T(e) \neq \bot\neq f_T(e')$ then
      $f_T(e) = f_T(e')\ \Rightarrow\ e\ \cnconf_0\ e'$. Since $f$ is an
      \raes{}-morphism, $f_T(e) = f_T(e')$ implies $e\#_0 e'$, i.e.,
      $e\nearrow e'$ and $e'\nearrow e$.
      Then, by \Cref{de:raestoracn}, $((e, e), e') \in I$ and
      $((e', e' ), e) \in I$. Hence, $e \cnconf e'$.

    \item We show that
      $\forall s_1\in S_1.\ \forall s_0, s_0'\in f_S^{-1}(s_1).$
      $s_0 \neq s_0'$ implies $\post{s_0}\ \cnconf_0\ \post{s_0'}$ or
      $\pre{s_0}\ \cnconf_0\ \pre{s_0'}$. We proceed by case analysis on the
      shape of $s_1$.
      \begin{itemize}
      \item $s_1=(\emptyset,\emptyset)$ or $s_1 = (\emptyset, e)$ and
        $f^{-1}(e) = \bot$. Then, $f_S^{-1}(s_1)$ is a singleton by
        \Cref{def:induced-morphism}; hence, the thesis follows vacuously.
      \item $s_1 = (\emptyset, e_1)$ with $e_1 = f(e_0)$ for some
        $e_0\in E_0$. Then, there exist $e_0, e_0' \in E_0$ such that
        $f(e_0) = f(e_0') = e_1$ and $s_0 = (\emptyset, e_0)$ and
        $s_0' = (\emptyset, e_0')$. Since $f$ is an \raes-morphism,
        $e_0\#_0 e_0'$. Therefore, $e_0\nearrow_0 e_0'$ and
        $e_0'\nearrow_0 e_0$. By \Cref{de:raestoracn}, both
        $((e_0, e_0), e_0')$ and $((e_0',e_0'), e_0)$ are in $I_0$. Hence,
        $e_0\ \cnconf_0\ e_0'$. Therefore,
        $\post{s_0}\ \cnconf_0\ \post{s_0'}$.
      \item $s_1 = ( e_1, e_1)$ with $e_1 = f(e_0)$ for some $e_0\in E_0$. By
        reasoning analogously to the previous case, it is shown that
        $\pre{s_0}\ \cnconf_0\ \pre{s_0'}$.
      \end{itemize}
    \noindent 
    \item We show that $\mrflt{f_S(\mathsf{m}_0)}= \mathsf{m}_1$. By
      \Cref{de:pre-acausal-net}(\ref{pcn:cond2}),
      $\mathsf{m}_i = S_i\setminus \post{T_i}$. By \Cref{de:raestoracn},
      \[
        \mathsf{m}_i = \setcomp{(\emptyset,e)}{e\in E_i}\ \cup\
        \setenum{(\emptyset,\emptyset)}
      \]
      Consequently,
      \[
        f_S(\mathsf{m}_0) = f_S(\setcomp{(\emptyset,e)}{e\in E_i})\cup
        f_S(\setenum{(\emptyset,\emptyset)})
      \]

      By \Cref{def:induced-morphism}(\ref{def:induced-morphism-1}),
      \[
        \hspace*{\leftmargini}
        f_S(\mathsf{m}_0) = \multisetcomp{(\emptyset,f(e))}{e\in E_0, f(e) \neq
        \bot} + \multisetcomp{(\emptyset, e)}{e\in E_1, f^{-1}(e) \neq
        \bot} + \multisetenum{(\emptyset,\emptyset)}
      \]
      Then,
      \[
        \hspace*{\leftmargini}
        \mrflt{f_S(\mathsf{m}_0)} = \setcomp{(\emptyset,f(e))}{e\in E_0, f(e)
          \neq \bot} \cup \setcomp{(\emptyset, e)}{e\in E_1, f^{-1}(e) \neq
          \bot} \cup \setenum{(\emptyset,\emptyset)}
      \]
      Note that the first set on the right-hand-side contains all places
      $(\emptyset, e)$ with $e\in E_1$ that are in the image of $f$, while the
      second one contains all places $(\emptyset, e')$ with $e'\in E_1$ that
      are not in the image in of $f$. Hence, they cover all the elements of
      $E_1$. Hence,
      \[
        \mrflt{f_S(\mathsf{m}_0)} = \setcomp{(\emptyset,e'))}{e'\in E_1}\
        \cup\ \setenum{(\emptyset,\emptyset)}\ = \ \mathsf{m}_1
      \]

    \end{enumerate}
  \item We check that $\forall \abwd\in \bwdset{0}$. If $\abwd\in \bwdset{0}$,
    then $\abwd = \un{\anr}$ with $u\in U_0$. Then, the corresponding forward
    transition is $t = u$ with $u\in U_0\subseteq E_0$. If
    $f_T(t) = f_T(u) \neq \bot$ then $f(t) = f(u) \neq \bot$ by
    \Cref{def:induced-morphism}(\ref{def:induced-morphism-2}).
    \begin{enumerate}
    \item By \Cref{def:induced-morphism}(\ref{def:induced-morphism-2}),
      $f_T(\abwd) = \un{f(\anr)} \neq \bot$. Since $f_T(t) = f(u)$,
      $\un{f_T(t)} = \un{f(u)}$ by \Cref{de:raestoracn}. Hence,
      $f_T(\abwd) = \un{f_T(\afwd)}$.
    \item Consider now $(s,f_T(\un t))\in I_1$, we have again to distinguish two
      cases:
      \begin{itemize}
      \item $s = (\emptyset, {e})$ and $e\prec f_T(\un{\anr})$. Since $f$ is
        an \raes-morphism, by
        \Cref{de:raes-morphisms}(\ref{de:raes-morphisms-2}),
        $\histtwo{\un{f({\anr})}}{\prec_1}\subseteq
        f(\histtwo{\un{\anr}}{\prec_0})$. Consequently, there exists
        $e_0\in\histtwo{\un{\anr}}{\prec_0}$ such that $f(e_0) = e$. Then,
        take $s' = (\emptyset, e_0)$ and note that
        $((\emptyset, e_0), \un{u})\in I_0$ by \Cref{de:raestoracn}.
      \item $ s =(f(e_0), f(e_0))$ and $f(\un{\anr})\lhd f(e)$. By
        \Cref{def:induced-morphism}, $(e,e) \in f_S^{-1}(s)$. Since $f$ is an
        \raes-morphism , $\un{\anr}\lhd e$, by
        \Cref{de:raes-morphisms}(\ref{de:raes-morphisms-3}). Then,
        $((e,e), \un{\anr}) \in I_0$, by \Cref{def:induced-morphism}. \qedhere
      \end{itemize}
    \end{enumerate}
  \end{enumerate}
\end{proof}
\fi

%% file: coreflection-proof.tex

\ifreport
\begin{proof}
  Let $\mathsf{H} = (E, U, <, \nearrow, \prec, \lhd)$ be an \raes and
  $\aestoacn{r}(\mathsf{H}) = \langle S, T, F, I, \mathsf{m}\rangle$ the
  associated \racn.
  By \Cref{pr:raes-same}, we have that
  $\mathsf{H} = \acntoaes{r}(\aestoacn{r}(\mathsf{H}))$ and we have the
  obvious identity mapping
  $\mathit{id} : \mathsf{H} \to \acntoaes{r}(\aestoacn{r}(\mathsf{H}))$.
  To prove the result, we show that for any \racn
  $V = \langle S_V, T_V, F_V, I_V, \mathsf{m}_V\rangle$, and any
  \raes-morphism $g : \mathsf{H} \rightarrow \acntoaes{r}(V)$, there exists a
  unique morphism $h = (h_S,h_T): \aestoacn{r}(\mathsf{H}) \rightarrow V$ such
  that the following diagram commutes.
  \begin{center}
    \centerline{\input{examples/theo1-schema}}
  \end{center}

  We begin by establishing the existence of the morphism. Let $h = (h_S, h_T)$
  be defined as follows:
  \[
    h_S(s_0,s_1) \text{ iff }
    \begin{cases}
      s_0 = (\emptyset, \emptyset) \wedge s_1 \in \mathsf{m}_V
      \land\ g(\post{s_1}) = \bot
      \\
      s_0 = (\emptyset, \setenum{e}) \wedge s_1 \in \mathsf{m}_V \land\ g(e)\in g(\post{s_1})
      \\
      s_0 = (\setenum{e},  \setenum{e}) \wedge s_1 \in \post{g(e)}
      \\
    \end{cases}
  \]
  and $h_T(e)=g(e)$ and $h_T(\un\anr) = \un{g(\anr)}$ if $u\in \anR$.

  We verify that $h$ is an \rpacn-morphism from $\aestoacn{r}(\mathsf{H})$ to
  $V$ by checking the conditions of \Cref{de:rev-morphism}.
  By definition of $\aestoacn{r}(\mathsf{H})$, note that $\fwdset{} = E$
  and $\bwdset{} = \setcomp{\un{\anr}}{\anr\in\anR}$.

  \begin{enumerate}
  \item By definition of $h$, $h_T(\fwdset{}) = g(\fwdset{}) = g(E)$. Since
    $g$ is an \raes-morphism, $g(E) \subseteq \fwdset{V}$. Similarly,
    $h_T(\bwdset{})= g(\setcomp{\un{\anr}}{\anr\in\anR})$. Since $g$ is an
    \raes-morphism, $g(\setcomp{\un{\anr}}{\anr\in\anR})\subseteq \bwdset{V}$.
  \item We need to show that
    $(h_S,h_T|_{E}) : \aestoacn{r}(\mathsf{H})_E \rightarrow
    \resarcn{V}{}{}_{\fwdset{V}}$ is an \paca-morphism. Hence, we check the
    conditions of \Cref{de:pca-morphism}.
    \begin{enumerate}
    \item for all $e\in E$ if $h_T(e) \neq \bot$ then
      \begin{enumerate}
      \item We show that $\pre{h_T(e)} = h_S(\pre{e})$ and
        $\post{h_T(e)} = h_S(\post{e})$
         \begin{align*}
           \pre{h_T(e)}
           & = \pre{g(e)} & \text{By def. of } h_T(e).\\
           & = \{(\emptyset, g(e))\} & \text{By def. of } F. \\
           & = h_S(\{ (\emptyset, e)\}) & \text{By def. of } h_S. \\
           & = h_S(\pre{e}) & \text{By def. of } F.
         \end{align*}

         Analogously,
         $\post{h_T(t)} =\post{g(e)} = \{ (g(e), g(e))\} = 
         h_S( \{ (e, e)\} ) =  h_S(\post{t})$.

       \item We now check that inhibitor arcs are reflected. Consider now $e\in E$ and
         $s \in S_V$ such that $(s,h_T(e))\in I_V$.
         \begin{enumerate}
         \item Case $\pre{s} = \emptyset$. By
           \Cref{lem:mapping-minimal-morphism}, $ h_S^{-1}(s) \neq \emptyset$.
           Hence, there exists $s'\in S$ such that $h_S(s') = s$. Moreover,
           $s' = (\emptyset, e')$ and $g(e') = \post s$. As
           $(s, h_T(e))\in I_V$ implies that $\post{s} \lessdot h_T(e)$, i.e.,
           $\post{s} = g(e') < g(e)$. As $g$ is an \raes-morphisms, $e' < e$.
           Consequently, $(s',e)\in I$;
         \item Case $\pre{s} \neq \emptyset$. For any $s'\in S$ such that
           $h_S(s') = s$, it should be $s' = (e', e')$ and $g(e') = \pre s$.
           As $(s, h_T(e))\in I_V$ implies that $\post{s} \nearrow h_T(e)$;
           hence $\post{s} = g(e') \nearrow g(e)$. As $g$ is an
           \raes-morphisms, $e' \nearrow e$. Consequently, $(s',e)\in I$;
         \end{enumerate}
      \end{enumerate}

    \item We show that $\forall e, e'\in E$ if $h_T(e) \neq \bot\neq h_T(e')$
      then $h_T(e) = h_T(e')\ \Rightarrow\ e\ \cnconf\ e'$. Note that
      $h_T(e) = h_T(e')$ implies $g(e) = g(e')$, by the definition of $h_T$.
      Since $g$ is an \raes{}-morphism, $g(e) = g(e')$ implies $e\# e'$, i.e.,
      $e\nearrow e'$ and $e'\nearrow e$.
      Then, by \Cref{de:raestoracn}, $\aestoacn{r}(\mathsf{H})$ is defined
      such that $((e, e), e') \in I$ and $((e', e' ), e) \in I$. Hence,
      $e\ \cnconf\ e'$.

    \item We show that
      $\forall s_1\in S_V.\ \forall s_0, s_0'\in h_S^{-1}(s_1).$
      $s_0 \neq s_0'$ implies $\post{s_0}\ \cnconf\ \post{s_0'}$ or
      $\pre{s_0}\ \cnconf\ \pre{s_0'}$. We proceed by case analysis on
      $s_1$ (according to the definition of $h_S$):
      \begin{itemize}
      \item $s_1 \in \mathsf{m}_V \land\ g(\post{s_1}) = \bot$: Hence,
        $h_S^{-1}(s_1)$ is the singleton $\setenum{(\emptyset,\emptyset)}$.
        Therefore, thesis follows vacuously.
      \item $s_1 \in \mathsf{m}_V \land\ g(\post{s_1}) \neq \bot$. Then, there
        exist $e, e'\in E$ such that $g(e) = g(e')$ and $s_0 = (\emptyset, e)$
        and $s_0' = (\emptyset, e')$. Since $g$ is an \raes-morphism,
        $e\# e'$. Therefore, $e\nearrow e$ and
        $e'\nearrow e$. By \Cref{de:raestoracn}, both
        $((e, e), e')$ and $((e',e'), e)$ are in $I$. Hence,
        $e\ \cnconf\ e'$. Therefore,
        $\post{s_0}\ \cnconf\ \post{s_0'}$.

      \item $s_1 \in \post{(g(e))})$. As in the previous case, there exist
        $e, e'\in E$ such that $g(e) = g(e')$ and $s_0 = (e, e)$ and
        $s_0' = (e', e')$. Since $g$ is an \raes-morphism, $e \# e'$.
        Therefore, $e\nearrow e$ and $e'\nearrow e$. By \Cref{de:raestoracn},
        both $((e, e), e')$ and $((e',e'), e)$ are in $I$. Hence,
        $e\ \cnconf\ e'$. Therefore, $\pre{s_0}\ \cnconf\ \pre{s_0'}$.
      \end{itemize}

    \item We prove that $\mrflt{h_S(\mathsf{m})}= \mathsf{m}_V$. By
      \Cref{de:raestoracn},
      \[
        \mathsf{m} = \setcomp{(\emptyset,e)}{e\in E}\ \cup\
        \setenum{(\emptyset,\emptyset)}
      \]
      Consequently,
      \[
        h_S(\mathsf{m}) = h_S(\setcomp{(\emptyset,e)}{e\in E})\cup
        h_S(\setenum{(\emptyset,\emptyset)})
      \]

      By the definition of $h_S$,
      \[
        h_S(\mathsf{m}) = \multisetcomp{s_1}{s_1 \in \mathsf{m}_V
          \land\ g(\post{s_1}) = \bot}
        + \multisetcomp{s_1}{s_1 \in \mathsf{m}_V \land\  g(\post{s_1}) \neq \bot}
      \]
      Note that both multisets on the right-hand-side are actually sets.
      Moreover, their union covers all places in $\mathsf{m}_V$. Hence,
      \[
        \mrflt{h_S(\mathsf{m}_0)} = \mathsf{m}_V
      \]
    \end{enumerate}
  \item $\forall \abwd\in \bwdset{}$. If $\abwd\in \bwdset{}$, then
    $\abwd = \un{\anr}$ with $u\in U$. Then, the corresponding forward
    transition is $t = u$ with $u\in U\subseteq E$. If
    $h_T(t) = h_T(u) \neq \bot$ then $g(t) = g(u) \neq \bot$ by the definition
    of $h$.
    \begin{enumerate}
    \item By the definition of $h_T$,
      $h_T(\abwd) = h_T(\un u) = \un{g(\anr)} = \un{h_T(u)}$.
    \item Consider $(s,h_T({\un{t}}))\in I_V$.
       \begin{enumerate}
       \item Case $\post{s}\cap \fwdset{V}\neq \emptyset$. If
         $h_T^{-1}(s)\neq \emptyset$ then there exists a
         $s' = (\emptyset, e')$ and $g(e') = \post s$. Since,
         $(s,\underline{t}) \in I_V$, it implies
         $post s = g(e') \prec g(\underline{t})$. Since $g$ is a
         \raes-morphism, $e \prec \un{t}$. Therefore,
         $(s', \underline{t}) \in I$
       \item Case $\pre{s}\cap \fwdset{V} = \emptyset$. Take
         $s'\in h_T^{-1}(s)$. Hence, $s' = (e, e)$ and  $g(e') = \pre s$.
         Since,
         $(s,\underline{t}) \in I_V$, it implies
         $\pre s =  g(\underline{t}) \lhd g(e')$. Since $g$ is a
         \raes-morphism, $ \un{t} \lhd e$. Therefore,
         $(s', \underline{t}) \in I$.
       \end{enumerate}
    \end{enumerate}
  \end{enumerate}

  \noindent 
  The uniqueness of the mapping is established by assuming the existence of
  another $h' = (h'_S, h'_T)$. As $\acntoaes{r}(h')$ makes the diagram
  commutative, we can conclude $g = h'_T$. Consequently, $h'_S$ must be the
  relation $h_S$, concluding the proof.
\end{proof}
\fi

%% file: examples/theo1-schema.tex
\scalebox{0.9}{\begin{tikzpicture}
\usetikzlibrary{decorations.pathmorphing}

\tikzset{snake it/.style={decorate, decoration=snake}}
\tikzstyle{tr}=[->,thick]
\tikzstyle{transition}=[rectangle, draw=none,thick,minimum size=5mm]

\node[transition] (g) at (0,0) {$\mathsf{H}$}
;

\node[transition] (eng) at (3,0) {$\acntoaes{r}(\aestoacn{r}(\mathsf{H}))$}
;

\node[transition] (ng) at (3,-2) {$\acntoaes{r}(V)$}

;

\draw[->] (g) -- (ng) node[midway, below]{$g$};
\draw[->] (g) -- (eng) node[midway, above]{$id$};
\draw[dashed,->] (eng) --(ng)node[midway, right] {$\acntoaes{r}(h)$};  

\end{tikzpicture}}

%% file: applications.tex

\section{Applications}\label{sec:app}

\paragraph{Reversible Debugging}
One of the successful applications of reversibility is causal-consistent
\emph{reversible debugging}~\cite{GiachinoLM14,Lanese0PV18}. Reversible
debuggers extend the classical ones with the possibility to get back in the
execution of a program, so to find the source of a bug in an effective way.
Causal-consistent reversible debuggers improve on reversible ones by
exploiting causal information while undoing computations. So far this
technique has been applied to message passing concurrent systems (e.g.,
actor-like languages), but not to shared-memory based concurrency.
Consider the next code snippet consisting of a two-threaded program that
accesses dynamically allocated memory. \input{exampleC} The behaviour of the
program can be thought in terms of events. Take $a$ as the event corresponding
to the initialisation of $x$ at line $13$, $b$ for the instruction at line $8$
and $c$ for the one at line $17$. It is clear that both $b$ and $c$ causally
depend on $a$ ($a < b$ and $a<c$); while $c$ can happen after $b$, $b$ cannot
happen after $c$, that is $b \nearrow c$. Moreover, the reversal of a complete
execution of the program should ensure that $c$ is reversed (i.e., the memory
is allocated) before $b$ is reversed, hence $\underline{b} \triangleleft c$.
Consider instead a version of the program in which $c$ is executed outside a
critical section (e.g., without acquiring and releasing the lock). In this
case, the execution may raise a \textit{segmentation fault} error. When
debugging such a faulty execution, the programmer would observe that the
execution violates $b \nearrow c$ because $b$ happened after $c$.
On the hand, an execution can be visualised in terms of events, i.e., the
programmer can be provided with a high-level description of the current state
of the system (a configuration of the event structure) along with the relevant
dependencies. On the other hand, the instrumented execution of the program and
of its reversal can be handled by the underlying operational model (i.e., a
reversible causal net).
Also, one could think of the undoing of an event as a backward breakpoint.
That is, one could trigger the undoing of an event from the net and then the
debugger will execute all the necessary backward steps in the code, to undo
such event. The seamless integration of \raes and \racn with causal-consistent
reversible debuggers can be a nice exploitation of our results, which we will
consider in future work.

Another way to exploit our results is to generate \emph{directly} a reversible
semantics for languages for shared memory \cite{castellan,JeffreyR19}. In
these works, weak memory models are interpreted as event structures. By
exploiting our results, we could use $\rpes$ to interpret such models and give
them a reversible debugging tool in terms of corresponding Petri net. For
example, let us consider a simple snippet (taken from \cite{castellan})
\[r \leftarrow a \,||\, a:= 2 \,||\, b:=1 \] where $r$ is a thread-local
variable (e.g., a register) while $a$ and $b$ are global shared variables, and
all the variables are initialised to $0$.
The interpretation of such a snippet in terms of event structures is depicted
in \Cref{es:castellan}, where the assignment of $2$ to $a$ (event
$\mathtt{Wr}_a^{(2)}$), causes the first thread to read the value $2$ (event
$\mathtt{Re}_{a}^{(r=2)}$). On the other hand, if the first thread reads the
value $0$, then the assignment of $2$ to $a$ happens. The event of the third
thread (i.e., the assignment of $1$ to $b$) is independent of the others,
hence it has no conflict neither causes. This interpretation works fine when
considering just the normal forward flow of computation, but it turns out to
be \emph{too strict} if one wants to reverse debug such a snippet, since
sequencing is seen as causation.
If we want to reverse debug such a snippet, we could reason as follows: the
event $\mathtt{Wr}_{b}^{(1)}$ can be reversed at any time since it has no
causes, neither it is caused by another event. Also, the undoing of
$\mathtt{Wr}_a^{(2)}$ should cause the undoing of the event
$\mathtt{Re}_{a}^{(r=2)}$. But then, the undoing of $\mathtt{Re}_{a}^{(r=0)}$
should not cause the undoing of $\mathtt{Wr}_a^{(2)}$, since the second event
is not a consequence of the first one. Hence, the right interpretation for
reversible debugging should be the $\aes$ depicted in \Cref{aes:castellan}.
Again, we leave as future work the adaptation of the results in
\cite{castellan,JeffreyR19} for reversible systems.

\begin{figure}[t]
   \begin{subfigure}{.40\textwidth}
	\centerline{\input{examples/ex_castellan}}
	\caption{$\es$ (from~\cite{castellan})}\label{es:castellan}
\end{subfigure}
\quad
   \begin{subfigure}{.40\textwidth}
	\centerline{\input{examples/ex_as_castellan}}
	\caption{$\aes$}\label{aes:castellan}
\end{subfigure}
\end{figure}

\paragraph{A speculative scenario}
We now show how our framework can be used to model a \emph{speculative
  scenario} borrowed from \cite{speculative}. In this scenario, value
speculation is used as a mechanism to boost parallelism by predicting values
of data dependencies between tasks. Whenever a value prediction is incorrect,
corrective actions must be taken in order to re-execute the data consumer code
with the correct data value. In this regard, as shown in \cite{GiachinoLMT17}
for a shared-memory setting, reversible execution can permit to relieve
programmers from the burden of properly undoing the actions subsequent to an
incorrect prediction. For simplicity, our scenario will involve a
\emph{producer} and a \emph{consumer}. The producer produces a value on which
the consumer has to perform some calculation. Since the production of such a
value may require some time, the consumer can try to guess (e.g. speculate)
the value and launch a computation on the predicted value. At the end of the
computation, the consumer compares the predicted value with the real one, and
if it has done a wrong guess then it reverts its computation and redoes the
calculation with the actual value. If the guess was right, then nothing has to
be reverted. Hence we can model such a scenario with the $\raes$
$\mathsf{H} = (E, \anR, <, \nearrow, \prec, \lhd)$ where:
\begin{align*}
&E = \setenum{prod, wait, pred, cT, cF} \\
 &\anR = \setenum{cF, pred}\\
&< = \setenum{(pred, cF), (pred, cT)} \\
 &\nearrow=\setenum{(wait, pred), (pred,wait), (cT,cF), (cF,cT), (prod,pred)}\\
 & \prec = \setenum{(cF,\un{cF}), (pred,\un{pred})}\\
 &\lhd = \setenum{(\un{pred}, cT), (\un{pred}, cF)}
\end{align*}

The sef of \emph{forward} events $E$ includes the following events: $prod$
(indicating the production of the value), $wait$ (indicating that the consumer
decided to wait for the real value), $pred$ (indicating that the consumer
predicts the value and hence speculates), $cT$ (the predicted value
coincides with the one generated by the producer) and $cF$ (the predicted
value is wrong). The only \emph{reversible} events (set $U$) are $cF$ and
$pred$, as the computation has to be reverted just in case of a wrong
prediction. The causality relation $<$ is as expected: predicting the value
causes its comparison with the real one. About conflicts, we have that
$pred \# wait$ and $cT \# cF$. Also we have that the consumer may decide to
predict the value and afterwards the producer produces the value. The opposite
cannon happen, as it is worthless to speculate if the value has been already
generated. Hence, $pred \nearrow prod$. Reverse causation $\prec$ is as
expected. The prevention relation $\lhd$ allows for preventing the undoing of
the prediction if the prediction was right, hence $\un{pred} \lhd cT$, and in
case the prediction was wrong to revert first the comparison and then the
prediction ($\un{pred} \lhd cF$). The net corresponding to the $\raes$
$\mathsf{H}$ is depicted in \Cref{fig:ex_spec}, where some inhibitor arcs (those induced by
the weak causality of forward events) are omitted.

\begin{figure}[t]
\input{examples/ex_speculative.tex}
\caption{Speculative scenario $\racn$}
\label{fig:ex_spec}
\end{figure}

%% file: exampleC.tex
\definecolor{mGreen}{rgb}{0,0.6,0}
\definecolor{mGray}{rgb}{0.5,0.5,0.5}
\definecolor{mPurple}{rgb}{0.58,0,0.82}
\definecolor{backgroundColour}{rgb}{0.95,0.95,0.92}

\lstdefinestyle{CStyle}{
    backgroundcolor=\color{backgroundColour},   
    commentstyle=\color{mGreen},
    keywordstyle=\color{magenta},
    numberstyle=\tiny\color{mGray},
    stringstyle=\color{mPurple},
    basicstyle=\tiny,
    breakatwhitespace=false,         
    breaklines=true,                 
    captionpos=b,                    
    keepspaces=true,                 
    numbers=left,                    
    numbersep=5pt,                  
    showspaces=false,                
    showstringspaces=false,
    showtabs=false,                  
    tabsize=2,
    language=C
}
\begin{lstlisting}[language=customc,numbers=left]
pthread_mutex_t m = PTHREAD_MUTEX_INITIALIZER;
int *x;
 
void thread(void *arg)
{
    pthread_mutex_lock(&m);
    if(x != NULL) 
    	doSomething(x);
    pthread_mutex_unlock(&m);
}
\end{lstlisting}
\begin{lstlisting}[language=customc,  firstnumber=11, numbers=left]
int main()
{
    x= malloc(sizeof(int));
    pthread_t t;
    pthread_create(&t, NULL, thread, NULL);
    pthread_mutex_lock(&m);
        free(x);
    pthread_mutex_unlock(&m);
    return 0;
}
\end{lstlisting}

%% file: examples/ex_castellan.tex
\scalebox{0.9}{\begin{tikzpicture}
\usetikzlibrary{decorations.pathmorphing}

\tikzset{snake it/.style={decorate, decoration=snake}}
\tikzstyle{cau}=[-,thick]
\tikzstyle{conf}=[snake it,thick]
\tikzstyle{transition}=[rectangle, draw=none,thick,minimum size=5mm]
\node[transition] (a) at (0,2)  {$\mathtt{Re}_a^{(r=2)}$}
;
\node[transition] (b) at (0,0) {$\mathtt{Wr}_a^{(2)}$}
edge[cau] (a)
;

\node[transition] (c) at (2,2)  {$\mathtt{Wr}_a^{(2)}$}
;
\node[transition] (d) at (2,0) {$\mathtt{Re}_a^{(r=0)}$}
edge[cau] (c)
edge[conf] (b);
;

\node[transition] (c) at (4,0)  {$\mathtt{Wr}_b^{(1)}$}
;
\end{tikzpicture}}

%% file: examples/ex_as_castellan.tex
\scalebox{0.9}{\begin{tikzpicture}
\usetikzlibrary{decorations.pathmorphing}

\tikzset{snake it/.style={decorate, decoration=snake}}
\tikzstyle{cau}=[-,thick]
\tikzstyle{conf}=[snake it,thick]
\tikzstyle{wcau}=[dashed,->,thick,draw=red]

\tikzstyle{transition}=[rectangle, draw=none,thick,minimum size=5mm]
\node[transition] (a) at (0,2)  {$\mathtt{Re}_a^{(r=2)}$}
edge[wcau, bend left = 30] (d)
;
\node[transition] (b) at (0,0) {$\mathtt{Wr}_a^{(2)}$}
edge[cau] (a)
;

\node[transition] (d) at (2,0) {$\mathtt{Re}_a^{(r=0)}$}
edge[wcau, bend left = 20] (b)
edge[wcau, bend left = 20] (a)

;

\node[transition] (c) at (4,0)  {$\mathtt{Wr}_b^{(1)}$}
;
\end{tikzpicture}}

%% file: examples/ex_speculative.tex
\begin{tikzpicture}[scale=.8]
\tikzstyle{inhibitorred}=[o-, draw=red,thick]
\tikzstyle{pre}=[<-,thick]
\tikzstyle{post}=[->,thick]
\tikzstyle{readblue}=[-, draw=blue,thick]
\tikzstyle{rev}=[-, draw=gray,thick]
\tikzstyle{prerev}=[<-,thick,draw=gray]
\tikzstyle{postrev}=[->,thick,draw=gray]

\tikzstyle{transition}=[rectangle, draw=black,thick,minimum size=5mm]
\tikzstyle{place}=[circle, draw=black,thick,minimum size=5mm]
\node[place,tokens=1] (p1) at (1,2.5) [label=above:$s_1$] {};
\node[place,tokens=1] (p2) at (3,2.5) [label=above:$s_2$] {};
\node[place,tokens=1] (p3) at (5,2.5) [label=above:$s_3$] {};
\node[place,tokens=1] (p4) at (8.5,2.5) [label=above:$s_4$] {};
\node[place,tokens=1] (p5) at (12,2.5) [label=above:$s_5$] {};

\node[place] (p6) at (1,0) [label=below:${s_6}$] {};
\node[place] (p7) at (3,0) [label=below:${s_7}$] {};
\node[place] (p8) at (5,0) [label=below:${s_{8}}$] {};
\node[place] (p9) at (8.5,0) [label=below:${s_{9}}$] {};
\node[place] (p10) at (12,0) [label=below:${s_{10}}$] {};

\node[transition] (t1) at (1,1.25)  {$prod$}
edge[pre] (p1)
edge[post](p6)
;
\node[transition] (t2) at (3,1.25)  {$wait$}
edge[pre] (p2)
edge[post](p7)
edge[inhibitorred] (p8)
;

\node[transition] (t3) at (5,1.25)  {$pred$}
edge[pre] (p3)
edge[post](p8)
edge[inhibitorred] (p7)
edge[inhibitorred] (p6)

;

\node[rev] (t4) at (6.5,1.25)  {$\underline{pred}$}
edge[postrev] (p3)
edge[prerev](p8)
edge[inhibitorred, bend right=40] (p10)
edge[inhibitorred] (p9)
edge[inhibitorred, bend right=15] (p3)

;

\node[transition] (t5) at (8.5,1.25)  {$cF$}
edge[inhibitorred, bend right=20] (p3)
edge[pre] (p4)
edge[post](p9)
edge[inhibitorred, bend right] (p10)

;

\node[rev] (t5) at (10,1.25)  {$\underline{cF}$}
edge[inhibitorred, bend right] (p4)
edge[postrev] (p4)
edge[prerev](p9)

;

\node[transition] (t6) at (12,1.25)  {$cT$}
edge[inhibitorred, bend right=40] (p3)
edge[pre] (p5)
edge[post](p10)
edge[inhibitorred] (p9)

;

\end{tikzpicture}

%% file: conclusions.tex

\section{Conclusions}\label{sec:conc}

In this paper we complete previous efforts~\cite{PhilippouP22a,PhilippouP22,lics} aimed at relating classes of reversible event structures with 
classes of Petri nets: Firstly, we account for the full class of \raeses instead of proper subclasses (being \raeses  the 
most general reversible event structures considered in the literature). Secondly, 
 the correspondence is established according to the standard technique of exhibiting a coreflection between suitable categories. 
 Regarding the philosophy behind our nets, it may be asked whether we adhere to
the \emph{individual token philosophy} or the \emph{collective token
philosophy}, as defined in~\cite{robindividual}. In the individual token
philosophy, each place can be uniquely marked, whereas in the collective token
philosophy, there can be multiple ways to place a token in a given place.
If we only consider the forward flow of our nets, they adhere to the
individual token philosophy. However, asymmetric conflicts suggest that the
\emph{past} of the transition putting a token in a given place may not be
unique.
Moreover, when considering the reversing flow, things become more complex,
particularly with out-of-causal order reversibility, as a forward transition
may only be executed because some reversing transition has been
executed.
To settle this question, it may be necessary to seek a different definition
for the individual token philosophy, perhaps by focusing on the forward
dependencies of the pre-asymmetric causal net in the \racn{}, rather than
considering the entire \racn.

Besides the theoretical relevance of establishing a correspondence between these two different models, such 
connection may be exploited in concrete scenarios as discussed in the previous section.

%% file: appendix.tex
In this appendix we prove that the notions of \raes{} and \acn{} we used are consistent to what is 
established in literature (\cite{lics}, \cite{rpes} and
\cite{GPY:categories})

\input{appen-prel}

\input{appen-pcnvspacn}

\input{appen-raes}

%% file: appen-prel.tex
We start fixing notation and introducing auxiliary definitions concerning nets that will be useful in the following. 
Given an \fs\ $\sigma$, $\mathcal{X}_{\sigma}$ is the set of all sequences of multisets of 
transitions that \emph{agree} with $\sigma$, namely the set 
$\setcomp{\theta \!}{\!\len{\theta}\! = \! \len{\sigma}\ \!\!\: \land\ \!\!\: (\sigma(i)\trans{\theta(i)}\sigma(i+1)\
\mathit{with}\ i<\len{\sigma})}$, and 
$X_{\sigma} = \setcomp{\sum_{i=1}^{\len{\theta}} \theta(i)}{\theta\in \mathcal{X}_{\sigma}}$
for the set of multisets of transitions associated to an \fs. 
Each multiset of $X_{\sigma}$ is a \emph{state} of the net and write
\(
   \states{N} = \bigcup_{\sigma\in\firseq{N}{\mathsf{m}}} X_{\sigma}
\) 
for the set of states of the net $N$, and 
$\theta\in \mathcal{X}_{\sigma}$ is an \emph{execution} of the net.

%% file: appen-pcnvspacn.tex

\subsection{Causal Nets and Asymmetric Causal Nets}
In \cite{lics} we introduced the notion of \emph{causal net} to show that it was the proper
kind of net corresponding to reversible prime event structures of \cite{rpes}. In that
paper we proved that each occurrence nets, the classical counterpart of prime event structures in 
net terms, could be seen as a causal net. 
Here we show that each causal net  can be seen as an asymmetric causal nets, and this implies that
the tight correspondence between causal nets and reversible prime event structures can be 
transferred to \racn. In that paper we did not considered causal nets in categorical terms.

\input{pcnversuspacn}

%% file: pcnversuspacn.tex
We first recall the definition of causal net from \cite{lics}.

\begin{defi}\label{de:pre-causal-net-old}
 Let $C = \langle S, T, F, I, \mathsf{m}\rangle$ be an \inet.
 $C$ is a \emph{causal} net (\ca) if the following conditions are
 satisfied:
 \begin{enumerate}
   \item $\forall t, t'\in T.\ \post{t}\cap\pre{t'} = \emptyset$;
   \item $\flt{\post T} = \post T$;
   \item $\forall t\neq t'\in T.\ {\pre{t}} \cap{\pre{t'}} \cap{\Inib{T}{}} = \emptyset$,
   \item $\forall t\in T$. $\inib{t}$ is finite;
   \item $\lessdot$ is an irreflexive partial order;
   \item $\forall t', t''\in \histtwo{t}{\lessdot}.\  t' \cnconf t'' 
         \Rightarrow\ t' = t''$; 
   \item $\forall t, t', t''\in T.\ t\cnconf t' 
         \land\
         t'\lessdot t''\ \Rightarrow\ t \cnconf t'' 
         $;      and
   \item $\mathsf{m} = \pre{T}$ and $\Inib{T}{}\subseteq \mathsf{m}$.
 \end{enumerate}
\end{defi}
\noindent 
The requirement that $\forall t\in T$. $\inib{t}$ is finite implies that $\histtwo{t}{\lessdot}$ is
finite and $\forall t', t''\in \histtwo{t}{\lessdot}.\  t' \cnconf t''\ \Rightarrow\ t' = t''$ is 
analogous to the requirement that the inverse prevention relation and the causality one are 
acyclic on $\histtwo{t}{\lessdot}$.

\begin{prop}\label{pr:pcntopacn}
  Let $C = \langle S, T, F, I, \mathsf{m}\rangle$ be a \ca{} and let $S_{\#} \subseteq S$ be the
  shared places, i.e.\  $S_{\#} = \setcomp{s\in S}{|\post{s}|>1}$. Define $\mathcal{W}(C) = 
  \langle S', T, F', I', \mathsf{m}'\rangle$ be the \inet{} where 
  \begin{enumerate}
   \item $S' = S\setminus S_{\#}$; 
   \item $F' = F \cap ((S'\times T) \cup (T'\times S'))$;
   \item $I' = I \cup \setcomp{(\pre{t},t')}{\exists s\in S_{\#}. t\neq t'\ \land\ t, t'\in\post{s}}$; and
   \item $\mathsf{m}' = \mathsf{m}\cap S'$.
  \end{enumerate}
  Then $\mathcal{W}(C)$ is an \aca{} and $\states{C} = \states{\mathcal{W}(C)}$.
\end{prop}

\ifreport
\begin{proof}
 $\mathcal{W}(C)$ is a \aca{} as $\forall t\in T$. $\inib{t}$ finite means that $\histtwo{t}{\lessdot}$
 is finite and requiring that $\histtwo{t}{\lessdot}$ does not contain conflicting transitions account
 to prescribe that the reverse prevention and the causality are acyclic on the set 
 of transitions $\histtwo{t}{\lessdot}$. The other requirements are trivial.
 
 To show that $\states{C} = \states{\mathcal{W}(C)}$ it is enough to observe that each transition can be executed 
 just once and that, given any marking $m \in \reachMark{C}$ and any multiset of transitions $A\in \mu T$
 such that $m\trans{A}$ there exists a marking $\widetilde{m}\in \reachMark{\mathcal{W}(C)}$ such that
 $\widetilde{m}\trans{A}$. Given a marking $m \in \reachMark{C}$, the associated marking 
 $\widetilde{m}$ is $m\cap S'$. Now consider  $A\in \mu T$ such that $m\trans{A}$, we have to show
 $\widetilde{m}\trans{A}$. Clearly $\pre{A}\subseteq m$ and this implies that 
 $\pre{A}\subseteq \widetilde{m}$, now assume that there is a transition $t\in A$ such that there exists a
 place $s\in S_{\#}$ such that 
 $s\in \pre{t}$. This implies that for all $t'\in\post{s}\setminus\setenum{t}$ $t'\not\in A$ and
 $\post{t'}\not\in m$, hence $\widetilde{m}\trans{A}$. It is straightforward to observe that if
 $m\trans{A}m'$ then $\widetilde{m}\trans{A}\widetilde{m'}$. 
 Consider now a marking $m\in \reachMark{\mathcal{W}(C)}$ and a step $A\in \mu T$ such that $m\trans{A}$, 
 we show that there exists a marking $\widehat{m}\in \reachMark{C}$ such that $\widehat{m}\trans{A}$.
 Consider $m\in \reachMark{\mathcal{W}(C)}$, the transitions executed to reach that marking
 are $T'\subseteq T$ such that for each $t\in T'$. $\post{t}\subseteq m$. 
 Define $\widehat{m} = m \cup \widehat{S_{\#}}$ where 
 $\widehat{S_{\#}} = \setcomp{s\in S_{\#}}{s\not\in\post{T'}}$. It is easy to see that
 if $m\trans{A}$ then $\widehat{m}\trans{A}$ reasoning as above, and observing that
 $\mathsf{m} = \widehat{\mathsf{m}'}$ we have that $\widehat{m}\in \reachMark{C}$.
 Observing that all the steps in 
 a \ca{} (and in a \aca{}) are indeed execution of set of transitions and that each transition is
 fired just once, we can conclude that $\states{C} = \states{\mathcal{W}(C)}$.
\end{proof}
\fi

As we have a precise correspondence between reachable markings of a \ca{} and of the associated
\aca{} and the states coincide, we can conclude that \acas are a proper generalisation of
\ca{s}.

%% file: appen-raes.tex
\subsection{Equivalence of the definitions of \raeses}
The definition of reversible asymmetric event structure given in \cite{rpes} and
\cite{GPY:categories} has just two relations which, as we already said, comprises 
both the forward and the reverse causality and prevention. 
To avoid confusion we call them \emph{standard reversible asymmetric event structures}.

\begin{defi}\label{de:raes-old}
 A \emph{standard reversible asymmetric event structures} (s\raes) is the quadruple 
 $\mathsf{K} = (E, \anR, \prec, \lhd)$ 
 where $E$ is the set of events and
 \begin{enumerate}
   \item\label{cond:1biso} $\anR \subseteq E$ is the set of \emph{reversible} events;
   \item\label{cond:3biso} $\prec \subseteq E \times (E\cup\un{U})$ is an irreflexive causation relation; 
   \item\label{cond:2biso} $\lhd \subseteq (E\cup\un{\anR}) \times E$ is an 
         irreflexive \emph{precedence} relation
         such that for all $\alpha\in E\cup\un{\anR}.\ \setcomp{e\in E}{e\prec \alpha}$ is finite and
         acyclic with respect to $\lhd\cup\prec$;
   \item\label{cond:4biso} $\forall \anr\in \anR.\ \anr \prec \un{\anr}$;
   \item\label{cond:5biso} for all $e\in E$ and $\alpha\in E\cup\un{\anR}$, if $e\prec \alpha$ 
         then not $\alpha\lhd e$; and
   \item\label{cond:6biso} the relation $\pprec$, 
        defined as $e \pprec e'$ when $e \prec e'$ and if $e = \anr$, 
         with $\anr\in \anR$, then 
         $\un{\anr}\lhd e'$, is such that
         \begin{itemize}
           \item $e\pprec e'$ implies $e\lhd e'$;
           \item it is a transitive relation; and
           \item if  $e \# e'$ and $e \pprec e''$ then $e' \# e''$, where $\# = \lhd\cap\rhd$.
         \end{itemize}
 \end{enumerate}
\end{defi}
\noindent 
In this definition $\prec$ comprises the forward causality (which we called \emph{causation})
and the reverse causality, and $\lhd$ comprises the weak causality (forward relation) 
and the prevention involving the undoing of events. 
Just observing that the relations on subsets of events reduces
always to the part concerning \emph{forward} relations, and that the last condition of the previous
definition  
can be rewritten just requiring that, 
when restricted to forward events only, 
sustained causation and the restriction of the prevention to these events is an \aes{},
it is straightforward to see that the two proposition below hold.

\begin{prop}\label{pr:raesdefcorruno}
 Let $\mathsf{K} = (E, \anR, \prec, \lhd)$ be an s\raes, then 
 $\mathsf{H} = (E, \anR, <_{\mathsf{H}}, \nearrow_{\mathsf{H}}, \prec_{\mathsf{H}}, \lhd_{\mathsf{H}})$ is an
 \raes, where $<_{\mathsf{H}} = \prec\cap(E\times E)$, $\nearrow_{\mathsf{H}} = \lhd\cap(E\times E)$,
 $\prec_{\mathsf{H}} = \prec\cap(E\times \un{\anR})$ and $\lhd_{\mathsf{H}} = \lhd\cap(\un{\anR}\times E)$.
\end{prop}

\begin{prop}\label{pr:raesdefcorrdue}
 Let $\mathsf{H} = (E, \anR, <, \nearrow, \prec, \lhd)$ be an \raes, then 
  $\mathsf{K} = (E, \anR, <\cup\prec, \nearrow\cup\lhd)$ is an
 s\raes.
\end{prop}
The proofs of both propositions are trivial and omitted.

%% file: main.bbl
\begin{thebibliography}{10}

\bibitem{revbook}
B.~Aman, G.~Ciobanu, R.~Gl{\"{u}}ck, R.~Kaarsgaard, J.~Kari, M.~Kutrib,
  I.~Lanese, C.~A. Mezzina, L.~Mikulski, R.~Nagarajan, I.~C.~C. Phillips, G.~M.
  Pinna, L.~Prigioniero, I.~Ulidowski, and G.~Vidal.
\newblock Foundations of reversible computation.
\newblock In I.~Ulidowski, I.~Lanese, U.~P. Schultz, and C.~Ferreira, editors,
  {\em Reversible Computation: Extending Horizons of Computing - Selected
  Results of the {COST} Action {IC1405}}, volume 12070 of {\em Lecture Notes in
  Computer Science}, pages 1--40. Springer, 2020.

\bibitem{BBCP:rivista}
P.~Baldan, N.~Busi, A.~Corradini, and G.~M. Pinna.
\newblock Domain and event structure semantics for {P}etri nets with read and
  inhibitor arcs.
\newblock {\em Theor. Comput. Sci.}, 323(1-3):129--189, 2004.

\bibitem{BCM01IC}
P.~Baldan, A.~Corradini, and U.~Montanari.
\newblock Contextual {P}etri nets, asymmetric event structures, and processes.
\newblock {\em Inf. Comput.}, 171(1):1--49, 2001.

\bibitem{flowEvent}
G.~Boudol.
\newblock Flow event structures and flow nets.
\newblock In {\em Semantics of Systems of Concurrent Processes, {LITP} Spring
  School on Theoretical Computer Science}, volume 469 of {\em Lecture Notes in
  Computer Science}, pages 62--95. Springer, 1990.

\bibitem{castellan}
S.~Castellan.
\newblock {Weak memory models using event structures}.
\newblock In J.~Signoles, editor, {\em {Vingt-septi{\`e}mes Journ{\'e}es
  Francophones des Langages Applicatifs (JFLA 2016)}}, Saint-Malo, France, Jan.
  2016.

\bibitem{rpi}
I.~Cristescu, J.~Krivine, and D.~Varacca.
\newblock A compositional semantics for the reversible $\pi$-calculus.
\newblock In {\em 28th Annual {ACM/IEEE} Symposium on Logic in Computer
  Science, {LICS}}, pages 388--397, 2013.

\bibitem{rccs}
V.~Danos and J.~Krivine.
\newblock Reversible communicating systems.
\newblock In {\em {CONCUR} 2004 - Concurrency Theory, 15th International
  Conference}, volume 3170 of {\em Lecture Notes in Computer Science}, pages
  292--307. Springer, 2004.

\bibitem{DanosK05}
V.~Danos and J.~Krivine.
\newblock Transactions in {RCCS}.
\newblock In {\em {CONCUR} 2005}, volume 3653 of {\em Lecture Notes in Computer
  Science}, pages 398--412. Springer, 2005.

\bibitem{GiachinoLM14}
E.~Giachino, I.~Lanese, and C.~A. Mezzina.
\newblock Causal-consistent reversible debugging.
\newblock In {\em Fundamental Approaches to Software Engineering - 17th
  International Conference, {FASE}}, volume 8411 of {\em Lecture Notes in
  Computer Science}, pages 370--384. Springer, 2014.

\bibitem{GiachinoLMT17}
E.~Giachino, I.~Lanese, C.~A. Mezzina, and F.~Tiezzi.
\newblock Causal-consistent rollback in a tuple-based language.
\newblock {\em J. Log. Algebraic Methods Program.}, 88:99--120, 2017.

\bibitem{GPY:categories}
E.~Graversen, I.~Phillips, and N.~Yoshida.
\newblock Towards a categorical representation of reversible event structures.
\newblock {\em J. Log. Algebraic Methods Program.}, 104:16--59, 2019.

\bibitem{GraversenPY21}
E.~Graversen, I.~C.~C. Phillips, and N.~Yoshida.
\newblock Event structure semantics of (controlled) reversible {CCS}.
\newblock {\em J. Log. Algebraic Methods Program.}, 121:100686, 2021.

\bibitem{JeffreyR19}
A.~Jeffrey and J.~Riely.
\newblock On thin air reads: Towards an event structures model of relaxed
  memory.
\newblock {\em Log. Methods Comput. Sci.}, 15(1), 2019.

\bibitem{Leeman86}
G.~B.~L. Jr.
\newblock A formal approach to undo operations in programming languages.
\newblock {\em {ACM} Trans. Program. Lang. Syst.}, 8(1):50--87, 1986.

\bibitem{landaurer}
R.~Landauer.
\newblock Irreversibility and heat generation in the computing process.
\newblock {\em IBM Journal of Research and Development}, 5(3):183--191, 1961.

\bibitem{LaneseLMSS13}
I.~Lanese, M.~Lienhardt, C.~A. Mezzina, A.~Schmitt, and J.~Stefani.
\newblock Concurrent flexible reversibility.
\newblock In {\em Programming Languages and Systems - 22nd European Symposium
  on Programming, {ESOP} 2013}, volume 7792 of {\em Lecture Notes in Computer
  Science}, pages 370--390. Springer, 2013.

\bibitem{LaneseMM21}
I.~Lanese, D.~Medic, and C.~A. Mezzina.
\newblock Static versus dynamic reversibility in {CCS}.
\newblock {\em Acta Informatica}, 58(1-2):1--34, 2021.

\bibitem{rhopi}
I.~Lanese, C.~A. Mezzina, and J.~Stefani.
\newblock Reversibility in the higher-order {\(\pi\)}-calculus.
\newblock {\em Theor. Comput. Sci.}, 625:25--84, 2016.

\bibitem{Lanese0PV18}
I.~Lanese, N.~Nishida, A.~Palacios, and G.~Vidal.
\newblock Cauder: {A} causal-consistent reversible debugger for {E}rlang.
\newblock In J.~P. Gallagher and M.~Sulzmann, editors, {\em Functional and
  Logic Programming {FLOPS} 2018, Nagoya, Japan, May 9-11, 2018, Proceedings},
  volume 10818 of {\em Lecture Notes in Computer Science}, pages 247--263.
  Springer, 2018.

\bibitem{MedicMPY20}
D.~Medic, C.~A. Mezzina, I.~Phillips, and N.~Yoshida.
\newblock A parametric framework for reversible \emph{{\(\pi\)}}-calculi.
\newblock {\em Inf. Comput.}, 275:104644, 2020.

\bibitem{lics}
H.~C. Melgratti, C.~A. Mezzina, and G.~M. Pinna.
\newblock A distributed operational view of reversible prime event structures.
\newblock In {\em 36th Annual {ACM/IEEE} Symposium on Logic in Computer
  Science, {LICS}}, pages 1--13. {IEEE}, 2021.

\bibitem{forte}
H.~C. Melgratti, C.~A. Mezzina, and G.~M. Pinna.
\newblock Relating reversible petri nets and reversible event structures,
  categorically.
\newblock In M.~Huisman and A.~Ravara, editors, {\em Formal Techniques for
  Distributed Objects, Components, and Systems - 43rd {IFIP} {WG} 6.1
  International Conference, {FORTE} 2023,}, volume 13910 of {\em Lecture Notes
  in Computer Science}, pages 206--223. Springer, 2023.

\bibitem{tocl}
H.~C. Melgratti, C.~A. Mezzina, and G.~M. Pinna.
\newblock A reversible perspective on petri nets and event structures.
\newblock {\em {ACM} Trans. Comput. Log.}, 25(4):1--38, 2024.

\bibitem{MelgrattiMP24}
H.~C. Melgratti, C.~A. Mezzina, and G.~M. Pinna.
\newblock A truly concurrent semantics for reversible {CCS}.
\newblock {\em Log. Methods Comput. Sci.}, 20(4), 2024.

\bibitem{revpt}
H.~C. Melgratti, C.~A. Mezzina, and I.~Ulidowski.
\newblock Reversing place transition nets.
\newblock {\em Log. Methods Comput. Sci.}, 16(4), 2020.

\bibitem{wg2}
C.~A. Mezzina, R.~Schlatte, R.~Gl{\"{u}}ck, T.~Haulund, J.~Hoey, M.~H.
  Cservenka, I.~Lanese, T.~{\AE}. Mogensen, H.~Siljak, U.~P. Schultz, and
  I.~Ulidowski.
\newblock Software and reversible systems: {A} survey of recent activities.
\newblock In I.~Ulidowski, I.~Lanese, U.~P. Schultz, and C.~Ferreira, editors,
  {\em Reversible Computation: Extending Horizons of Computing - Selected
  Results of the {COST} Action {IC1405}}, volume 12070 of {\em Lecture Notes in
  Computer Science}, pages 41--59. Springer, 2020.

\bibitem{MezzinaTY25}
C.~A. Mezzina, F.~Tiezzi, and N.~Yoshida.
\newblock Checkpoint-based rollback recovery in session programming.
\newblock {\em Log. Methods Comput. Sci.}, 21(1):2, 2025.

\bibitem{MR:CN}
U.~Montanari and F.~Rossi.
\newblock Contextual nets.
\newblock {\em Acta Informatica}, 32(6):545--596, 1995.

\bibitem{NielsenPW79}
M.~Nielsen, G.~D. Plotkin, and G.~Winskel.
\newblock {P}etri nets, event structures and domains.
\newblock In {\em Semantics of Concurrent Computation}, volume~70, pages
  266--284. Springer, 1979.

\bibitem{PhilippouP22a}
A.~Philippou and K.~Psara.
\newblock A collective interpretation semantics for reversing {Petri} nets.
\newblock {\em Theor. Comput. Sci.}, 924:148--170, 2022.

\bibitem{PhilippouP22}
A.~Philippou and K.~Psara.
\newblock Reversible computation in nets with bonds.
\newblock {\em J. Log. Algebraic Methods Program.}, 124:100718, 2022.

\bibitem{rpes}
I.~Phillips and I.~Ulidowski.
\newblock Reversibility and asymmetric conflict in event structures.
\newblock {\em J. Log. Algebraic Methods Program.}, 84(6):781--805, 2015.

\bibitem{PhillipsUY13}
I.~Phillips, I.~Ulidowski, and S.~Yuen.
\newblock Modelling of bonding with processes and events.
\newblock In G.~W. Dueck and D.~M. Miller, editors, {\em Reversible Computation
  - 5th International Conference, {RC} 2013,}, volume 7948 of {\em Lecture
  Notes in Computer Science}, pages 141--154. Springer, 2013.

\bibitem{ccsk}
I.~C.~C. Phillips and I.~Ulidowski.
\newblock Reversing algebraic process calculi.
\newblock {\em J. Log. Algebraic Methods Program.}, 73(1-2):70--96, 2007.

\bibitem{speculative}
P.~Prabhu, G.~Ramalingam, and K.~Vaswani.
\newblock Safe programmable speculative parallelism.
\newblock In {\em Proceedings of the 2010 {ACM} {SIGPLAN} Conference on
  Programming Language Design and Implementation, {PLDI} 2010}, pages 50--61.
  {ACM}, 2010.

\bibitem{robindividual}
R.~J. van Glabbeek.
\newblock The individual and collective token interpretations of {P}etri nets.
\newblock In M.~Abadi and L.~de~Alfaro, editors, {\em {CONCUR} 2005 -
  Concurrency Theory, 16th International Conference, {CONCUR} 2005, San
  Francisco, CA, USA, August 23-26, 2005, Proceedings}, volume 3653 of {\em
  Lecture Notes in Computer Science}, pages 323--337. Springer, 2005.

\bibitem{VassorS18}
M.~Vassor and J.~Stefani.
\newblock Checkpoint/rollback vs causally-consistent reversibility.
\newblock In {\em Reversible Computation - 10th International Conference, {RC}
  2018}, volume 11106 of {\em Lecture Notes in Computer Science}, pages
  286--303. Springer, 2018.

\bibitem{Vidal23}
G.~Vidal.
\newblock From reversible computation to checkpoint-based rollback recovery for
  message-passing concurrent programs.
\newblock In J.~C{\'{a}}mara and S.~Jongmans, editors, {\em Formal Aspects of
  Component Software - 19th International Conference, {FACS} 2023}, volume
  14485 of {\em Lecture Notes in Computer Science}, pages 103--123. Springer,
  2023.

\bibitem{Win:ES}
G.~Winskel.
\newblock Event structures.
\newblock In W.~Brauer, W.~Reisig, and G.~Rozenberg, editors, {\em {P}etri
  Nets: Central Models and Their Properties, Advances in {P}etri Nets 1986,
  Part II, Proceedings of an Advanced Course, Bad Honnef, Germany, 8-19
  September 1986}, volume 255 of {\em Lecture Notes in Computer Science}, pages
  325--392. Springer, 1986.

\bibitem{Win:PNAMC}
G.~Winskel.
\newblock {P}etri nets, algebras, morphisms, and compositionality.
\newblock {\em Inf. Comput.}, 72(3):197--238, 1987.

\bibitem{YokoyamaG07}
T.~Yokoyama and R.~Gl{\"{u}}ck.
\newblock A reversible programming language and its invertible
  self-interpreter.
\newblock In G.~Ramalingam and E.~Visser, editors, {\em Proceedings of the 2007
  {ACM} {SIGPLAN} Workshop on Partial Evaluation and Semantics-based Program
  Manipulation}, pages 144--153. {ACM}, 2007.

\end{thebibliography}
